\documentclass[usenatbib]{mn2e}
\usepackage{graphicx}
\usepackage{rotating}
\usepackage{txfonts}
\usepackage[british]{babel}
\usepackage{dcolumn}
\newcolumntype{d}{D{.}{.}{-1}}
\def\dhead#1{\multicolumn{1}{c}{#1}}
\def\twolines#1#2{$\kern-6pt\Big\{ {\textrm{#1\hfill}\atop\textrm{#2\hfill}}$}
\def\vpad{{\Large$\mathstrut$}}
\usepackage{multirow}
\usepackage{supertabular}
\usepackage{lscape}
\usepackage{hyperref}
\graphicspath{{.}{./FIGS/}}
\hypersetup{pdfauthor={I. H. Whittam, J. M. Riley, D. A. Green, M. J. Jarvis and M. Vaccari},
               pdftitle={The faint radio source population at 15.7~GHz -- II. Multi-wavelength properties},
               pdfkeywords={galaxies: active, radio continuum: galaxies, catalogues, surveys},
               bookmarksnumbered=true}
\hypersetup{colorlinks=true,
            linkcolor=blue,
            citecolor=blue,
            filecolor=blue,
            urlcolor=blue}
\volume{{\rm in press}}

\title[The faint radio source population at 15.7~GHz -- II. Multi-wavelength properties]{The faint radio source population at 15.7~GHz -- II. Multi-wavelength properties}

\author[I.~H.~Whittam et al.]{I.~H.~Whittam$^{1,2}$\thanks{email:
\texttt{i.whittam@mrao.cam.ac.uk}}, J.~M.~Riley$^2$, D.~A.~Green$^2$, M. J. Jarvis$^{1,3}$ and M. Vaccari$^{1,4}$\\
   $^{1}$Physics Department, University of the Western Cape, Bellville 7535, South Africa\\
   $^{2}$Astrophysics Group, Cavendish Laboratory, 19 J.~J.~Thomson Avenue, Cambridge CB3 0HE\\
   $^{3}$Astrophysics, University of Oxford, Denys Wilkinson Building, Keble Road, Oxford, OX1 3RH\\
   $^{4}$INAF - Istituto di Radioastronomia, via Gobetti 101, 40129 Bologna, Italy}

\date{Accepted ---; received ---; in original form ---}

\pagerange{\pageref{firstpage}--\pageref{lastpage}}

\pubyear{2015}
\begin{document}

\label{firstpage}

\maketitle

\begin{abstract}
A complete, flux density limited sample of 96 faint ($> 0.5$~mJy) radio sources is selected from the 10C survey at 15.7~GHz in the Lockman Hole. We have matched this sample to a range of multi-wavelength catalogues, including SERVS, SWIRE, UKIDSS and optical data; multi-wavelength counterparts are found for 80 of the 96 sources and spectroscopic redshifts are available for 24 sources. Photometric reshifts are estimated for the sources with multi-wavelength data available; the median redshift of the sample is 0.91 with an interquartile range of 0.84. Radio-to-optical ratios show that at least 94 per cent of the sample are radio loud, indicating that the 10C sample is dominated by radio galaxies. This is in contrast to samples selected at lower frequencies, where radio-quiet AGN and starforming galaxies are present in significant numbers at these flux density levels. All six radio-quiet sources have rising radio spectra, suggesting that they are dominated by AGN emission. These results confirm the conclusions of Paper I that the faint, flat-spectrum sources which are found to dominate the 10C sample below $\sim 1$~mJy are the cores of radio galaxies. The properties of the 10C sample are compared to the SKADS Simulated Skies; a population of low-redshift starforming galaxies predicted by the simulation is not found in the observed sample.
\end{abstract}


\section{Introduction}\label{section:intro}

Most studies of the faint ($\lesssim 1$~mJy) radio sky have focused on frequencies around 1.4~GHz, due to the increased telescope time required to survey a field to an equivalent depth at a higher frequency. This means that while the composition of the lower-frequency radio sky is well constrained (e.g.\ \citealt{2009ApJ...694..235P,2010A&ARv..18....1D,2013MNRAS.436.1084M}), the faint radio sky at higher frequencies is relatively unstudied. The Tenth Cambridge (10C; \citealt{2011MNRAS.415.2708D,2011MNRAS.415.2699F}) survey at 15.7~GHz has covered $\approx 27$~deg$^2$ in ten different fields to a completeness limit of 1~mJy and a further 12~deg$^2$ to a completeness limit of 0.5~mJy. The 10C survey therefore provides the ideal starting point from which to study the faint, high-frequency sky. This survey has recently been extended to even fainter flux densities in two of the 10C fields by Whittam et al. (in prep), who calculated the 15~GHz source counts down to 0.1~mJy. 

To investigate the nature of the extragalactic radio source population multi-wavelength studies are required, as the power-law nature of radio spectra means that radio data alone is not sufficient to classify source types or estimate redshifts. In the first paper in this series studying the faint radio source population at 15.7~GHz (\citealt{2013MNRAS.429.2080W}, hereafter Paper~I) we discuss a sample of sources selected from the Lockman Hole for which data over a wide range of frequencies are available. The Lockman Hole is a region of the sky centred near 10$^{\rm h}$45$^{\rm m}$, +58$^{\circ}$ (J2000 coordinates, which are used throughout this work) with exceptionally low \textsc{Hi} column density \citep{1986ApJ...302..432L}. The low infrared background (0.38~MJy~sr$^{-1}$ at 100~$\muup$m; \citealt{2003PASP..115..897L}) in this area of the sky makes it ideal for deep extragalactic infrared observations. As a result, as part of the \emph{Spitzer} Wide-area Infrared Extragalactic survey (SWIRE; \citealt{2003PASP..115..897L}) sensitive infrared observations of $\approx$ 14 deg$^2$ of the Lockman Hole area have been made. The availability of deep infrared observations in the Lockman Hole has triggered deep observing campaigns at optical, X-ray and radio wavelengths (e.g.\ \citealt{2006MNRAS.371..963B,2008A&A...479..283B,2001PASJ...53..445I,2003PASP..115..897L,2012PASP..124..714M,2010AJ....140.1868W}). The availability of data at such a wide range of frequencies makes the Lockman Hole a particularly good area for a multi-wavelength study of the faint radio source population. 

In Paper~I we selected a sample of 296 sources from the 10C sample in the Lockman Hole. By matching this catalogue to several lower frequency surveys we have investigated the radio properties of the sources in this sample; all but 30 of the 10C sources are matched to sources in one or more of these surveys. We found a significant increase in the proportion of flat-spectrum sources at flux densities below $\approx$1 mJy -- the median spectral index between 15.7~GHz and 610~MHz changes from $\alpha =0.75$ for flux densities greater than 1.5~mJy to $\alpha =0.08$ for flux densities less than 0.8~mJy (the convention $S \propto \nu ^{-\alpha}$, for a source with flux density $S$ at frequency $\nu$, is used throughout this work). This suggests that a population of faint, flat-spectrum sources is emerging at flux densities $\lesssim 1 \textrm{ mJy}$ in the high-frequency sky.

In Paper~I the spectral index distribution of this sample of sources selected at 15.7~GHz was compared to those of two samples selected at 1.4~GHz from FIRST and NVSS. This showed that there is a significant flat-spectrum population present in the 10C sample which is missing from the samples selected at 1.4~GHz. The 10C sample was compared to a sample of sources selected from the Square Kilometre Array Design Studies (SKADS) Simulated Sky (S$^3$) by \citet{2008MNRAS.388.1335W,2010MNRAS.405..447W} and we found that this simulation fails to reproduce the observed spectral index distribution and significantly under-predicts the number of sources in the faintest flux density bin. It is likely that the observed faint, flat-spectrum sources are a result of the cores of Fanaroff and Riley type I (FRI; \citealt{1974MNRAS.167P..31F}) sources becoming dominant at high frequencies. These results highlight that the faint, high-frequency source population is poorly understood and therefore the importance of further study of this population.

In this paper, we select all sources from the 10C sample studied in Paper I which which have deep 1.4-GHz observations available (from the \citealt{2012rsri.confE..22G}, \citealt{2006MNRAS.371..963B} or \citealt{2008AJ....136.1889O} surveys, see Section \ref{section:radio-data} for full details) and have flux densities above the 10C completeness limit in the given region. This complete sample of 96 sources is matched to optical and infrared data available in the field. These data enable us to distinguish between different source types and estimate photometric redshifts for the objects. Crucially, by using radio-to-optical ratios we are able to separate radio-loud AGN, which we know dominate the extragalactic radio source population at higher flux densities ($S_{15~\rm GHz} \gtrsim~10~\rm mJy$) from radio-quiet AGN and starforming galaxies, which are predicted to begin to contribute to the source population at lower flux density levels (e.g.\ \citealt{2008MNRAS.388.1335W}). The range of multi-wavelength data available in the Lockman Hole is described in Section~\ref{section:multi-data} and the methods used to match these catalogues to the 10C catalogue are described in Section~\ref{section:fused-matching}. Photometric redshifts are calculated in Section~\ref{section:lephare} and combined with available spectroscopic redshifts to produce a final redshift catalogue. In Section~\ref{section:R} the radio-to-optical ratio is estimated for all sources in this sample. The properties of the sample are discussed in light of the redshift information in Section~\ref{section:properties_z}. The 10C sample is compared to the S$^3$ simulation in Section~\ref{section:s3_z} and to other observational studies in Section~\ref{section:other-studies}.

\section{Data used}\label{section:multi-data}

\subsection{Radio data}\label{section:radio-data}

The work in Paper I is based on a sample of sources selected from the 10C survey at 15.7~GHz. The 10C survey was observed with the Arcminute Microkelvin Imager (AMI; \citealt{2008MNRAS.391.1545Z}) which has a resolution of 30~arcsec. Full details of the 10C survey can be found in \citet{2011MNRAS.415.2708D} and \citet{2011MNRAS.415.2699F}. This work uses the Lockman Hole field of the survey, which consists of 4.64~deg$^2$ complete to 1~mJy and 1.73~deg$^2$ complete to 0.5~mJy. In Paper~I we investigated the radio properties of this sample by matching the catalogue to several lower-frequency catalogues available in the field; a deep Giant Meterwave Radio Telescope (GMRT) survey at 610~MHz \citep{2008MNRAS.387.1037G,2010BASI...38..103G}, a Westerbork Synthesis Radio Telescope (WSRT) survey at 1.4~GHz \citep{2012rsri.confE..22G}, two deep Very Large Array (VLA) surveys at 1.4~GHz by \citealt{2006MNRAS.371..963B} (BI2006) and \citealt{2008AJ....136.1889O} (OM2008), the National Radio Astronomy Observatory (NRAO) VLA Sky Survey (NVSS; \citealt{1998AJ....115.1693C}) and Faint Images of the Radio Sky at Twenty centimetres (FIRST; \citealt{1997ApJ...475..479W}). These radio data are summarised in Table \ref{tab:radio-data}. Lower-frequency counterparts were found for 266 out of the 296 sources in the sample, allowing radio spectral indices to be calculated for these 266 sources and upper limits to be placed on the spectral indices of the remaining 30 sources. 

\begin{figure}
\centerline{\includegraphics[width=\columnwidth,angle=270]{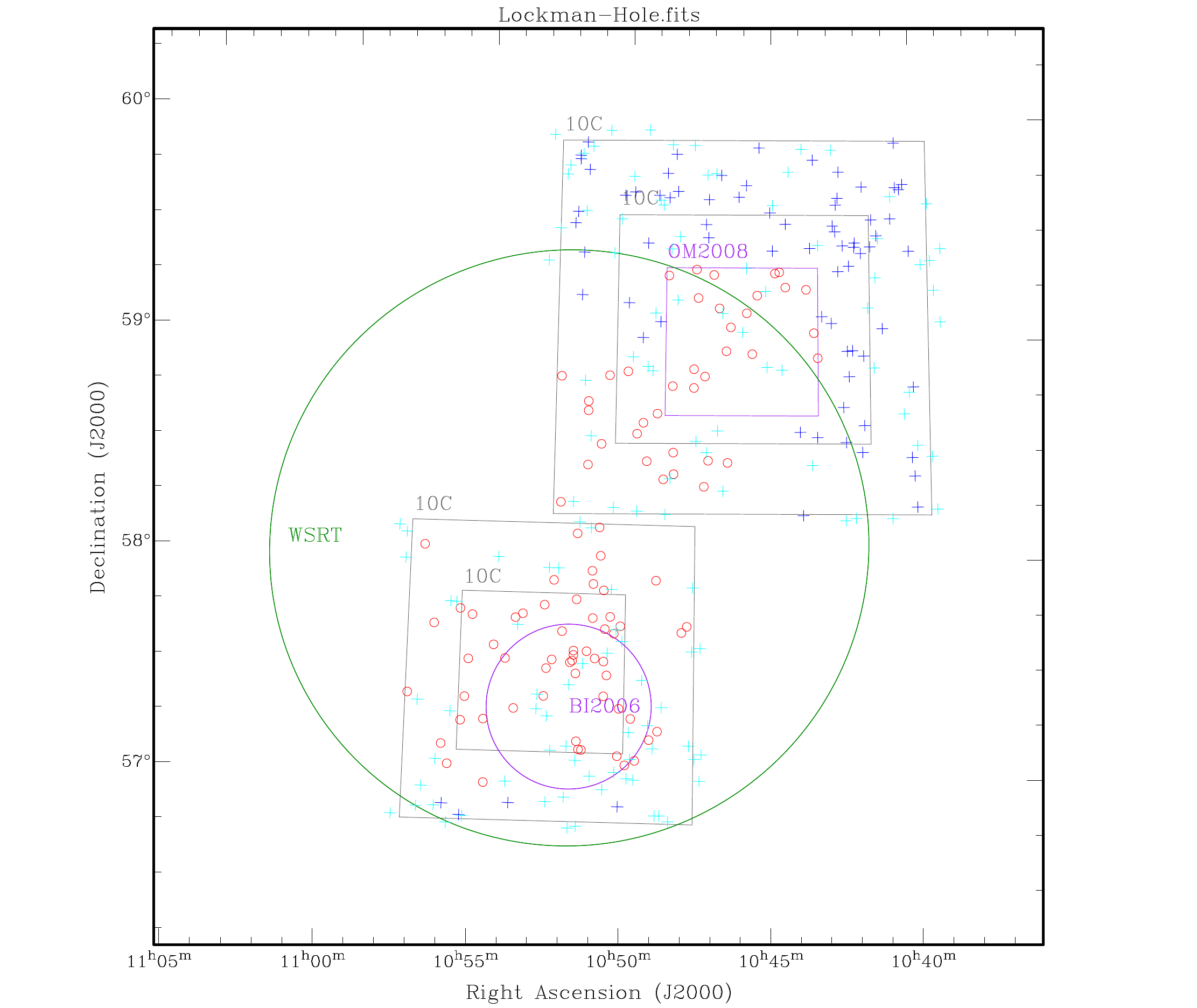}}
\caption{Positions of all 296 sources in the 10C Lockman Hole sample. The positions of the 96 sources in the sample used in this paper are shown as small circles (red in the online version), and the remaining 200 sources which are not in this sample are shown as crosses (blue in the online version). The sources which fall below the 10C completeness limit, and are therefore not included in the sample used in this paper, as shown as pale blue crosses. The 10C complete areas are shown by the rectangles (the large rectangles indicate the regions complete to 1~mJy and small rectangles contained within these show the regions complete to 0.5~mJy). The WSRT survey area is shown by the large (green) circle and the BI2006 and OM2008 survey areas are shown by the (purple) small circle and small square.}\label{fig:sampleW}
\end{figure}

\begin{table*}
\caption{Radio catalogues in the Lockman Hole used in Paper~I.}\label{tab:radio-data}
\bigskip
\begin{tabular}{lllddl}\hline
\vpad
 Catalogue & Reference(s) & Epoch of observation & \dhead{Frequency} & \dhead{Beam size} & rms noise \\
 &  &  & \dhead{/GHz} & \dhead{/arcsec} & /mJy \\\hline
  10C -- shallow  & \twolines{\citet{2011MNRAS.415.2699F}}{\citet{2011MNRAS.415.2708D}} & Aug 2008 -- June 2010  & 15.7 & 30 & 0.1  \\
  10C -- deep     & \twolines{\citet{2011MNRAS.415.2699F}}{\citet{2011MNRAS.415.2708D}} & Aug 2008 -- June 2010  & 15.7 & 30 & 0.05\\
  GMRT    & \twolines{\citet{2008MNRAS.387.1037G}}{\citet{2010BASI...38..103G}} & Jul 2004 -- Oct 2006 & 0.610 & \dhead{$6\times5$} & 0.06\\
  WSRT    & \citet{2012rsri.confE..22G}                     & Dec 2006 -- Jun 2007   & 1.4   & \dhead{$11\times9$}  & 0.011\\
  OM2008  & \citet{2008AJ....136.1889O}                 & Dec 2001 -- Jan 2004   & 1.4   & 1.6 & 0.0027 \\
  OMK2009 & \citet{2009AJ....137.4846O}                 & Feb 2006 -- Jan 2007   & 0.324 & 6   & 0.07 \\
  BI2006  & \citet{2006MNRAS.371..963B}                 & Jan 2001 -- Mar 2002   & 1.4   & 1.3 & 0.0046  \\
  FIRST   & \citet{1997ApJ...475..479W}                 & 1997 -- 2002   & 1.4   & 5   & 0.15   \\
  NVSS    & \citet{1998AJ....115.1693C}                 & 1997  & 1.4   & 45  & 0.45   \\\hline
\end{tabular}
\end{table*}

\begin{figure*}
\centerline{\includegraphics[width=\columnwidth]{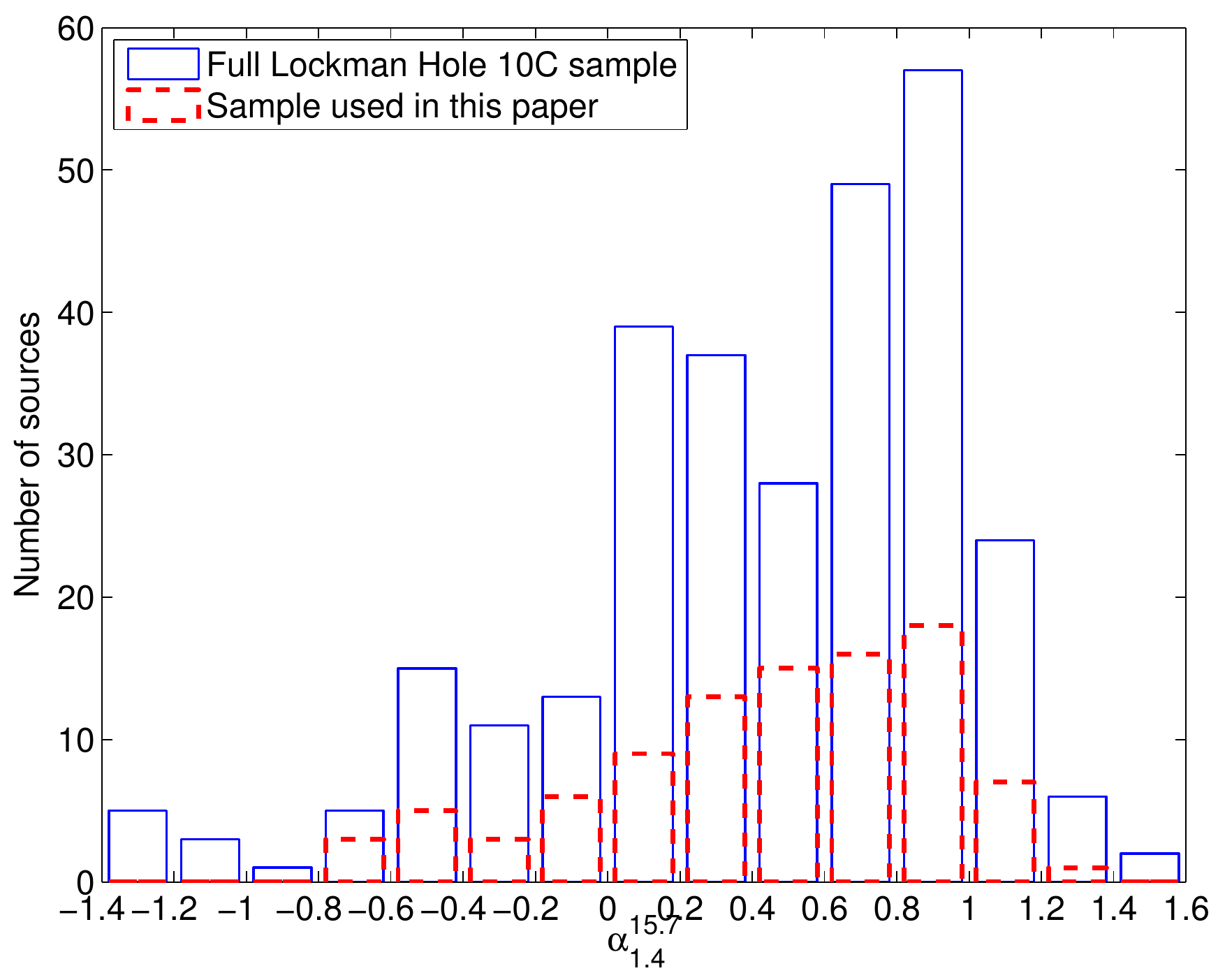}
            \quad
            \includegraphics[width=\columnwidth]{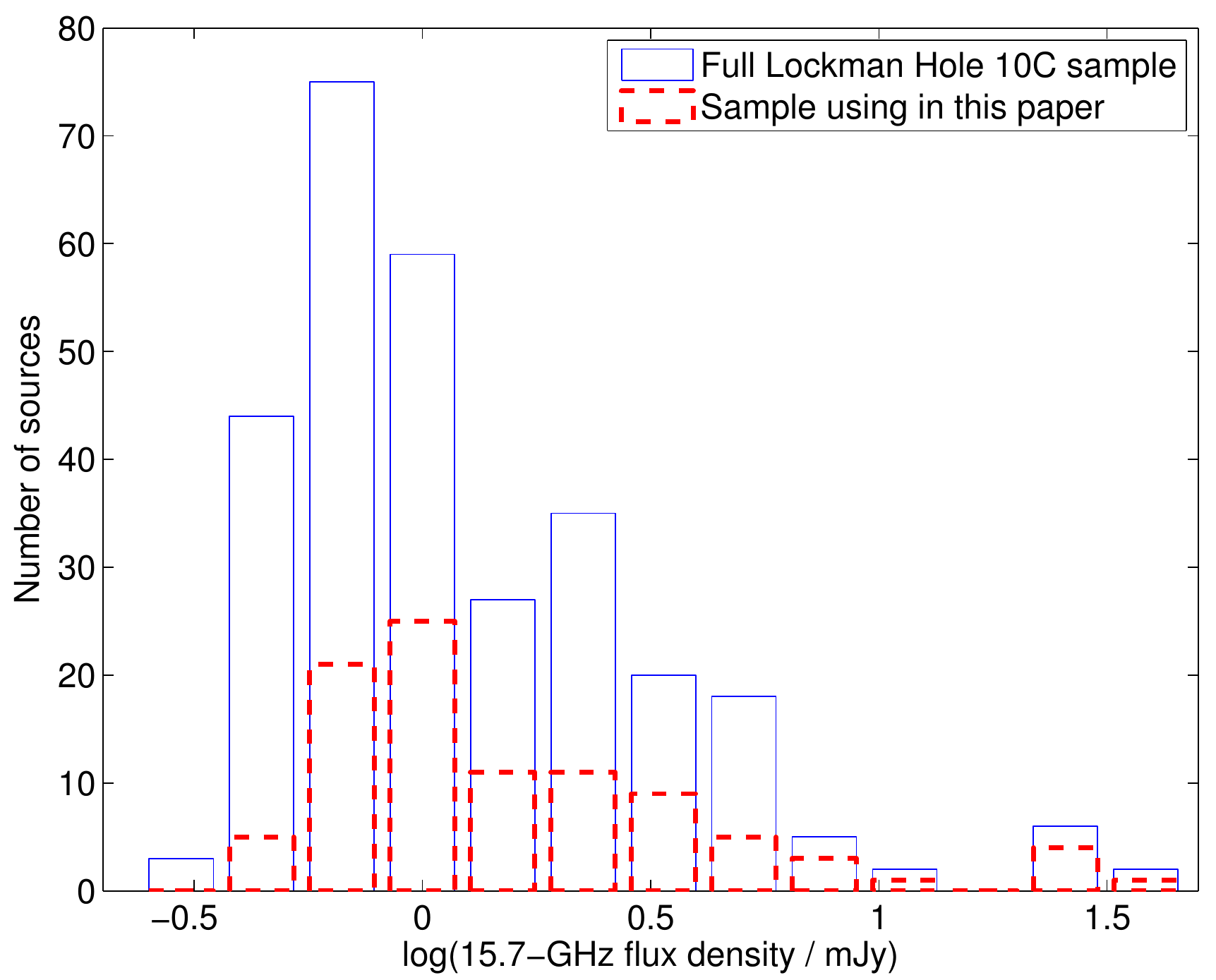}}
\caption{Spectral index and flux density distributions for sources in the full sample of 296 sources used in Paper~I and sample of 96 sources discussed in this paper. Left: spectral index, right: 15.7-GHz flux density. Note that the spectral index distribution for the full sample includes 30 sources with upper limits on their spectral indices.}\label{fig:sample_properties}
\end{figure*}

As many of these lower frequency catalogues have higher resolutions than the 10C survey, they also provide more accurate positions which are vital when searching for multi-wavelength counterparts for the 10C. It is therefore useful to define a complete sub-sample of sources which have 1.4-GHz data available. For this purpose, we used the BI2006, OM2008 and WSRT surveys as they are the deepest in the field. OM2008 and BI2006 have rms noises of 2.7 and 6.0 $\muup$Jy/beam respectively and all the 10C sources in these fields are detected at 1.4~GHz. The majority of the \citeauthor{2012rsri.confE..22G} WSRT map has an rms noise of $< 15~\muup$Jy/beam and parts have an rms noise of 11 $\muup$Jy/beam. This means that it should be possible to detect the faintest sources in the complete 10C sample (with $S_{15.7 \rm{GHz}} \geqslant 0.5~$mJy) in the WSRT map provided they have spectral indices $\alpha^{15.7}_{1.4} > -1$  (where $S \propto \nu ^{-\alpha}$), assuming a 3$\sigma$ WSRT detection. Sources with spectra which rise as steeply as this are very rare so the vast majority of the 10C sources in the WSRT map are detectable -- in fact, all but one of the sources in the complete 10C catalogue are detected in the WSRT map. We therefore define the sample as all 10C sources in the complete catalogue in the OM2008, BI2006 or WSRT deep survey areas. This sample contains 96 sources and accurate positions and spectral index information are therefore available for all but one of the sources. The positions of the sources in the sample are shown in Fig.~\ref{fig:sampleW}. This sample of 96 sources with deep 1.4 GHz data available is the subject of this paper.

The spectral indices and flux density distributions of the full sample of 296 sources studied in Paper~I and the sub-sample of 96 sources studied in this paper are plotted in Fig.~\ref{fig:sample_properties}. The spectral index distributions of the two samples are relatively similar, although all nine of the very steeply rising sources, with $\alpha^{15.7}_{1.4} < -0.8$, in the full sample are not included in the sample studied in this paper. Seven of these nine sources are below the 10C completeness limits (0.5~mJy in the deep regions and 1~mJy in the shallow regions) and are not detected at 1.4~GHz so the spectral indices are upper limits calculated from the $3\sigma$ noise in the WSRT map. The flux density distributions of the full sample and sub-sample used in this paper are also very similar, although none of the faintest sources in the full sample appear in the sample studied here, as this sample only contains sources above the 10C completeness limits (0.5 and 1~mJy in the deep and shallow regions respectively).

\subsection{The Lockman Hole SERVS Data Fusion}\label{section:FUSED-data}

The \emph{Spitzer} Extragalactic Representative Volume Survey  (SERVS; \citealt{2012PASP..124..714M}) is a warm Spitzer survey which imaged $\approx~18~\rm deg^2$ using the 3.6 and 4.5 $\muup$m IRAC bands down to an rms noise of $0.4~\muup$Jy. The SERVS Data Fusion is a multi-wavelength infrared-selected catalogue compiled by Vaccari et al. (in preparation, \texttt{www.mattiavaccari.net/df}) containing most of the multi-wavelength public photometry and spectroscopy for SERVS sources within the Lockman Hole and other SERVS fields. For inclusion in the SERVS Data Fusion, a source must be detected at either 3.6 or 4.5 $\muup$m in the SERVS images, and ancillary datasets are matched against the SERVS position using a search radius of 1 arcsec. In the Lockman Hole, the SERVS Data Fusion includes optical photometry by \citet{2011MNRAS.416..927G} (GS11), near-infrared photometry from the United Kingdom Infrared Telescope (UKIRT) Infrared Deep Sky Survey  (UKIDSS; see \citealt{2007MNRAS.379.1599L}) and mid-infrared and far-infrared photometry from the Spitzer Wide-Area Infrared Extragalactic survey (SWIRE; see \citealt{2003PASP..115..897L}).

The GS11 deep optical data were taken with the Wide Field Camera (WFC) at the Isaac Newton Telescope (INT) and the Mosaic-1 camera on the Mayall 4-meter Telescope at the Kitt-Peak National Observatory (KPNO) in $g$, $r$, $i$ and $z$-bands. The average magnitude limits in the $g$, $r$, $i$ and $z$-bands are 24.5, 24.0, 23.3 and 22.0 (AB, 5$\sigma$ for a point-like object measured in a 2-arcsec aperture). Information about the optical morphology of the objects is included in the full published GS11 catalogue but is not in the SERVS Data Fusion catalogue. We therefore matched the full GS11 catalogue to the Data Fusion catalogue to include this information. A match radius of 1 arcsec was used and all of the objects with optical information in the Data Fusion catalogue have matches to the GS11 catalogue.

UKIDSS \citep{2007MNRAS.379.1599L} used the UKIRT Wide Field Camera (WFCAM) to map 7500 deg$^2$ in five different surveys in $J$, $H$ and $K$ bands. The Lockman Hole is part of the Deep Extragalactic Survey (DXS), which has a limiting $K$ magnitude of 21 (vega). The Data Fusion catalogue contains data from UKIDSS Data Release 9, which includes only $J$ and $K$ bands in the Lockman Hole.

SWIRE \citep{2003PASP..115..897L} is a wide-field high galactic latitude survey covering nearly 50 deg$^2$ in six different fields, one of which is the Lockman Hole. These fields have been surveyed by the \emph{Spitzer Space Telescope} using both the Infrared Array Camera (IRAC) and the Multi-Band Imaging Photometer (MIPS) far-infrared camera. The Data Fusion catalogue contains data from IRAC, which made observations at 3.6, 4.5, 5.8 and 8.0 $\muup$m with $5\sigma$ sensitivities of 3.7, 5.4, 48 and 37.8 $\muup$Jy respectively (the longer wavelength MIPS photometry is not included here).

\begin{table*}
\tabcolsep=2pt
\caption{A summary of the multi-wavelength information used in this work.}\label{tab:multi-data}
\medskip
\centering
\small
\begin{tabular}{llll}\hline
Survey & Reference & Band & Flux density limit ($5\sigma$)\\\hline
 SWIRE  & \citet{2003PASP..115..897L} & 3.6, 4.5, 5.8, 8.0 $\muup$m & 3.7, 5.4, 48, 37.8 $\muup$Jy\\
 SERVS  & \citet{2012PASP..124..714M} & 3.6, 4.5 $\muup$m 		  & 1.3, 1.5 $\muup$Jy\\
 UKIDSS & \citet{2007MNRAS.379.1599L} & $J$, $K$					  & 21 (AB)\\
 GS11   & \citet{2011MNRAS.416..927G} & $g$, $r$, $i$, $z$		  & 24.5, 24.0, 23.3, 22.0 (AB)\\
 \hline
\end{tabular}
\end{table*}

These catalogues are summarised in Table \ref{tab:multi-data}. There are a maximum of ten photometric bands available for each source. The data fusion catalogue also contains the spectroscopic redshifts from a number of different catalogues available in the field. The spectroscopic redshifts used in this paper and their references are listed in table \ref{tab:specz}.

\subsection{Fotopoulou et al. photometric redshift catalogue (F12)}\label{section:F12}

\citet{2012ApJS..198....1F} (F12) produced a deep photometric redshift catalogue covering $0.5~\rm deg^2$ contained within the southern 10C Lockman Hole field using the \textsc{Le Phare} photometric redshift code \citep{1999MNRAS.310..540A,2006A&A...457..841I}. This catalogue contains 187,611 objects and has up to 21 bands available, ranging from far-ultraviolet to mid-infrared. The far-ultraviolet (FUV) and near-ultraviolet (NUV) observations used in F12 were made by the \emph{Galaxy Evolution Explorer (GALEX)}, with limiting magnitudes of 24.5 in both bands. At optical wavelengths, they use data from the Large Binocular Telescope (LBT), Subaru, and the Sloan Digital Sky Survey (SDSS). The LBT data consists of five bands, $U,~B,~V,~Y~\rm{and}~z'$, and covers about $0.25~\rm deg^2$, the Subaru data contain $R_c,~I_c~\rm{and}~z'$-band observations and SDSS contains data in $u',~g',~r',~i'~\rm{and}~z'$-bands. F12 also use the UKIDSS and \emph{Spitzer} data which are used in this work and described in more detail in Section~\ref{section:FUSED-data}. X-ray observations are available for $0.2~\rm deg^2$ of the field, and the 388 X-ray-detected sources, presumed to be AGN, are treated differently in the fitting process. 

\subsection{Revised SWIRE Photometric Redshift Catalogue (RR13)}\label{section:RR13}

\citet{2013MNRAS.428.1958R} (RR13) produced an updated SWIRE photometric redshift catalogue, which is a revised version of the redshift catalogue produced by \citet{2008MNRAS.386..697R}. The revised catalogue uses the Data Fusion multi-wavelength catalogue (Section~\ref{section:FUSED-data}), which provides deeper optical data and more photometric bands than the catalogue used in \citet{2008MNRAS.386..697R}. The redshifts are estimated using a two-pass template method based on six galaxy and three AGN templates in the first pass and 11 galaxy and three AGN templates in the second. AGNs are identified by their optical morphology -- only objects which appear point-like are fitted with AGN templates. This reduces the risk of catastrophic outliers occurring when normal galaxies are erroneously fitted with AGN templates, but does mean that some AGN will be misclassified as normal galaxies. The main difference between the RR13 catalogue and the previous 2008 version is the treatment of AGNs, as dust torus emission is now included in the quasar templates. The RR13 catalogue only contains redshift values for objects which were included in the original SWIRE photometric redshift catalogue \citep{2008MNRAS.386..697R}, so does not include redshifts for every object in the Data Fusion catalogue. There are also a small number of objects in the RR13 catalogue which are not in the Data Fusion catalogue as these appear in the original SWIRE redshift catalogue.

RR13 contains redshift estimates for 1,009,607 sources in all eight of the SWIRE fields and covers all but a small section of the 10C Lockman Hole field.

\section{Matching the catalogues}\label{section:fused-matching}

\subsection{Morphology of the radio sources}

Due to the high density of sources in the Data Fusion catalogue it is necessary to take into account the structure of the radio sources when matching the catalogues, as there may be several optical objects within the radio contours of extended sources. 

In order to determine whether or not a source is extended the ratio of total flux to peak flux density ($C=S_{\rm int}/S_{\rm peak}$) was calculated for all sources with a match in either the FIRST, GMRT or WSRT surveys or the 324~MHz VLA survey by \citet{2009AJ....137.4846O} (OMK2009). Flux density values from these lower-frequency catalogues are used instead of the 10C catalogue as they have a higher resolution than the 10C catalogue (which has a beam size of 30~arcsec). FIRST, GMRT and OMK2009 all have a synthesised beam of $\approx$5~arcsec, while WSRT has a larger beam of $\approx$10~arcsec. (OM2008 and BI2006 have significantly smaller beam sizes, $\approx$1~arcsec, meaning that values of $C$ from these catalogues cannot be directly compared to the other $C$ values so they are not used in this analysis). FIRST and GMRT cover the whole field while WSRT and OMK2009 only cover part of the field. Those sources with a match in either FIRST or GMRT were classified as extended if either $C_{\rm GMRT}$ or $C_{\rm FIRST} > 1.2$, otherwise they were classified as compact for the purpose of matching. Sources without a match in FIRST or GMRT but with a match to OMK2009 were classified as extended if $C_{\rm OMK2009} > 1.2$ and compact otherwise. All sources with a match in WSRT and not in the other three catalogues used here were classified as extended for the purpose of matching as the resolution of the WSRT map is not high enough to ensure these sources are not extended. 

There are six sources which are not classified as extended or compact because they do not have a counterpart in any of the four catalogues used when classifying the sources but which do have a match in OM2008 or BI2006, and therefore an accurate position. These sources all have angular sizes less than 3 arcsec in the OM2008 or BI2006 catalogues so were considered to be compact for this purpose. 

There are therefore 38 sources which are compact and 57 sources which have been classified as extended for the purpose of matching but which may in fact not be significantly extended on these angular scales. These two groups of sources are treated separately when identifying optical matches.

\subsection{Matching the catalogues}\label{section:matching-details}

\begin{figure}
\centerline{\includegraphics[width=\columnwidth]{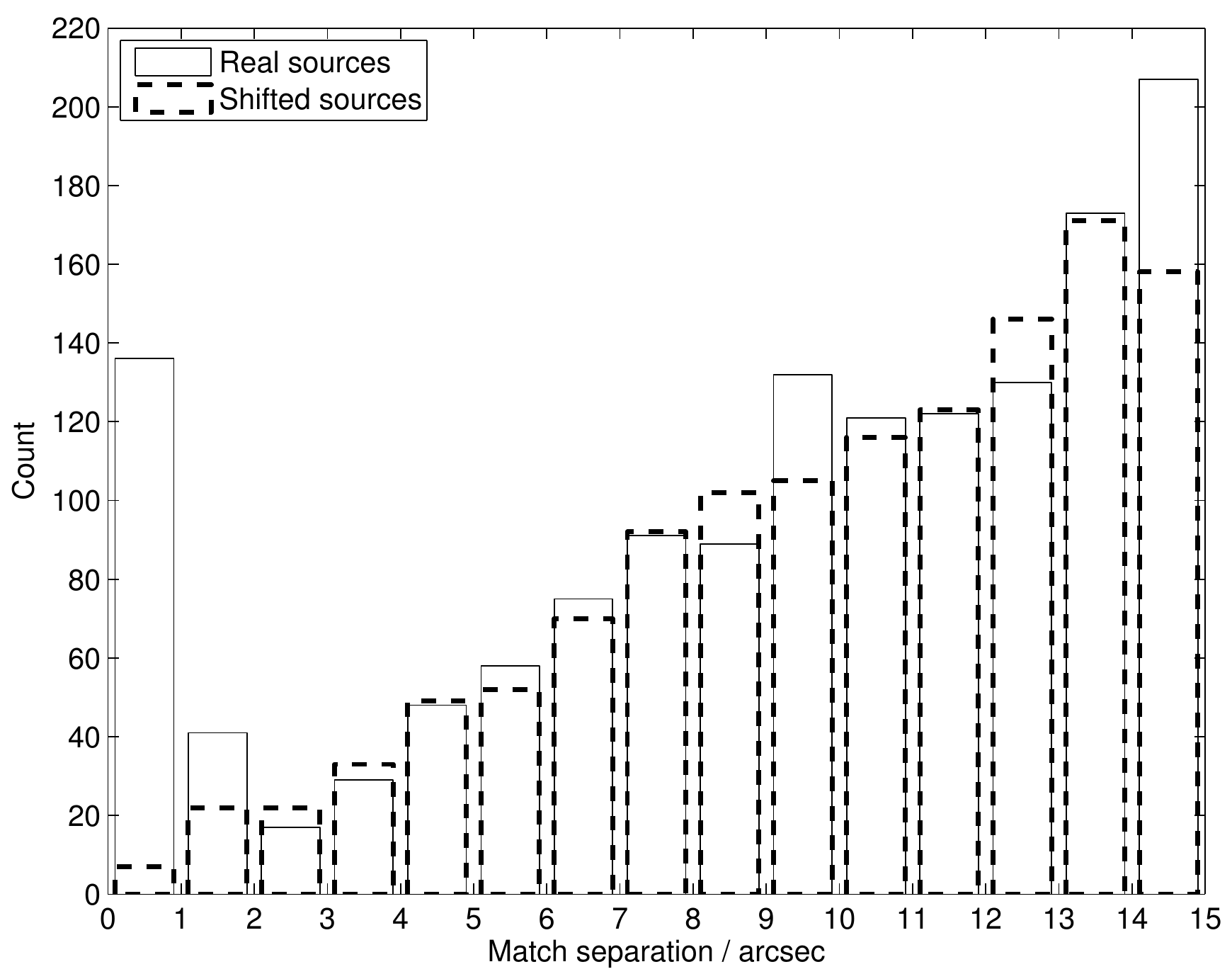}}
\caption{The separation distribution when the radio sources in the 10C catalogue and the simulated catalogue are matched to the Data Fusion catalogue, taking all matches to the radio sources within 30 arcsec. Note that the full 10C sample is matched here.}\label{fig:matches}
\end{figure}

\begin{figure*}
\centerline{\includegraphics[width=7.5cm]{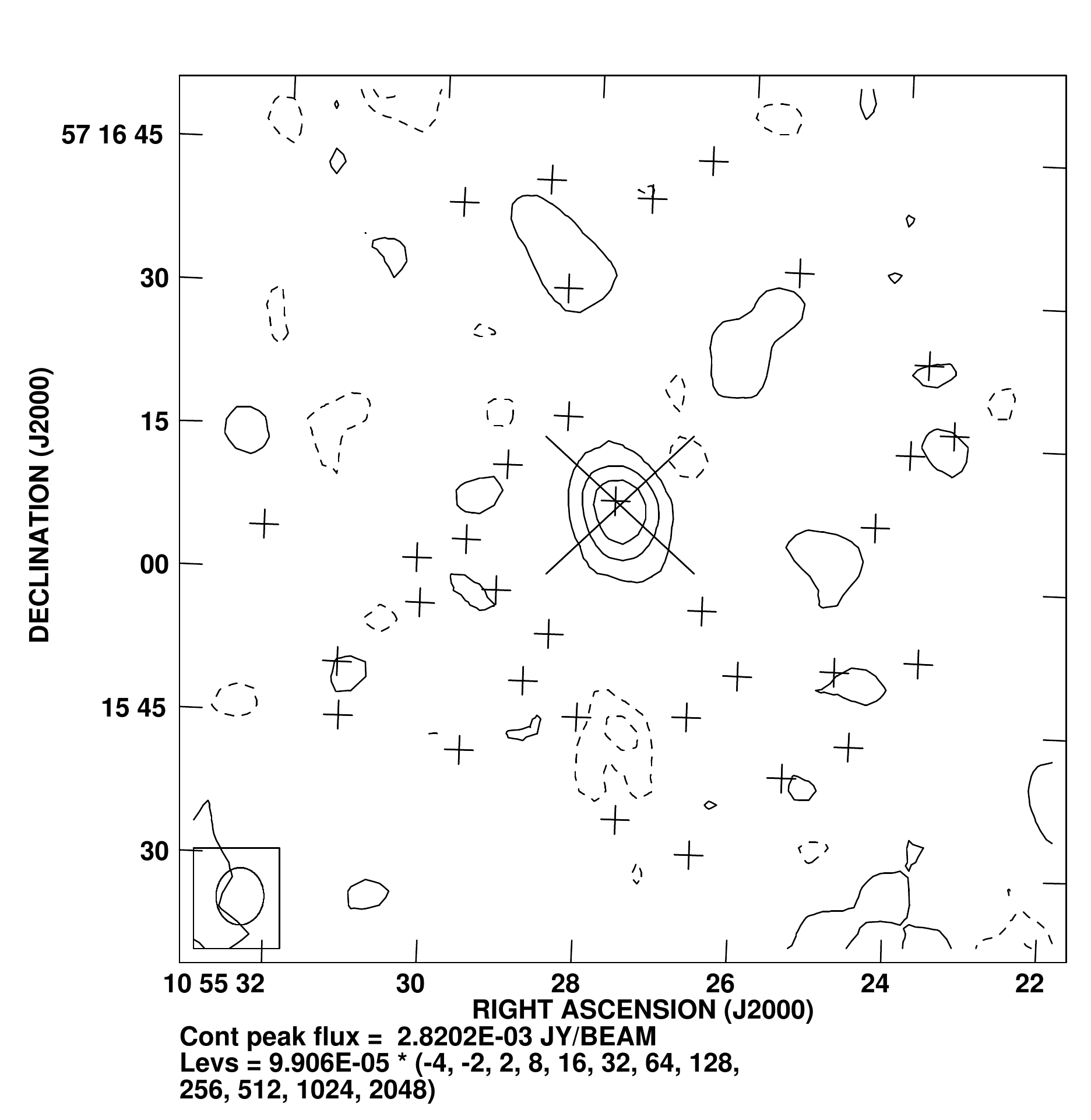}
            \quad
      \includegraphics[width=7.5cm]{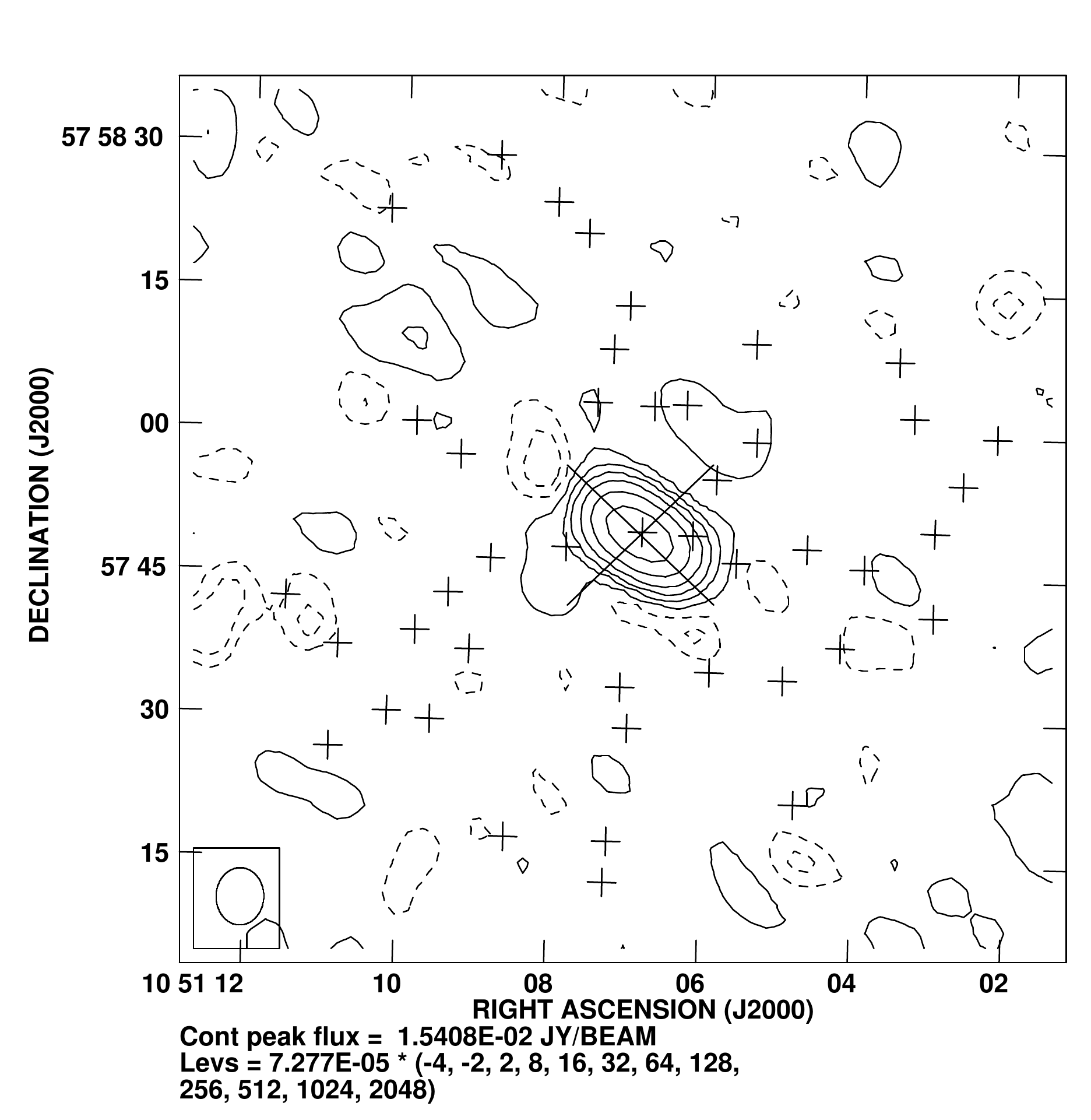}}
\bigskip
\centerline{\includegraphics[width=7.5cm]{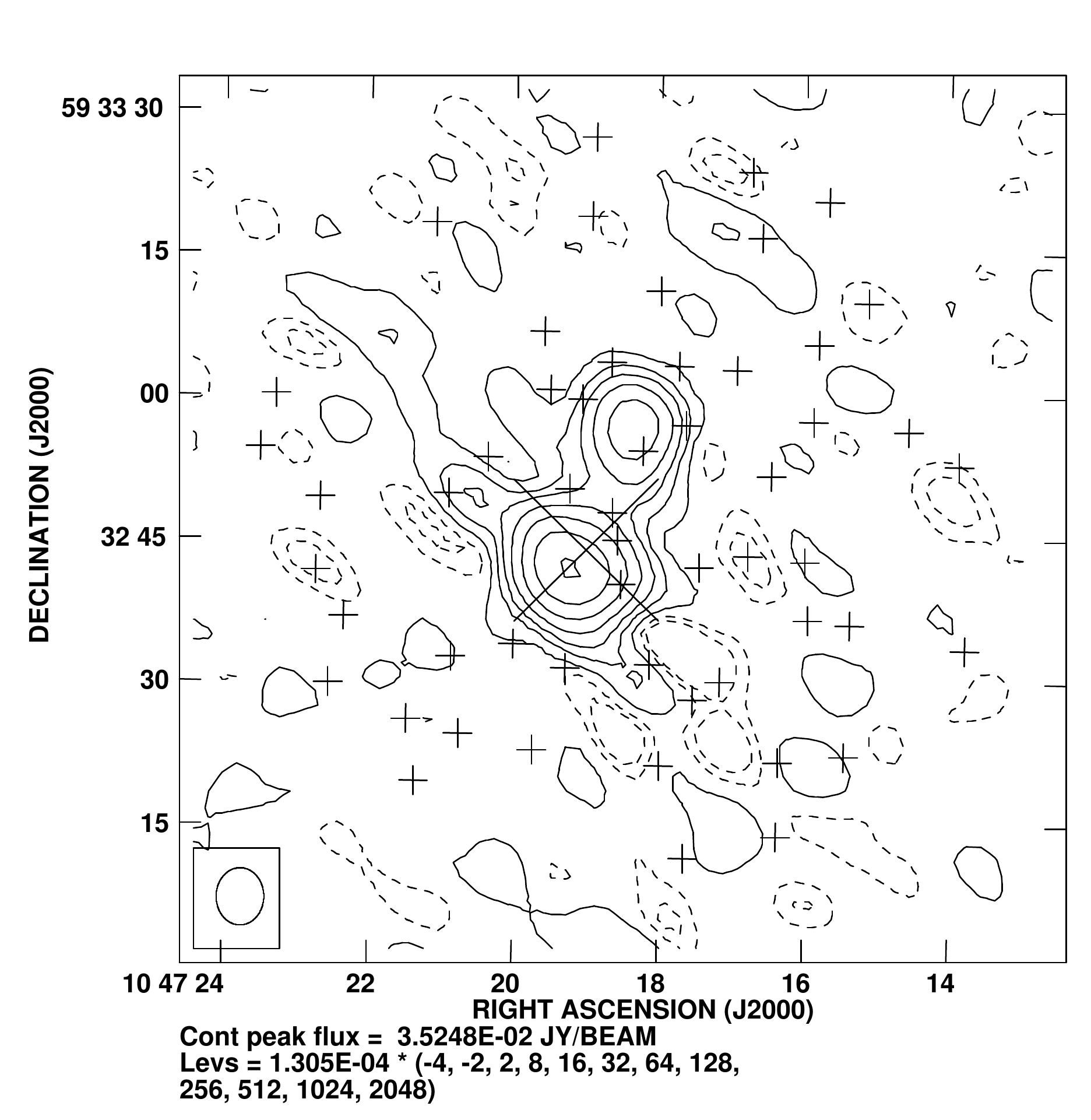}
            \quad
      \includegraphics[width=7.5cm]{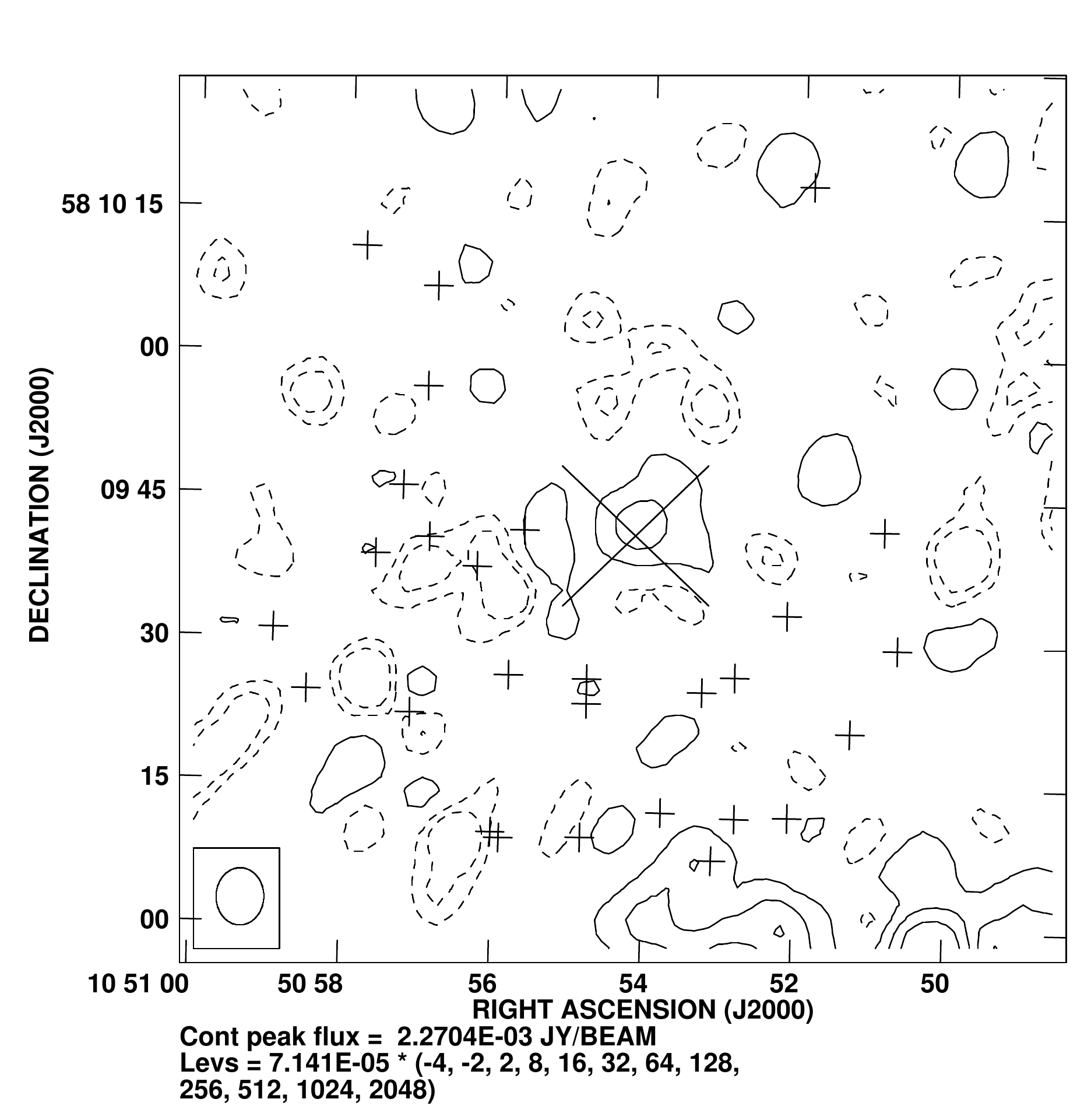}}
\caption{Examples of extended sources assigned each of the four flags. Top left = 1 (probable match), top right = 2 (possible match), bottom left = 3 (confused) and bottom right = 4 (no match). The large cross ($\times$) marks the position of the radio source in the 10C catalogue and the contours are taken from the GMRT image. The smaller pluses ($+$) mark the positions of objects in the Data Fusion catalogue.}\label{fig:ext-matching}
\end{figure*}

The different catalogues were matched using the \textsc{topcat}\footnote{see: http://www.starlink.ac.uk/topcat/} software package. The density of sources in the Data Fusion catalogue is high compared to the potential error in the 10C source positions ($\approx 6$~arcsec) so we therefore use the more accurate positions from the lower frequency radio catalogues (typical error $\approx1$ arcsec). When there are several positions available for a source, they are used in the following order of preference: FIRST, GMRT, BI2006/OM2008, WSRT, 10C. For sources which are resolved into multiple components in FIRST or GMRT, the position from the 10C catalogue was used instead as this gives a best estimate of the centre of the flux. For the three sources which have two separate components listed in the original 10C catalogue (see Paper~I for details), the average of the two 10C positions is used. 

The match radius needs to be chosen carefully to avoid false matches while still maximising the number of real matches. The 10C sources were shifted by 0.2 degrees in declination to produce a simulated sample of randomly positioned sources. Both this simulated sample and the real sample were matched to the Data Fusion catalogue, and all Data Fusion objects within 30 arcsec of each source were noted. The separation between the matches is shown in Fig.~\ref{fig:matches}; it is clear that beyond 2 arcsec the number of real and random matches becomes comparable. (Note that the full 10C sample is used in this plot to provide better statistics.)

For those sources classified as compact, the nearest match within 2 arcsec was accepted. If there was no match within 2 arcsec then the source in question was considered to have no optical counterpart. In total, 36 of the compact sources have a match within 2 arcsec and two do not.

For sources classified as extended the positions of all the optical sources were plotted on top of the radio contours (GMRT maps were used for those sources with a GMRT match, WSRT maps were used for the remaining sources). These images were then examined and the sources were also assigned one of the following flags: 

\begin{enumerate}

\item probable match -- only source within the $3 \sigma$ radio contours (21 sources);

\item possible match -- looks likely but there are other sources within the $3 \sigma$ radio contours (23 sources);

\item confused -- several sources within the $3 \sigma$ radio contours so cannot identify the correct match (8 sources);

\item no match -- no sources within the $3 \sigma$ radio contours (5 sources).

\end{enumerate}

\noindent Examples of sources assigned each of these flags are given in Fig.~\ref{fig:ext-matching}.  Table \ref{tab:optical-matches} contains a summary of the number of matches to the Data Fusion catalogue and the flags used in the catalogue.

\begin{figure}
\centerline{\includegraphics[width=\columnwidth]{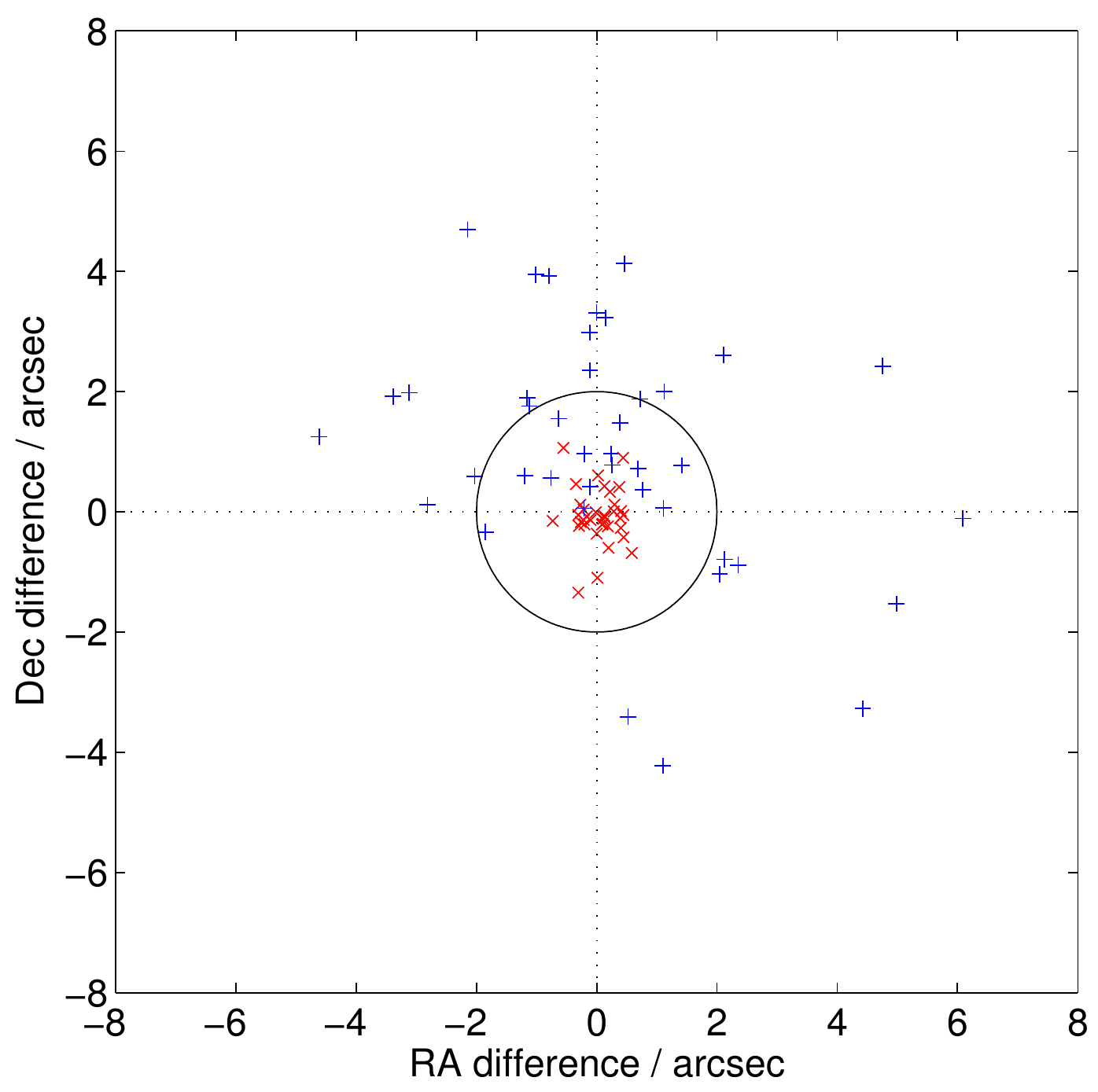}}
\caption{Separation between the best radio position for each source and the position of its counterpart in the Data Fusion catalogue. Sources which are considered extended for the purposes of matching are shown by `$+$' (blue in the online version), compact sources are shown by `$\times$' (red in the online version). The circle indicates a separation of 2~arcsec (the match radius for compact sources).}\label{fig:best_fused_sepn}
\end{figure}

\begin{table}
\caption{A summary of the matches to the Data Fusion catalogue found for the 96 10C sources studied in this paper.}\label{tab:optical-matches}
\centering
\medskip
\begin{tabular}{lr}\\\hline
Description & No. of sources \\\hline
Extended -- probable match & 21\\
Extended -- possible match & 23\\
Extended -- confused       & 8\\
Extended -- no match       & 5\\
Compact -- match within 2 arcsec & 36\\
Compact -- no match within 2 arcsec & 2\\
10C position only -- no matching attempted & 1\\\hline
\end{tabular} 
\end{table}

In summary, we have identified possible counterparts for 80 out of the 96 sources (83 percent). A table listing the multi-wavelength counterparts for each source is included in table \ref{tab:all-sources} in the appendix. 24 of these 80 sources have a spectroscopic redshift available in the Data Fusion catalogue, these redshifts and their references are listed in table \ref{tab:specz}. Fig.~\ref{fig:best_fused_sepn} shows the separation between the radio position of each source and the Data Fusion object associated with it. The contour plot of the one source in the sample without an accurate position available was examined by eye; there are no possible counterparts within the radio contours. This source is therefore included in the group of sources without a match in future discussions (giving a total of eight sources without a match, and a further eight confused sources). 

SERVS 3.6-$\muup$m and 4.5-$\muup$m images of the eight sources without a match were examined by eye, and in one case (source 10CJ105040+573308) there was a source visible in the SERVS images. This source lies close to a very bright ($S_{3.6~\muup \rm m} = 1.07 \times 10^{4} ~\muup \rm Jy$) source so is not included in the SERVS catalogue. The 3.6~$\muup$m-flux density for this source was estimated from the image, and although we do not have enough information to calculate a photometric redshift for this source, this value is used in Section~\ref{section:R} when calculating radio-to-infrared ratios.

\begin{figure*}
\centerline{\includegraphics[width=7cm]{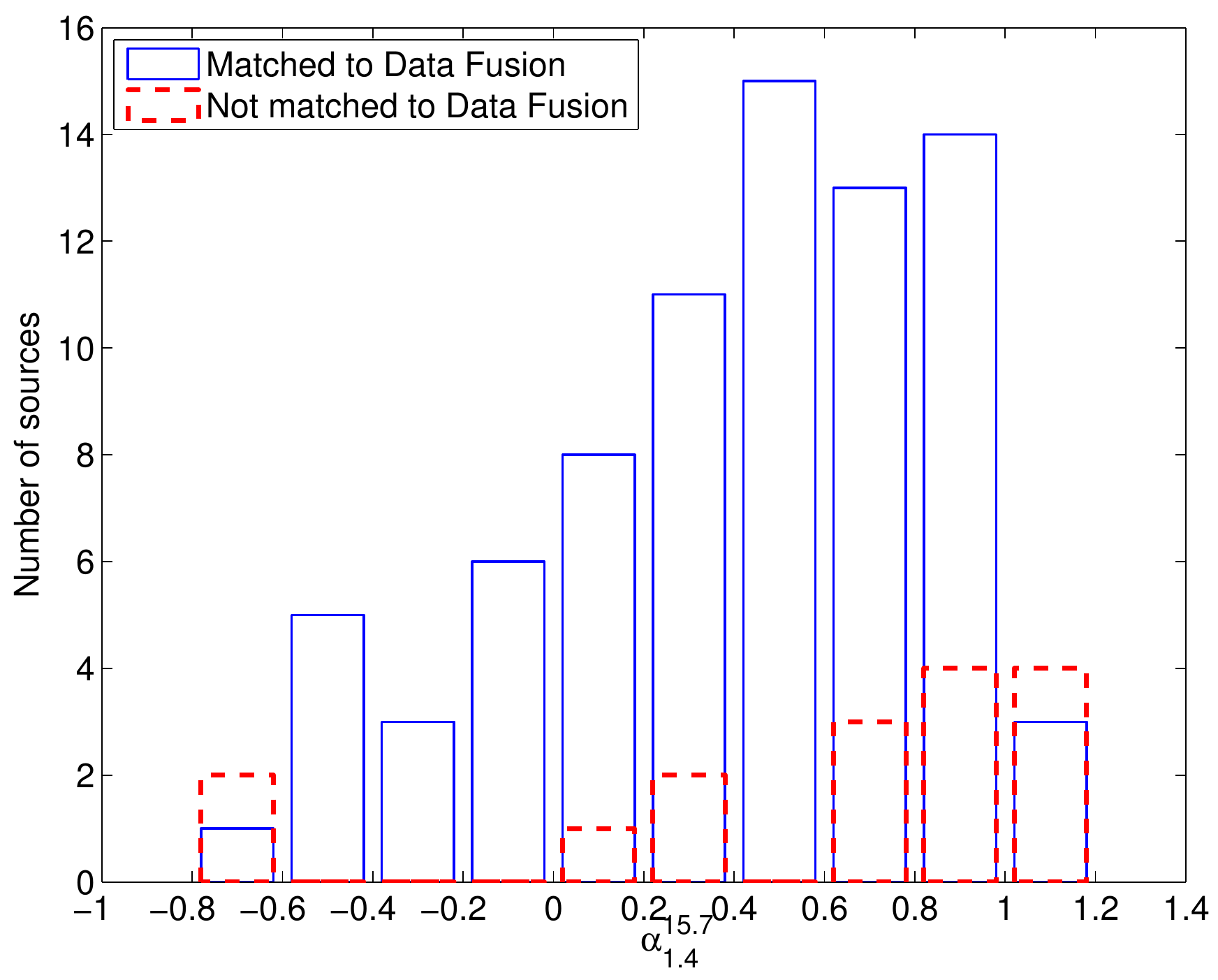}
      \quad
      \includegraphics[width=7cm]{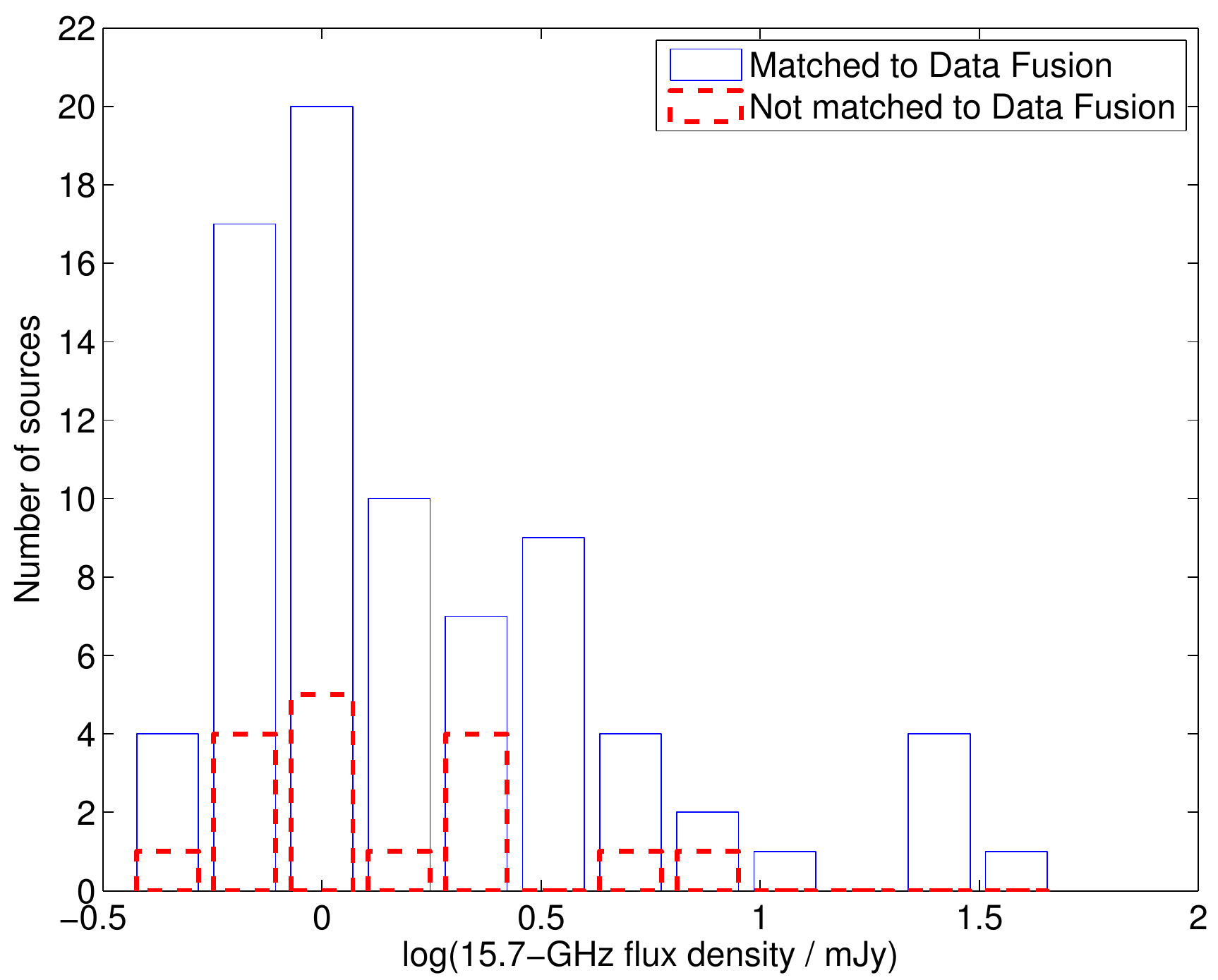}}
\caption{A comparison of the spectral index and 15.7-GHz flux density distributions of sources which have a match in the Data Fusion catalogue and those which do not.}\label{fig:alpha_matched}
\end{figure*}

The spectral index and flux density distributions of the sources with and without a counterpart are compared in Fig.~\ref{fig:alpha_matched}. The spectral index distributions of the two groups of sources (those with and without a match) are broadly similar, except that a higher proportion of the very steep ($\alpha^{15.7}_{1.4} > 0.8$) sources are unmatched (32 percent of sources with $\alpha^{15.7}_{1.4} > 0.8$ are unmatched compared to 11 percent of sources with $\alpha^{15.7}_{1.4} < 0.8$). This is probably because these very steep sources tend to be significantly extended  (all of the sources with $\alpha > 1$ have angular sizes $> 20$~arcsec, Section~\ref{section:sizes}) and are therefore more likely to be classified as confused when matching. The 15.7-GHz flux density distributions of the two groups of sources are similar, although all six of the brightest 10C sources have counterparts in the multi-wavelength catalogue. 

\subsection{Matching to other photometric redshift catalogues}\label{section:matching-other}

The sample was matched to the F12 and RR13 photometric redshift catalogues. For the 80 sources which have a counterpart in Data Fusion the catalogues were matched using the position from the Data Fusion catalogue and the nearest match within 1.5 arcsec is accepted (the shifting procedure described in Section~\ref{section:matching-details} was repeated for each of the catalogues to choose these match radii). This gave a total of 53 matches to RR13 and 20 to F12. 

\begin{table}
\caption{A summary of matches to multi-wavelength catalogues for 10C sources used in this work.}\label{tab:other-matches}
\begin{center}
\begin{tabular}{cc}\hline
Catalogue & Number of matches to 10C sources\\\hline
Data Fusion & 80\\
Spec $z^a$ & 24\\
GS11  & 59\\
RR13  & 53\\
F12   & 20\\\hline
\end{tabular} 
\end{center}
Notes:\\
a) Spectroscopic redshift from the Data Fusion catalogue.\\
\end{table}

\subsection{Possible matches for confused sources}\label{section:confused-matches}

There are eight sources which were classified as confused when matching to the Data Fusion catalogue. Although it is not possible to identify a single counterpart for these sources, some useful information about their nature can be gained by looking at all possible counterparts within the radio contours. Therefore, all objects within one tenth of the peak flux in the GMRT sub-image were selected as possible counterparts for each 10C source. For the one source without a GMRT image, all objects within the $3\sigma$ contour in the WSRT map were selected instead. In total, 30 possible counterparts were identified for the eight sources. These are included in later discussion.


\section{Photometric redshift fitting}\label{section:lephare}

The two photometric redshift catalogues described in Section~\ref{section:multi-data} do not contain redshift values for all the sources in this sample with multi-wavelength data available, so we performed our own photometric redshift fitting. The publicly available photometric redshift code \textsc{Le Phare}\footnote{http://www.cfht.hawaii.edu~arnouts/LEPHARE/lephare.html} \citep{1999MNRAS.310..540A,2006A&A...457..841I} was used to compute photometric redshifts for the sources in this sample with counterparts in the Data Fusion catalogue. The code takes an input library of spectral energy distribution (SED) templates, which are assumed to represent the SEDs of the observed sample, and shifts them to a range of redshift values. These templates are then fitted to the photometric data, and a least-squares minimisation is used to select the best-fitting SED template for each source. The redshift of the best-fitting template is then adopted as the redshift estimate. The photometric data used here was from the Data Fusion catalogue (see Section~\ref{section:FUSED-data}), which has up to ten photometric bands available for each source. Data at 3.6 and 4.5~$\muup$m are available from both SWIRE and SERVS for some sources, in which case values from SERVS were used for the photometric fitting, as these observations have a better signal to noise.

Each source was fitted to two different template libraries, the first containing galaxy templates and the second containing AGN templates. These two libraries, and the extinction laws applied, are the same as those used in F12. The galaxy templates used are the library produced by \citet{2009ApJ...690.1236I}. These include nine templates generated by \citet{2007ApJ...663...81P} -- three elliptical galaxy SEDs and six spiral galaxy SEDs (S0, Sa, Sb, Sc, Sd, Sdm) -- and 12 starburst galaxy SEDs generated using the \citet{2003MNRAS.344.1000B} models (with starburst ages ranging from 3 to 0.03 Gyr). This gives a total of 21 SED templates. \citet{2009ApJ...690.1236I} linearly interpolated between some of the \citeauthor{2007ApJ...663...81P} templates to refine the sampling in colour-redshift space, resulting in a total of 31 templates. For templates Sb to SB3 (template IDs 11 to 23) extinction is applied according to the \citet{1984A&A...132..389P} Small Magellanic Cloud (SMC) law, while for SB4 to SB11 (template IDs 24 to 31) the \citet{2000ApJ...533..682C} laws are applied. No additional extinction is applied for templates earlier than Sb. The intrinsic galactic absorption is calculated with values of $E(B - V) = 0.00, 0.05, 0.10, 0.15, 0.20, 0.25, 0.30, 0.40$ and 0.50. Emission lines were added to the templates using the option in the \textsc{Le Phare} code as this has been shown to give better results, even in the case of broadband photometry \citep{2009ApJ...690.1236I}.

The AGN template library was taken from \citet{2009ApJ...690.1250S}. This library of 30 templates contains galaxy and AGN templates, as well as a number of hybrid templates. These hybrid templates contain contributions from both galaxy and AGN templates in proportions ranging from 10\%:90\% to 90\%:10\% in steps of 10\% (see \citealt{2009ApJ...690.1250S} for full details). For this template library extinction is allowed to vary from $E(B - V) = 0$ to 0.5 in steps of 0.05.

For both libraries the templates are calculated at redshifts 0--6 in steps of $\Delta z = 0.01$ and in steps of $\Delta z = 0.02$ for redshifts 6--7. The absolute magnitude was restricted to the range $-28 < M < -8$, where $M$ is the absolute magnitude in the $K$-band.

\subsection{Selecting a final redshift value}\label{section:lephare-z-used}

Each of the 80 matched objects in the sample has been fitted to two SED template libraries (the galaxy template library and the AGN template library), so there are two possible redshift values produced for each source. Throughout this section the redshift value resulting from the best-fitting template from the galaxy library is referred to as $z_{\rm GAL}$ and the redshift value resulting from the AGN template library is referred to as $z_{\rm AGN}$.  

To decide which value is the most appropriate one to use for each source the SEDs which provided the best fit to the photometric data from each of the two libraries were plotted for each source and examined by eye. Both the galaxy and AGN template fits were then qualitatively assigned one of the following template flags to characterise the fit:

1 = good fit;

2 = possible fit;

3 = poor fit.

\noindent One of the possible two redshift values was then selected for each source. $z_{\rm GAL}$ was selected if the fit to the galaxy template is better than the fit to the AGN template (i.e.\ galaxy template flag $<$ AGN template flag). If the fit to the galaxy and AGN templates were judged to be equally good (i.e.\ galaxy template flag $=$ AGN template flag), then $z_{\rm GAL}$ was selected if the object's optical morphology indicated that it was extended in the optical observations, and $z_{\rm GAL}$ was also selected if there was no optical morphology information available. $z_{\rm AGN}$ was selected if the fit to the AGN template was better than the fit to the galaxy template (i.e.\ AGN template fitting flag $<$ galaxy template fitting flag), or if the two fits appeared equally good and the object appeared point-like in the optical. Any source with fewer than three photometric bands available was not assigned a redshift value.

In the final photometric redshift catalogue 59 sources have redshift values from the galaxy templates library and 11 have redshift values from the AGN templates library. This gives a total of 70 sources with redshift values from the \textsc{Le Phare} photometric redshift fitting process (the remaining ten sources with matches to the Data Fusion catalogue had fewer than four photometric band available so no redshift estimate is included for these sources).

\subsection{Comparison with other redshift catalogues}\label{section:z-cat-comparison}

\begin{figure*}
\centerline{\includegraphics[width=\columnwidth]{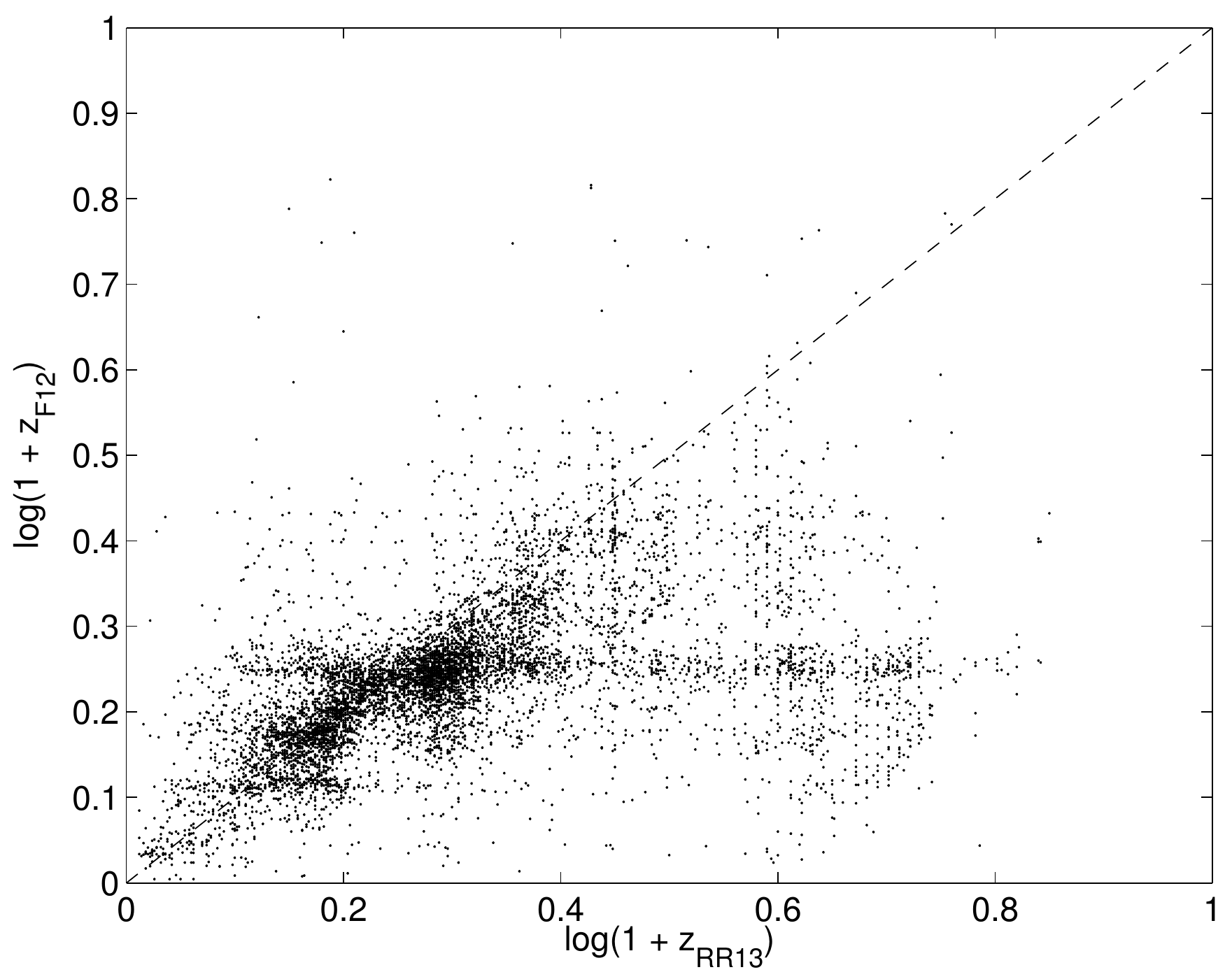}
      \quad
      \includegraphics[width=\columnwidth]{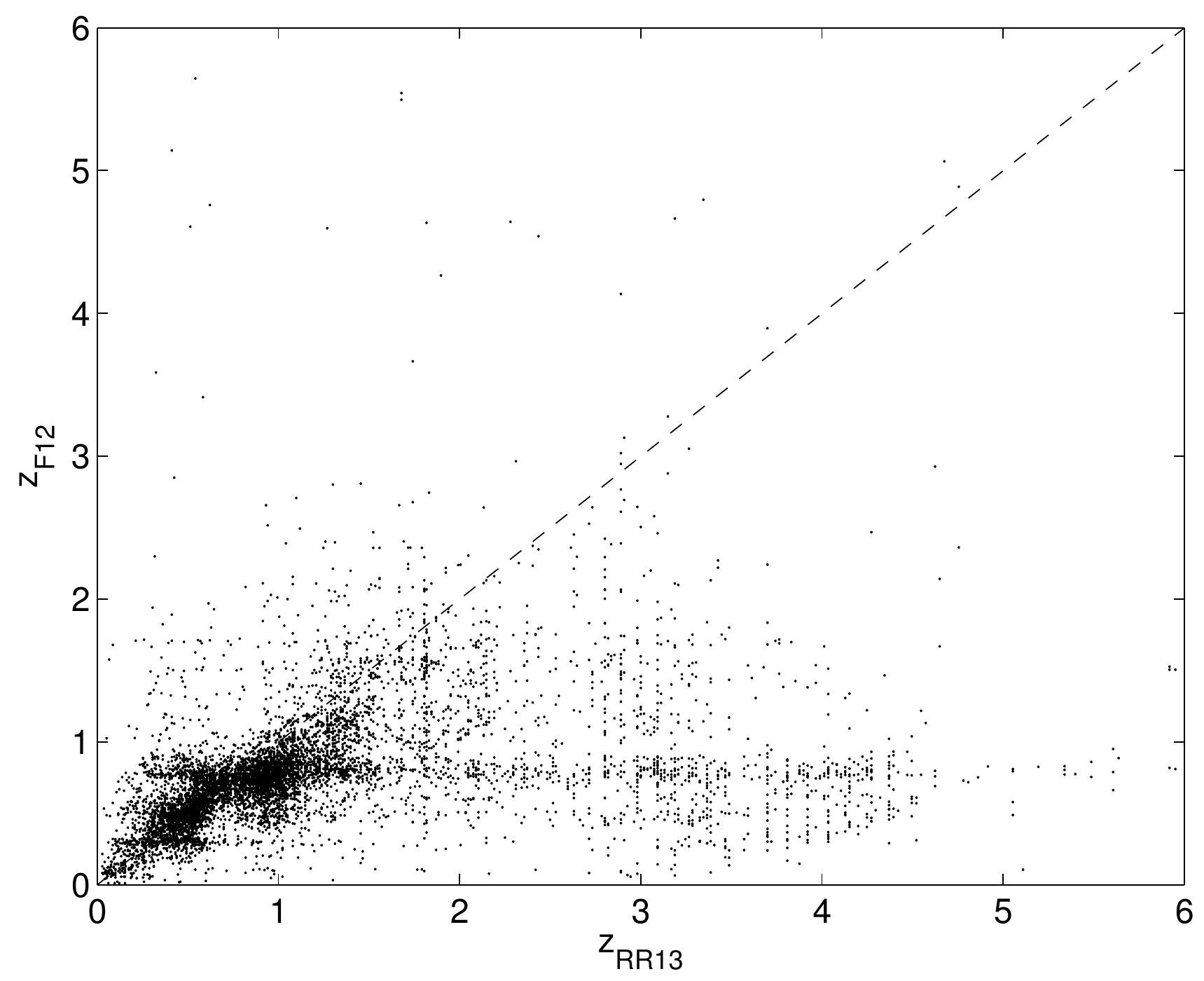}}
\caption{Comparison of the redshift values from the RR13 and F12 catalogues. The dashed line indicates where the redshift values from the two catalogues are equal. The same values are plotted in the two panels -- in the left-hand panel log(1 + $z$) is plotted to make it easier to see the high density of points, and the right-hand panel shows $z$ on a linear scale for easier comparison with Fig.~\ref{fig:lephare_RR13_F12}.}\label{fig:z_RR13_F12}
\end{figure*}

\begin{figure}
\centerline{\includegraphics[width=\columnwidth]{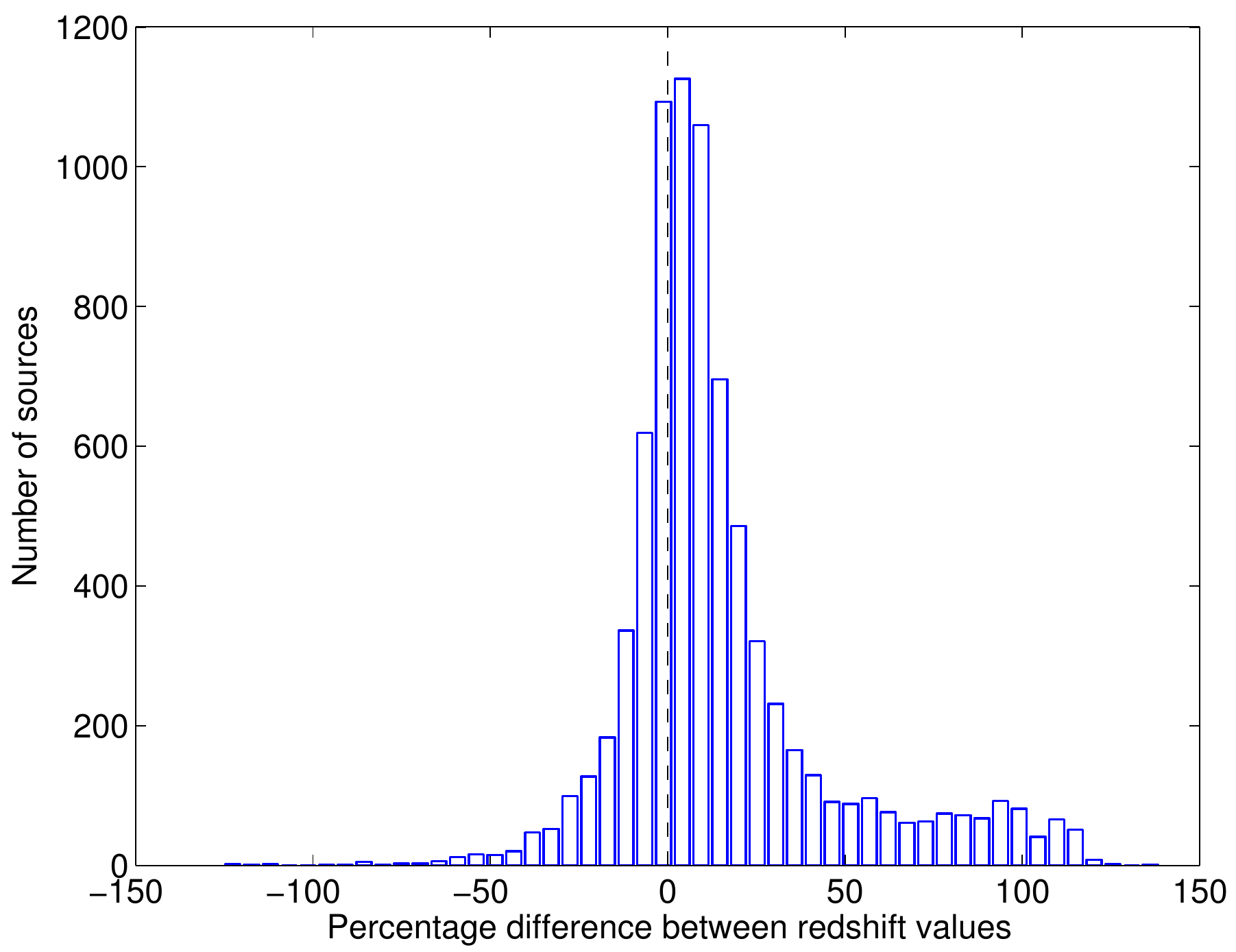}}
\caption{Percentage difference between the redshift values from the RR13 and F12 catalogues (i.e.\ $100 \times (z_{\rm RR13} - z_{\rm F12})/(1+z)$ where $z = (z_{\rm RR13} + z_{\rm F12}) / 2$).}\label{fig:z_comparison_hist}
\end{figure}

\begin{figure*}
\centerline{\includegraphics[width=5.5cm]{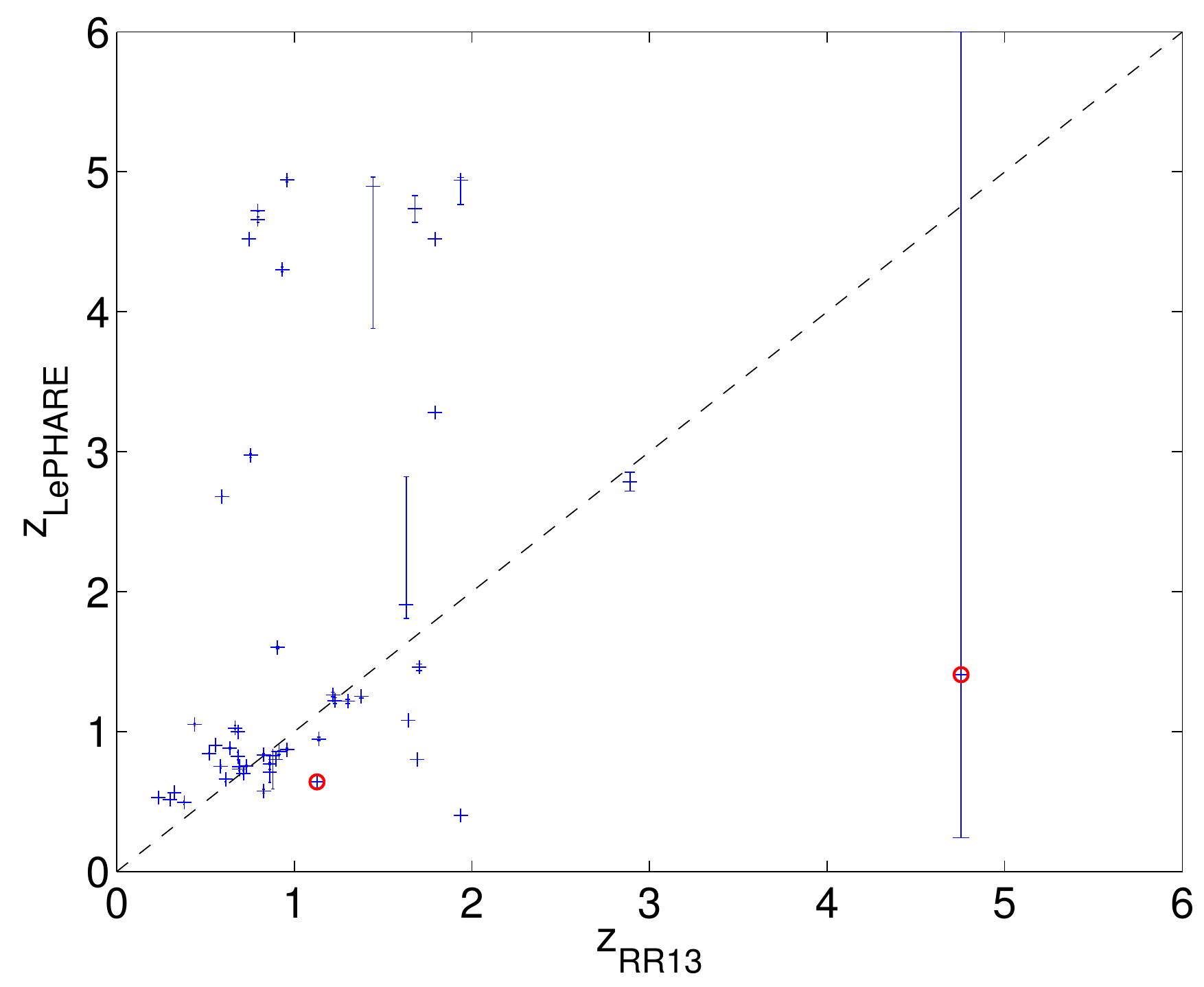}
            \quad
            \includegraphics[width=5.5cm]{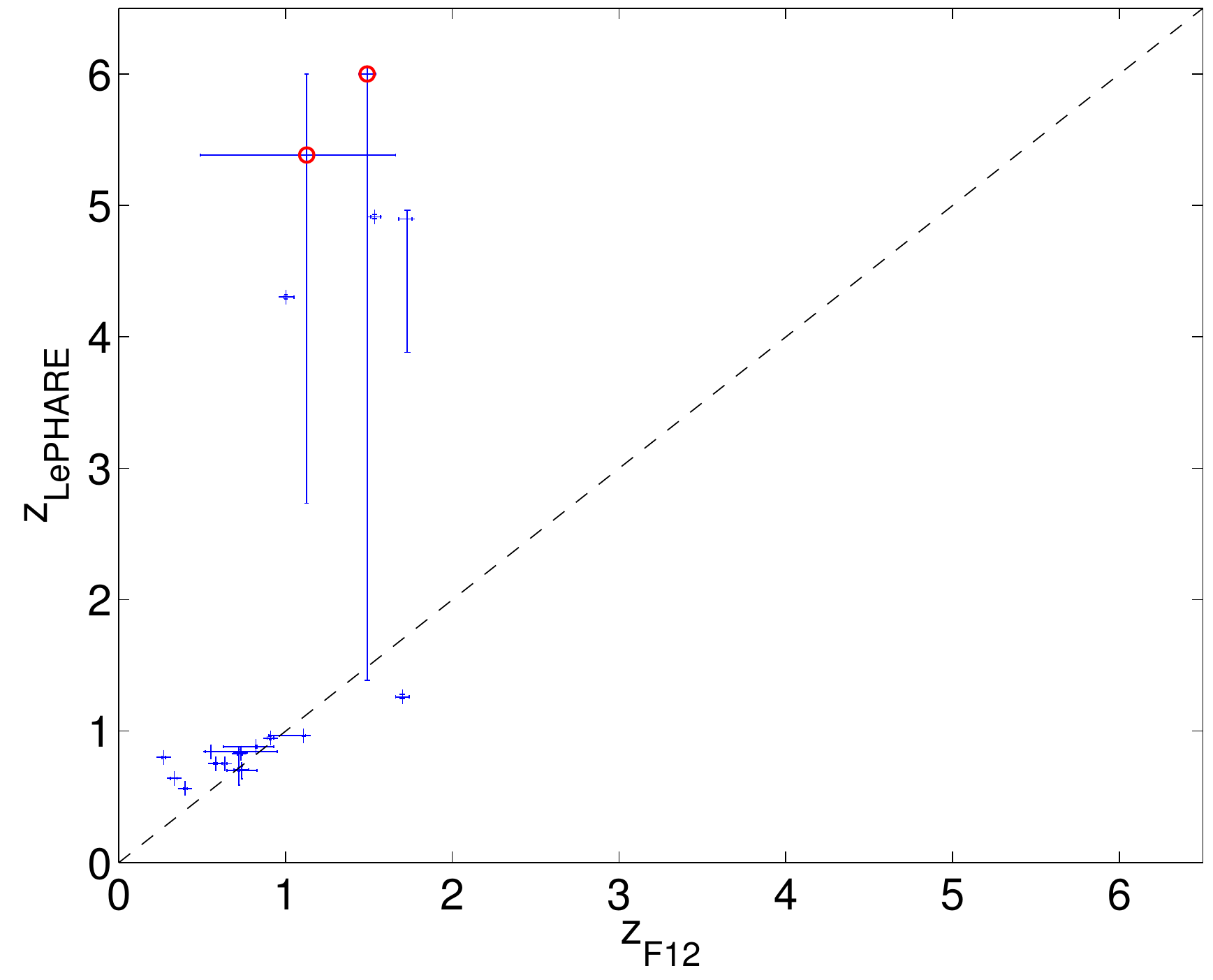}
            \quad
            \includegraphics[width=5.5cm]{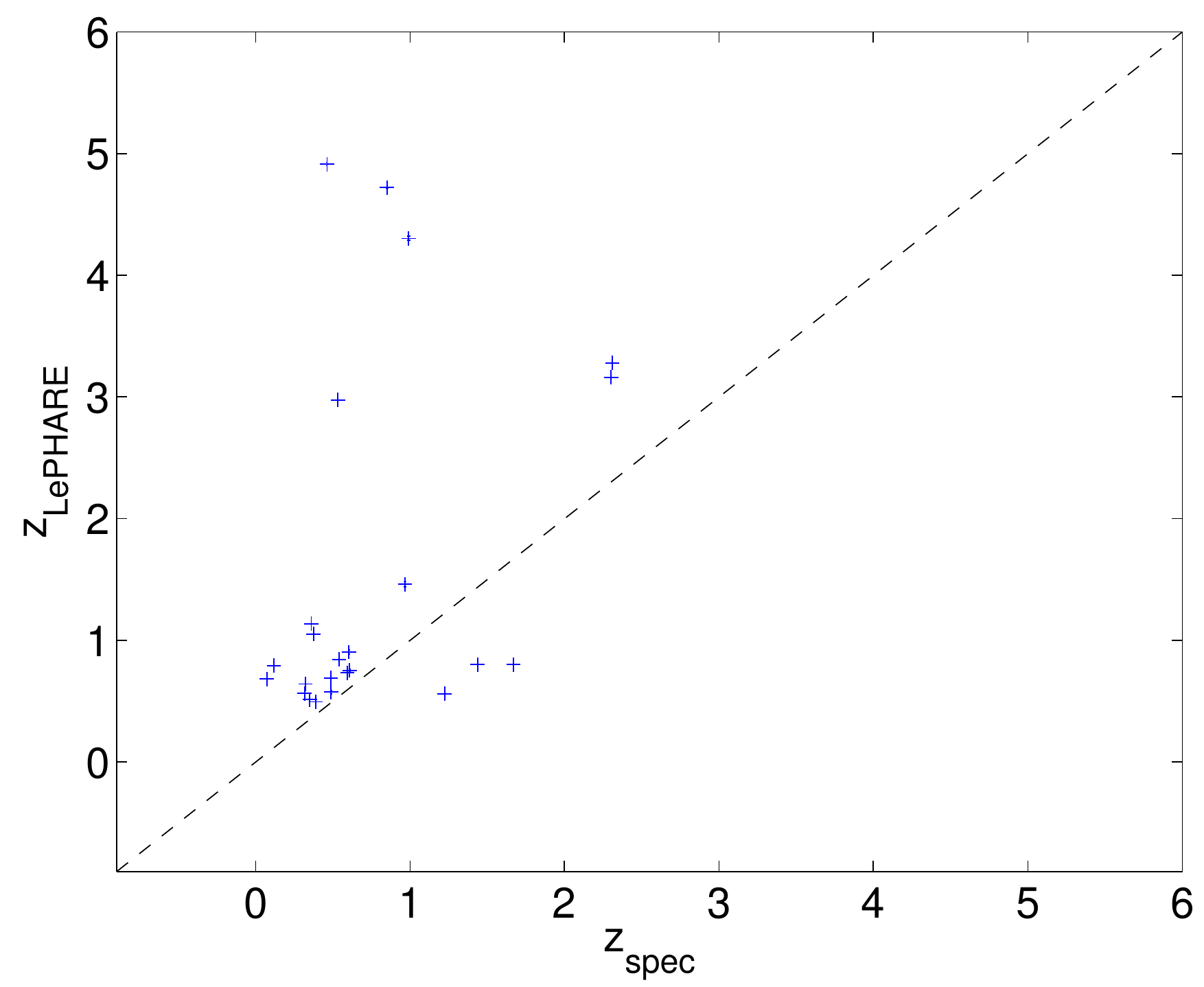}}
\caption{Comparison of the redshift values from our work using \textsc{Le Phare} with those from the RR13 (left panel) and F12 (middle panel) catalogues and the spectroscopic redshift values (right panel). The dashed line indicates where the redshift values from the two catalogues are equal. $1 \sigma$ error bars are plotted for sources in the \textsc{Le Phare} catalogue with a final value from the galaxy library and values from the F12 catalogue. Sources with only three photometric bands available for the \textsc{Le Phare} fitting are circled (in red in the online version); these values are not included in the final catalogue.}\label{fig:lephare_RR13_F12}
\end{figure*}

The redshift values obtained from the \textsc{Le Phare} photometric redshift fitting were compared to the F12 and RR13 photometric redshift catalogues. First the two published catalogues were compared to each other. The two catalogues were matched using a match radius of 1.5 arcsec, giving 7895 matches; the redshift values of these matched sources from the two catalogues are compared in Fig.~\ref{fig:z_RR13_F12}. This plot shows significant scatter, demonstrating that there are some significant disagreements between the two catalogues and highlighting the difficulties in achieving reliable redshifts using photometric methods for these sources. This is demonstrated further in Fig.~\ref{fig:z_comparison_hist} which shows the percentage difference between the redshift values from the two catalogues. The percentage of `catastrophic' outliers, defined as when $z_{2} - z_{1}/(1 + z) > 0.3$, where $z = (z_1 + z_2)/2$, is 20 per cent. Redshift values derived in this work are compared to the F12 and RR13 catalogues in Fig.~\ref{fig:lephare_RR13_F12}. For the majority of the sources there is a good agreement between the \textsc{Le Phare} values calculated here and the values from RR13 and F12. However, for several sources there are significant differences between the \textsc{Le Phare} values and the values from the other two catalogues, with the percentage of catastrophic outliers being 24 per cent when compared to the F12 catalogue and 30 per cent when compared to the RR13 catalogue. Many of the sources do not agree within the error bars, although note that these error bars simply quantify the goodness of the fit of the photometric data to the chosen template, which does not necessarily mean that the chosen template is the correct one. The right panel of Fig.~\ref{fig:lephare_RR13_F12} shows the \textsc{Le Phare} photometric redshifts compared to the spectroscopic values for the 24 sources with spectroscopic values available. There is a reasonable agreement for the majority of the sources, with a catastrophic outlier percentage of 29 per cent.

\subsection{Compiling the final redshift catalogue}\label{section:z-used}

There are spectroscopic redshifts available for 24 sources from the Data Fusion catalogue, which are listed in table \ref{tab:specz} in the appendix. For the remaining sources a photometric redshift is selescted in the following order of preference: 1) photometric redshift from RR13 (38 sources), 2) photometric redshift from F12 (6 sources) and 3) photometric redshift from \textsc{Le Phare} fitting (9 sources). RR13 values were given preference over F12 values because they are available for a greater number of sources as the RR13 catalogue covers the whole 10C survey area, giving greater consistency. A full catalogue of the redshifts is given in table \ref{tab:all-sources} in the appendix.

\begin{figure}
\centerline{\includegraphics[width=\columnwidth]{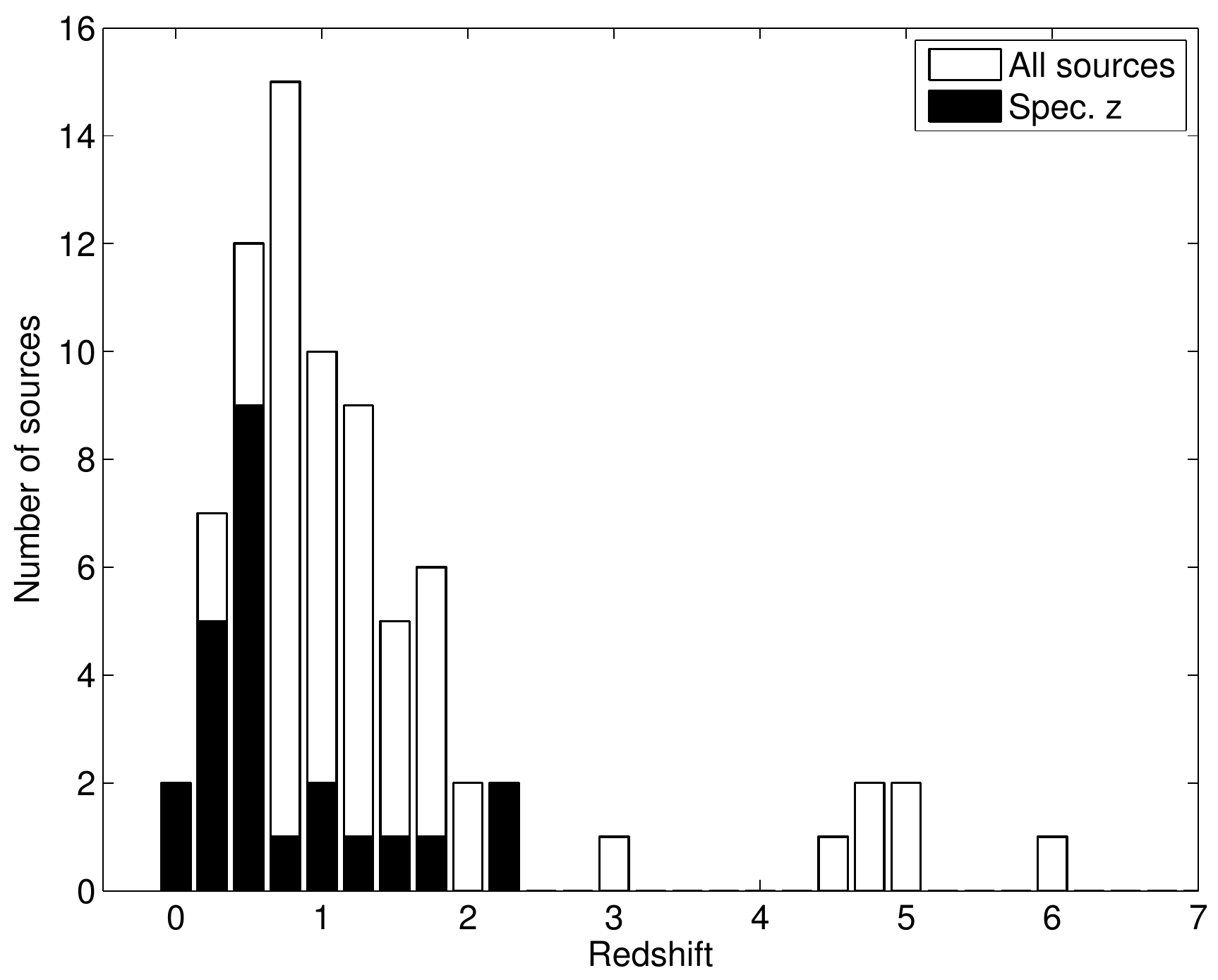}}
\caption{Redshift distribution for all sources with a redshift estimate. Sources with spectroscopic redshift values are shown separately.}\label{fig:z-dist}
\end{figure}

The redshift distribution of all 77 sources with a redshift estimate (or value) is shown in Fig.~\ref{fig:z-dist}. The median redshift is 0.91 with an interquartile range of 0.84. The redshift distribution of the sample is discussed in Section~\ref{section:properties_z}.

\subsection{Confused sources}\label{section:confused}

\begin{figure}
\centerline{\includegraphics[width=\columnwidth]{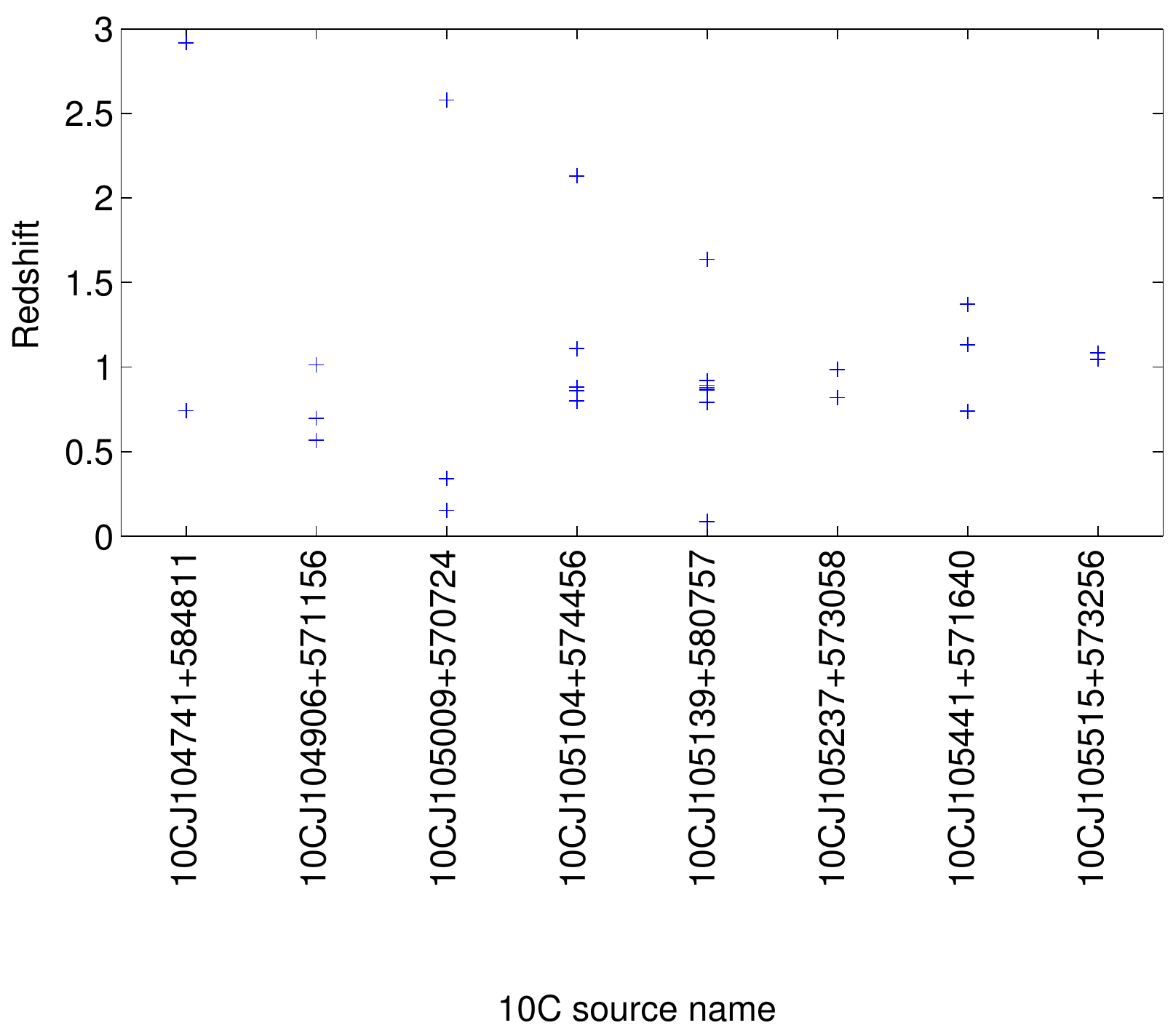}}
\caption{Redshift values for the possible counterparts for each 10C source classified as confused.}\label{fig:z-confused_poss}
\end{figure}

In Section~\ref{section:confused-matches}, a total of 30 possible counterparts were identified for the eight `confused' sources. The \textsc{Le Phare} photometric code was run on these possible counterparts in exactly the same way as detailed in Section~\ref{section:lephare}. The resulting redshift values for all the possible counterparts to each `confused' 10C source are shown in Fig.~\ref{fig:z-confused_poss}.

\section{Radio to optical ratio}\label{section:R}

The ratio of the radio and the optical flux densities of a source provides useful information about its nature, since for the same optical magnitude radio galaxies have much higher radio flux densities than either radio-quiet AGN or starforming galaxies. The radio-to-optical ratio $R$ was defined following \citet{1980ApJ...242..894C}, using:
\begin{equation}
R = S_{1.4~\rm GHz} \times 10^{0.4(m-12.5)}
\label{eqn:R}
\end{equation}
where $S_{1.4~\rm GHz}$ is the flux density at 1.4~GHz in mJy and $m$ is the optical magnitude in the $i$-band. Sources with radio-to-optical ratios $R > 1000$ are considered to be radio loud \citep{1999ApJS..123...41M}, while those with smaller values of $R$ are classified as radio quiet (and are therefore either radio-quiet AGN or starforming sources). $i$-band magnitudes are used here, as they are the best match to the $I$-band magnitudes used by \citeauthor{1999ApJS..123...41M}.

\begin{figure}
\centerline{\includegraphics[width=\columnwidth]{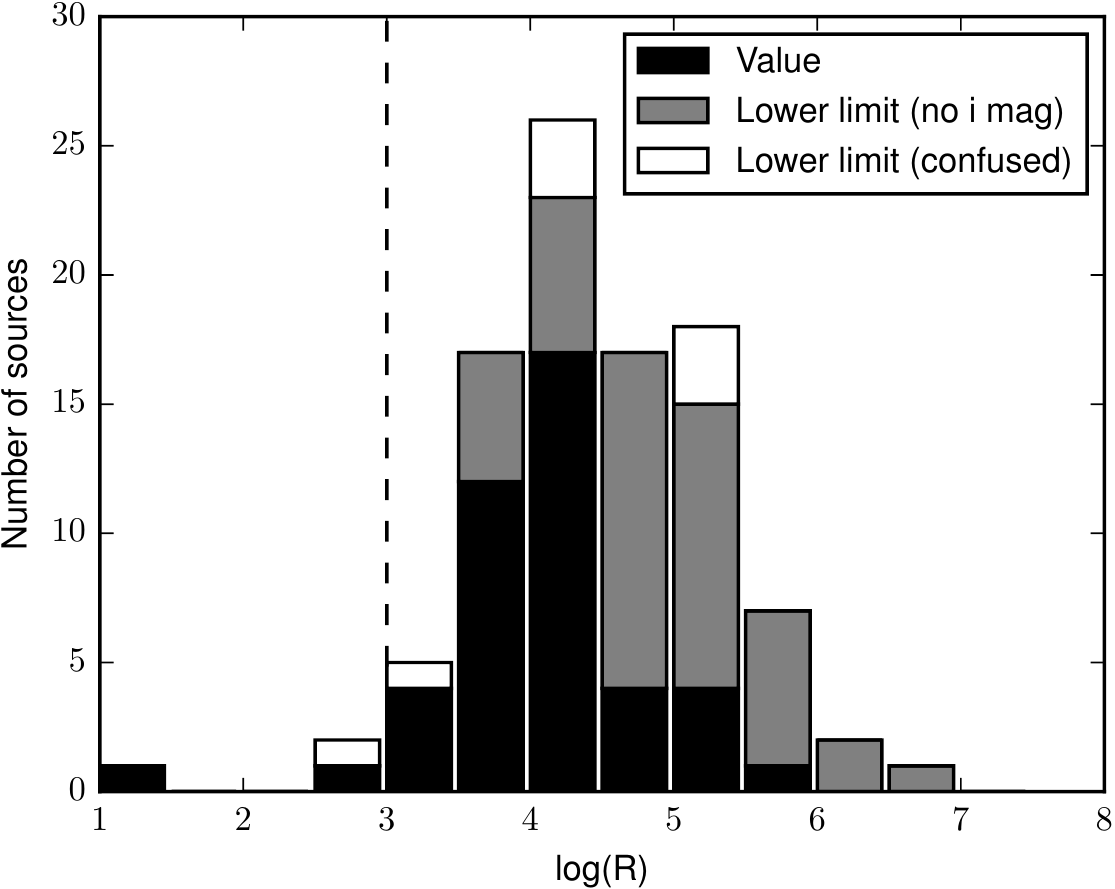}}
\caption{The distribution of radio-to-optical ratio, $R$. Sources with an $i$-band detection are in black. Lower limits are included for those sources without an $i$-band detection (grey) and for those classified as confused (white); these sources could move to the right on this diagram. Sources with $R > 1000$, indicated by the vertical dashed line, are considered radio loud.}\label{fig:ro_ratio}
\end{figure}

The radio-to-optical ratio, $R$, was calculated using equation \ref{eqn:R} for the sources which have an $i$-band magnitude available. Of the 80 sources with a match in the Data Fusion catalogue, 44 have an $i$-band magnitude, whilst 36 sources are not detected in the $i$-band.  For the sources without an $i$-band magnitude, including the eight sources which are unmatched in the Data Fusion catalogue, lower limits on $R$ are calculated using a lower limit on the $i$-band magnitude of 23.3 (the limiting magnitude of the $i$-band observations). For the eight sources classified as confused when matching, $R$ was calculated using the brightest of the possible counterparts identified in Section~\ref{section:confused-matches}, which serves as a lower limit on $R$.

The distribution of the radio-to-optical ratio of all sources is shown in Fig.~\ref{fig:ro_ratio}, including lower limits for unmatched and confused sources. Based on the study out to $z \sim 0.2$, \citet{1999ApJS..123...41M} found that 98 percent of normal and starburst galaxies had $R<1000$, so this is used as the cut-off point between radio-loud and radio-quiet objects. Fig.~\ref{fig:ro_ratio} shows that only three sources have $R<1000$, so using this criterion the other 93 are radio loud. 

There are far more detections in the SERVS mid-infrared bands than in the optical $i$-band traditionally used to calculate radio-to-optical ratios. It is therefore useful to calculate a `radio-to-optical' ratio based on 3.6-$\muup$m flux densities (referred to as the radio-to-infrared ratio, $R_{3.6~\rm \muup m}$, from now on) from SERVS as there are far fewer lower limits. There is the additional advantage that the 3.6-$\muup$m band remains longwards of the 4000~$\AA$ break out to redshifts of $z \sim 8$. By contrast, the $i$-band samples below the break at $z \gtrsim 1$; variations in starformation rate and absorption by dust have strong effects in that part of the spectrum and can introduce uncertainties in interpreting the $R$ values based on $i$-band magnitudes at higher redshifts \citep{2008MNRAS.386.1695S}. Calculating 3.6-$\muup$m radio-to-infrared ratios therefore also enables us to check that this is not having a major effect on our results.

We therefore define a radio-to-infrared ratio based on 3.6-$\muup$m flux densities ($R_{3.6~\rm \muup m}$) as follows:

\begin{equation}
R_{3.6~\rm \muup m} = \frac{S_{1.4~\rm GHz}}{S_{3.6~\rm \muup m}}
\end{equation}

\noindent 80 sources have a 3.6-$\muup$m flux density available, including one source where the flux density is not included in the SERVS catalogue and is estimated from the map (see Section~\ref{section:matching-details}). Lower limits on $R_{3.6~\rm \muup m}$ were calculated for the eight sources without a 3.6-$\muup$m band detection using the SERVS 3.6-$\muup$m 5$\sigma$ detection limit as an upper limit on the 3.6-$\muup$m flux density of the sources. Lower limits on $R_{3.6~\rm \muup m}$ were calculated for the eight confused sources using the highest 3.6-$\muup$m flux density of any of the possible counterparts for each confused source. 

\begin{figure}
\centerline{\includegraphics[width=\columnwidth]{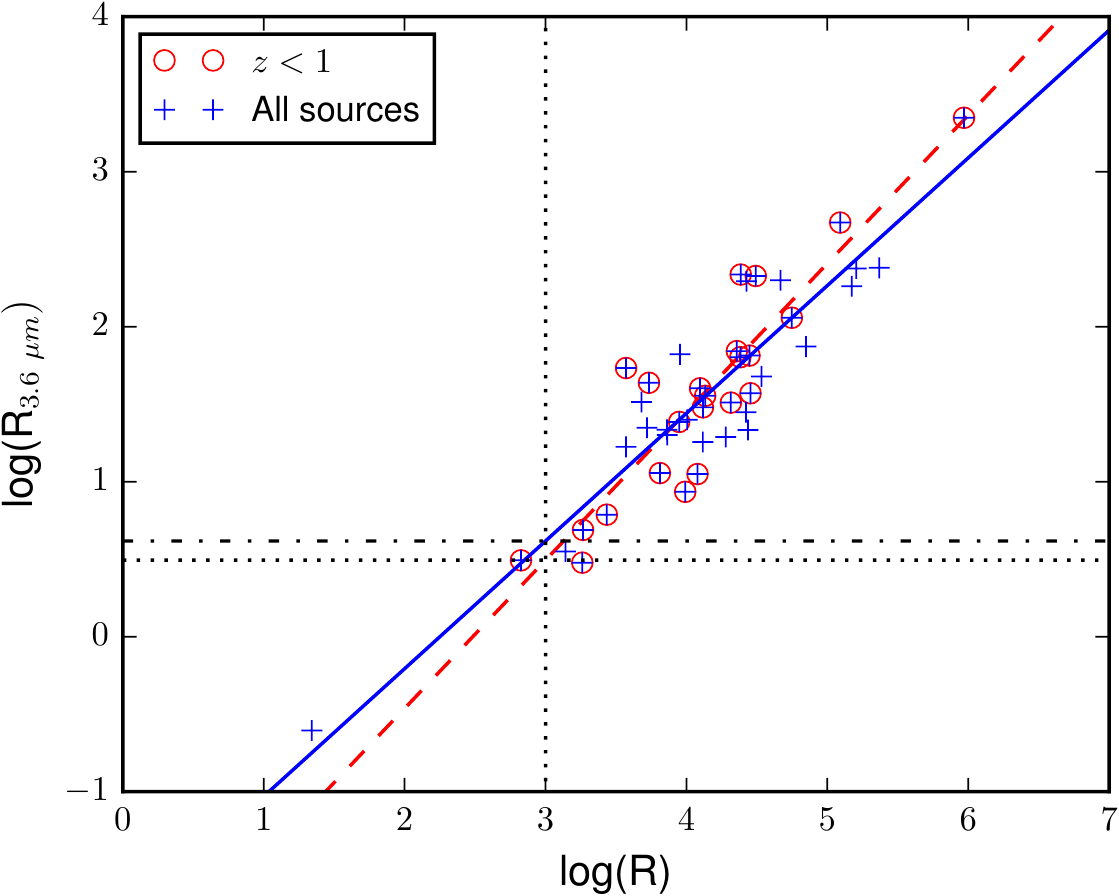}}
\caption{Radio-to-optical and infrared ratios for $i$-band and 3.6~$\muup$m respectively. All sources with values of both $R$ and $R_{3.6~\rm \muup m}$ available are shown as (blue) '+', those which are at redshifts of $z<1$ are circled (in red in the online version). The line of best fit to all sources is shown as a (blue) solid line, the line of best fit to the sources with $z<1$ is shown as a (red) dashed line. The vertical dotted line is at $R=1000$, the value used to distinguish between radio-loud and radio-quiet sources. The horizontal dotted line shows the $R_{3.6~\rm \muup m}$ dividing value derived from the best fit line to sources with $z<1$ and the horizontal dot-dashed line shows the $R_{3.6~\rm \muup m}$ cut-off derived from the best fit line to all sources with both $R$ values available.}\label{fig:R_fit}
\end{figure}

\begin{figure}
\centerline{\includegraphics[width=\columnwidth]{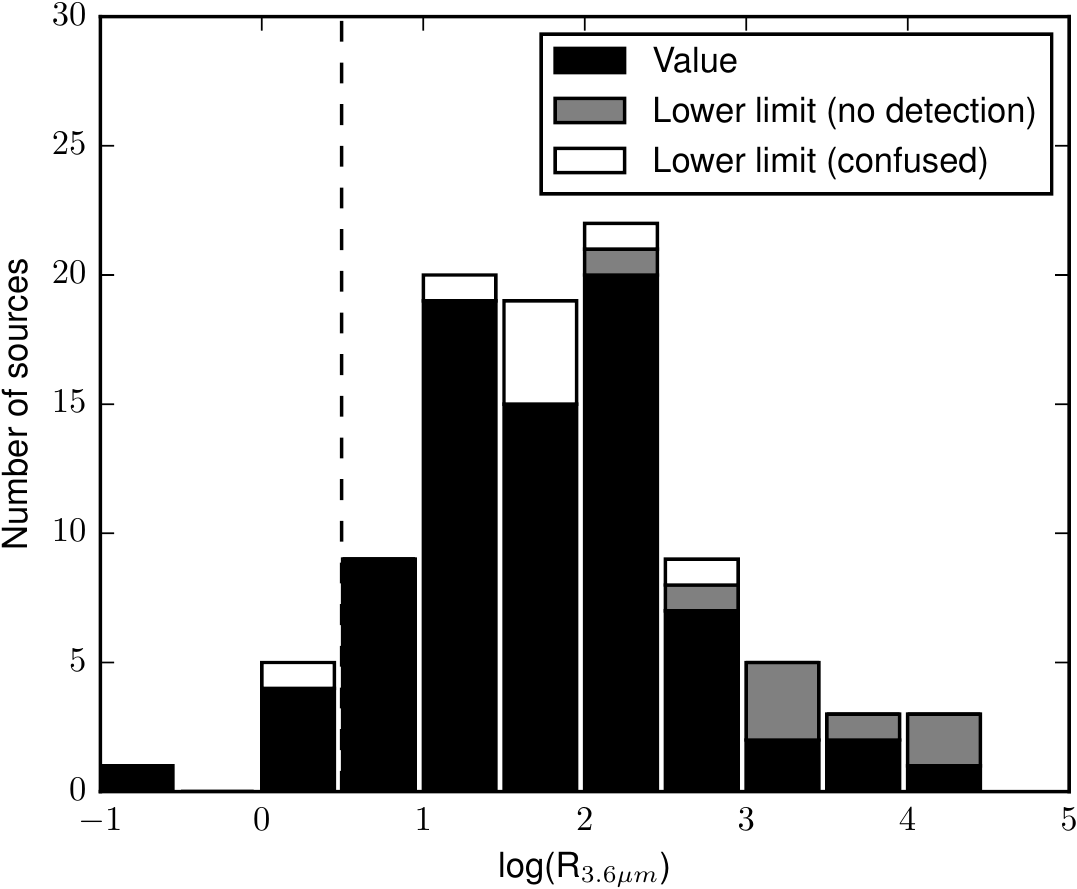}}
\caption{The distribution of radio-to-infrared ratio, $R_{3.6 \muup \rm m}$. Sources with a 3.6~$\muup$m-band detection are in black. Lower limits are included for those sources without an 3.6~$\muup$m-band detection (grey) and for those classified as confused (white); these sources could move to the right on this diagram. Sources with $R_{3.6 \muup \rm m} > 3.1$, indicated by the vertical dashed line, are considered radio loud.}\label{fig:r36_ratio}
\end{figure}

The sources with both $R$ and $R_{3.6~\rm \muup m}$ values available (43 sources) were used to define the value of $R_{3.6~\rm \muup m}$ used as the cut-off between radio-quiet and radio-loud sources. A linear best fit line was fitted to the data, and this was used to estimate an  $R_{3.6~\rm \muup m}$ cut-off equivalent to $R=1000$, as shown in Fig.~\ref{fig:R_fit}. This gave a cut-off value of $R_{3.6~\rm \muup m} = 4.1$. In order to check that the sources with redshifts greater than $z \approx 1$, where there could be uncertainties due to sampling below the 4000~$\AA$ break in the $i$-band, were not having a significant effect on this choice of cut-off the analysis was repeated using sources with redshifts of $z < 1$ only. The redshifts estimated in Section~\ref{section:lephare} were used to perform the redshift cut, which resulted in 20 sources being excluded from the analysis. Both fits and the cut-offs derived from them are shown in Fig.~\ref{fig:R_fit}. Excluding the sources with $z>1$ resulted in a lower cut-off of $R_{3.6~\rm \muup m} = 3.1$. This value is used to distinguish between radio-quiet and radio-loud sources in the following analysis. A histogram of the $R_{3.6~\rm \muup m}$ values is shown in Fig.~\ref{fig:r36_ratio}.

\begin{figure}
\centerline{\includegraphics[width=\columnwidth]{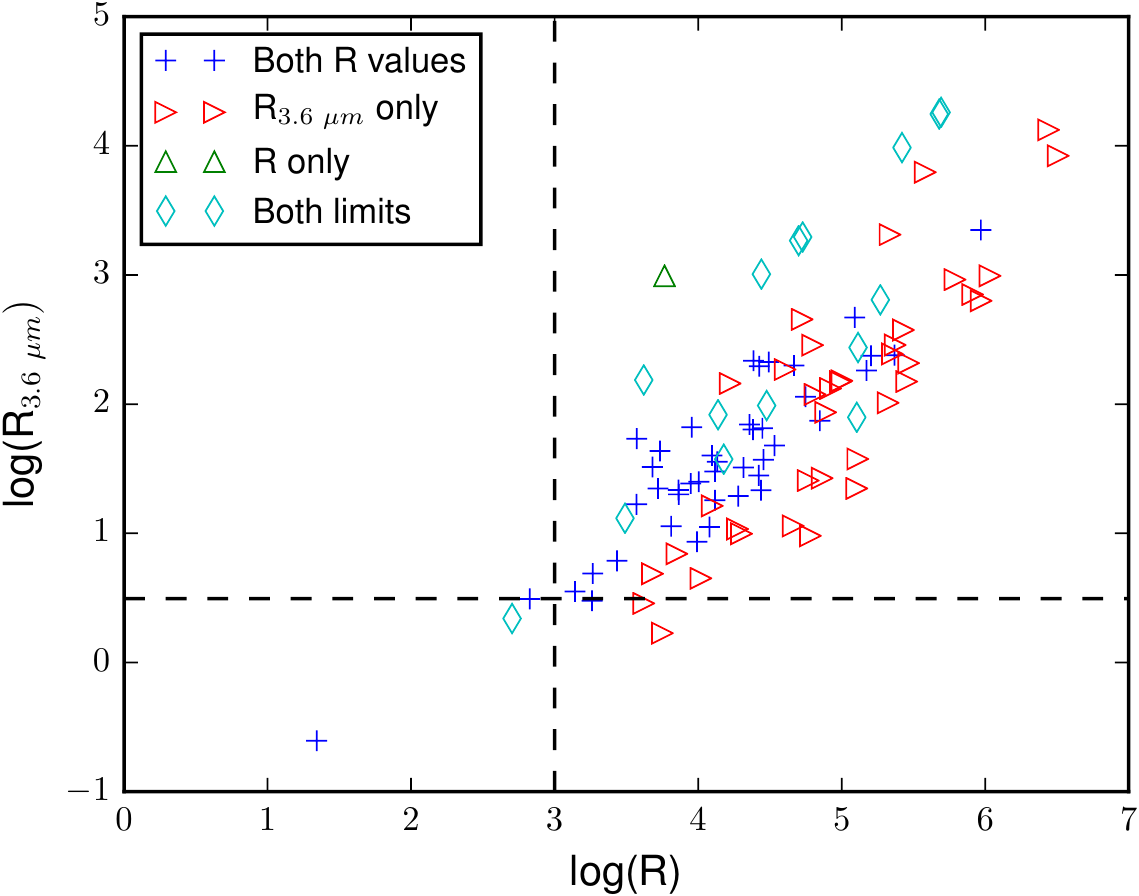}}
\caption{Radio-to-optical and radio-to-infrared ratios for all sources. Sources which only an infra-red $R_{3.6~\rm \muup m}$ value have a lower limit on $R$ and could therefore move to the right. Sources which only have an optical $R$ value could move up. Sources with a lower limit on both values of $R$ could move up or to the right. The dashed lines show the values of $R$ used to distinguish between radio-quiet and radio-loud sources. }\label{fig:R_all}
\end{figure}

Fig.~\ref{fig:R_all} shows the radio-to-optical and radio-to-infrared ratios for all 96 sources in the sample, including lower limits where necessary. It is clear from this figure that only one source is significantly radio quiet, while several other sources lie close to the radio-loud/radio-quiet boundary. There are six sources which could be classified as radio quiet using one of the two ratios; three sources are classified as radio quiet using the $R_{3.6~\rm \muup m}$ criterion only and three are radio quiet using both criteria (one of these classifications is based on lower limits on both ratios so the source could actually be radio loud). Therefore at least 90 out of the 96 sources (94 per cent) in this sample are radio loud; in fact, given how close to the boundary five of the remaining six sources are, it is likely that they are also dominated by AGN activity. This backs up the conclusions of \citet{2014MNRAS.440...40W}, in which we used VLBI data to show that at least 65 per cent of the 10C sources are associated with an AGN. Thus at $\sim 1$~mJy the faint, high-frequency sky is still dominated by radio-loud AGN rather than by starforming galaxies or radio-quiet AGN.

\subsection{Far-infrared -- radio correlation}

Radio-loudness can also be defined in terms of the far-infrared -- radio correlation (e.g.\ \citealt{2008MNRAS.386..953I,2015arXiv150701144L}) which provides a useful comparison as it gives a view of radio loudness not affected by the obscuration of the AGN in the optical. The far-infrared -- radio correlation is often quantified using the far-infrared -- radio ratio $q_{IR}$ (e.g.\ \citealt{2004ApJS..154..147A,2009MNRAS.394..105G}), where
\begin{equation}
q_{IR} = \rm log_{10}(S_{24~\muup m} / S_{1.4~\rm GHz})
\end{equation}
Starforming sources and radio-quiet AGN are expected to have positive $q_{IR}$ values, for example, \citet{2007ApJ...663..218M} found that the typical value of $q_{IR}$ for starforming galaxies was $0.83 \pm 0.31$. Radio-loud objects tend to have much lower values of $q_{IR}$, with typical values $-0.6$ to $-1.2$ \citep{2010arXiv1008.4918P}. 

\begin{figure}
\centerline{\includegraphics[width=\columnwidth]{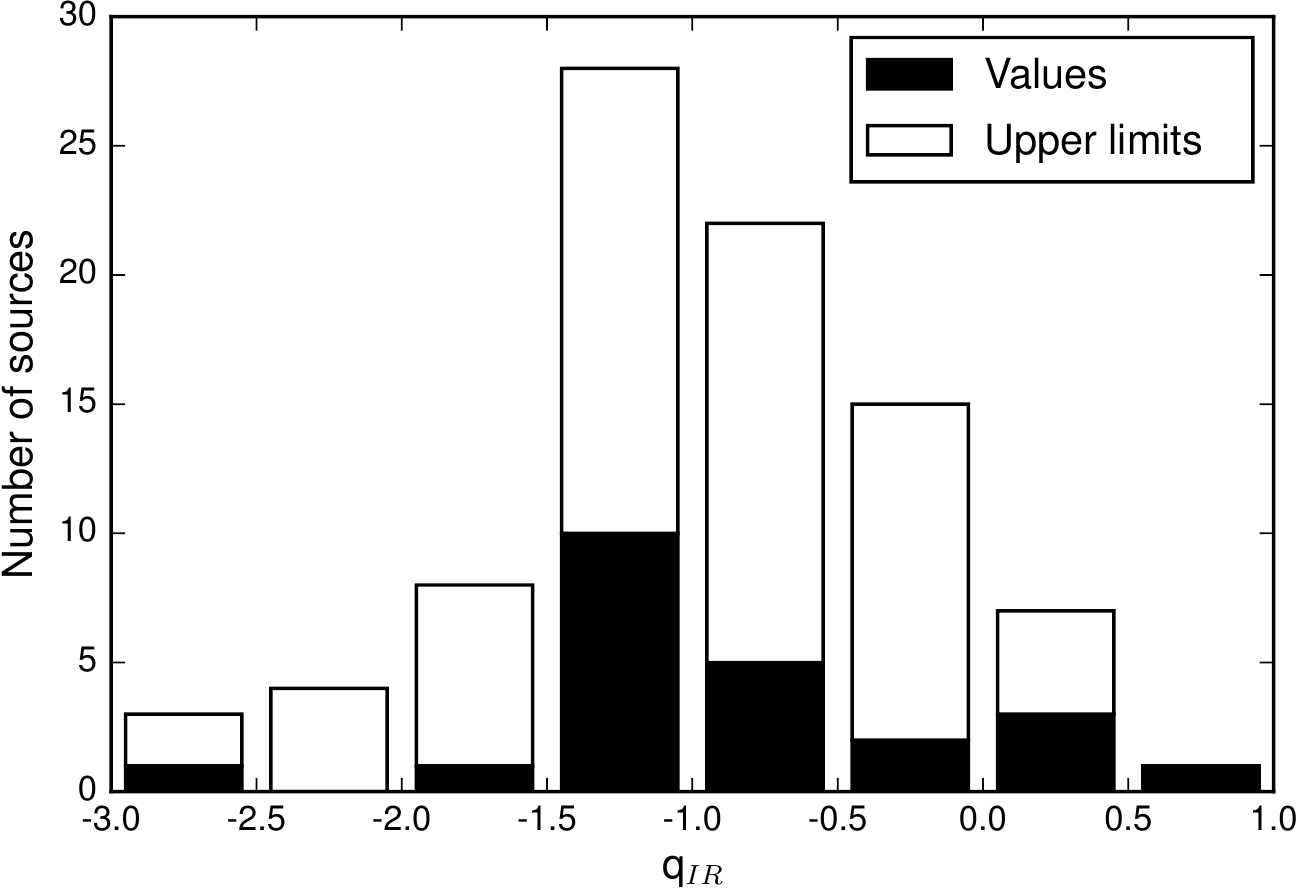}}
\caption{Far-infrared -- radio correlation parameter $q_{IR}$. Upper limits are shown in white and could move to the left.}\label{fig:qIR}
\end{figure}

23 sources have a 24~$\muup$m-flux density available from the SWIRE catalogue and these were used to calculate $q_{IR}$ values, which are shown in Fig.~\ref{fig:qIR}. For the undetected sources, the $5\sigma$ detection limit of 450~$\muup$Jy was used to calculate an upper limit on $q_{IR}$. Fig.~\ref{fig:qIR} shows that the majority of the sources have $q_{IR}$ values consistent with those expected for radio-loud AGN. Only eight sources have $q_{IR} > 0$, four of which are upper limits so could have much lower values, and only one source has $q_{IR} > 0.5$ (this is the same source that was the the only source to be identified as significantly radio quiet in Fig.~\ref{fig:R_all}). This provides further evidence that this sample is entirely dominated by radio-loud sources, and shows that this result is not affected by obscuration of the AGN in the optical.

\subsection{Correlations between $R$ and spectral index}\label{section:radio-to-optical-properties}

\begin{figure*}
\centerline{\includegraphics[width=\columnwidth]{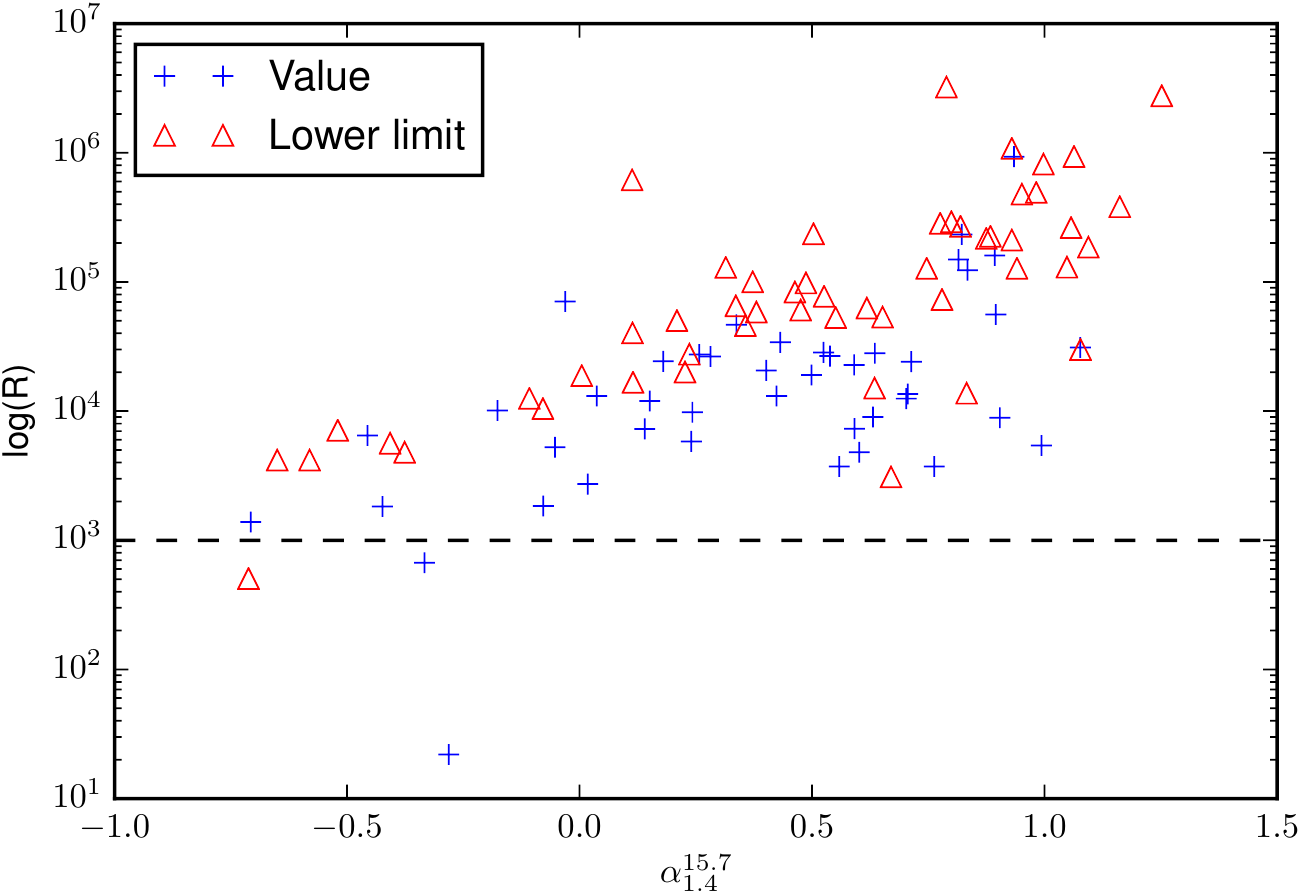}
            \quad
            \includegraphics[width=\columnwidth]{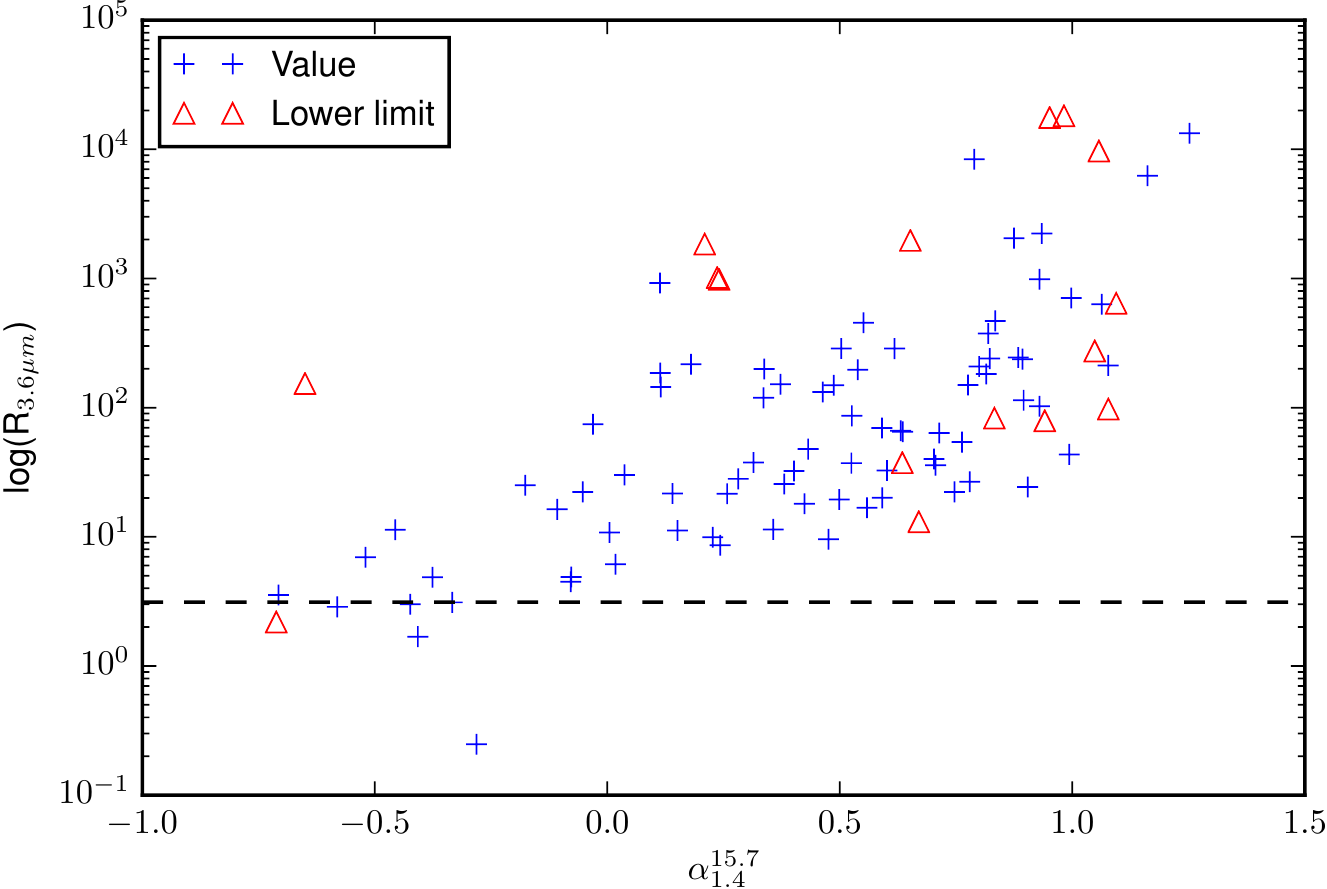}}
\caption{Radio-to-optical and radio-to-infrared ratios as a function of spectral index. Left-hand panel shows $R$ and right hand panel shows $R_{3.6~\rm \muup m}$. Triangles (red in the online version) indicate lower limits on the ratio. The dashed line indicates the cut-off used to distinguish between radio-loud and radio-quiet sources ($R = 1000$ in the left panel and $R_{3.6~\rm \muup m} = 3.1$ in the right panel).}\label{fig:r_alpha}
\end{figure*}

\begin{figure*}
\centerline{\includegraphics[width=\columnwidth]{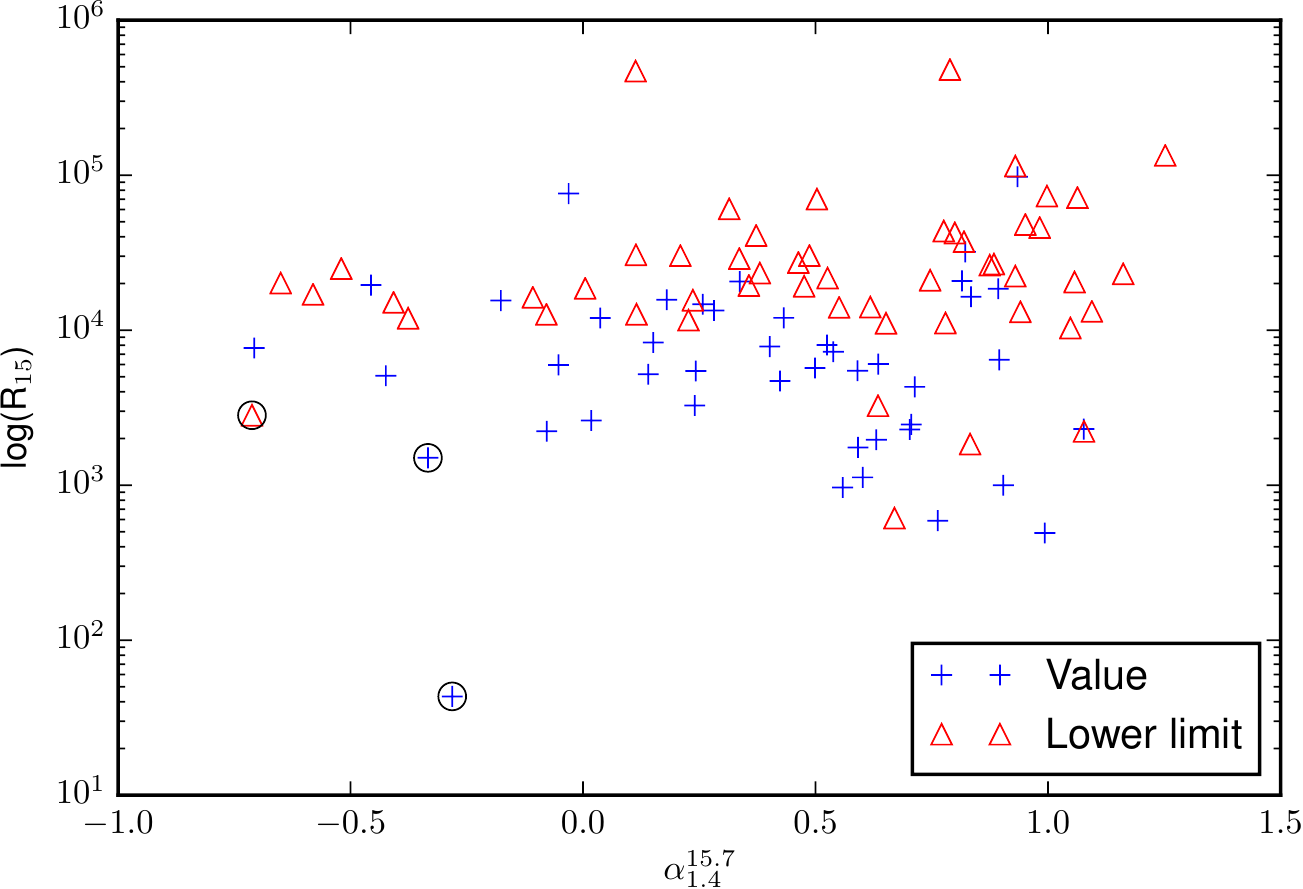}
            \quad
            \includegraphics[width=\columnwidth]{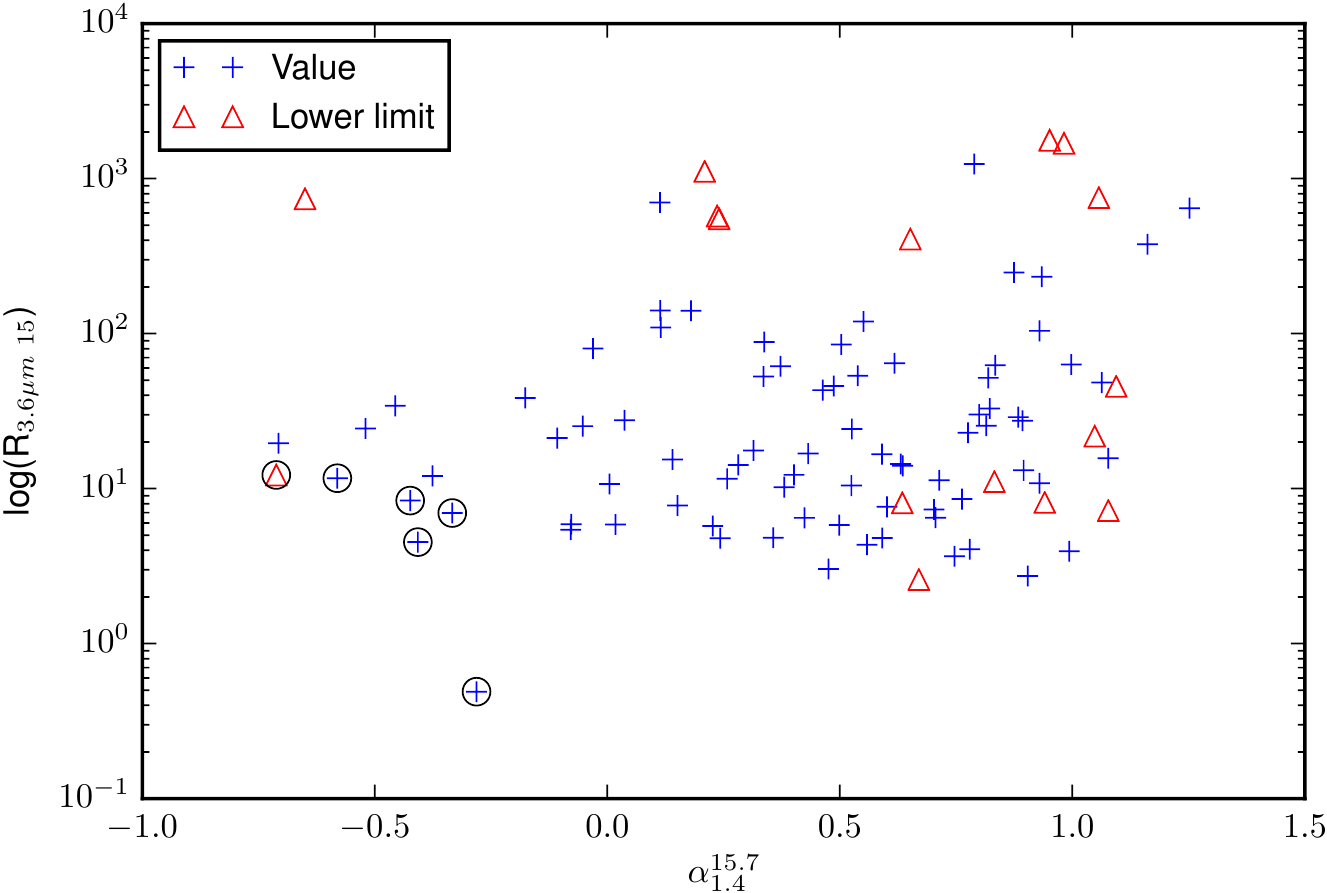}}
\caption{Radio-to-optical and radio-to-infrared ratios calculated using 15.7-GHz flux densities as a function of spectral index. Left-hand panel shows $R_{15}$ and right hand panel shows $R_{3.6~\rm \muup m ~ 15}$. Triangles (red in the online version) indicate lower limits on ratios. The three sources which are classified as radio-quiet based on their $R$ value are circled in the left panel and the six sources classified as radio-quiet based on their $R_{3.6~\rm \muup m}$ values are circled in the right-hand panel.}\label{fig:r15_alpha}
\end{figure*}

Fig.~\ref{fig:r_alpha} shows the radio-to-optical and radio-to-infrared ratios as a function of radio spectral index. There appears to be a positive correlation between both ratios and spectral index; this is, however, a selection effect as the sample is selected at 15.7~GHz but 1.4-GHz flux densities are used to calculate the ratios. To test for any real correlation between the ratios and spectral index radio-to-optical and radio-to-infrared ratios were calculated using 15.7-GHz flux density, as follows,
\begin{equation}
R_{15} = S_{15.7~\rm GHz} \times 10^{0.4(m-12.5)}\\
R_{3.6~\rm \muup m ~ 15} = \frac{S_{15.7~\rm GHz}}{S_{3.6~\rm \muup m}}
\label{eqn:radio-to-optical15}
\end{equation}
These ratios are shown as a function of spectral index in Fig.~\ref{fig:r15_alpha}. The positive correlation is no longer seen, confirming that this effect was due to the sources being selected at 15.7~GHz. 

Fig.~\ref{fig:r_alpha} shows that all six sources classified as radio quiet using either or both ratios have rising spectra ($\alpha < 0$). This rising spectral shape means these sources have lower 1.4-GHz flux densities than the majority of the 10C sample, which could explain why they have small radio-to-optical and radio-to-infrared ratios. The three sources classified as radio quiet based on their radio-to-optical ratios and the six sources classified as radio quiet based on their radio-to-infrared ratios are circled in black on the left and right panels of Fig.~\ref{fig:r15_alpha} respectively. As for the ratios determined at 1.4~GHz, only one of these radio-quiet sources has 15.7-GHz radio-to-optical and radio-to-infrared ratios significantly lower than the rest of the sample; the other radio-quiet sources all have 15.7-GHz ratios similar to the rest of the sample. This significantly radio-quiet source (10C J105028+574522) has the lowest redshift in the sample, with $z=0.072$ (spectroscopic redshift) and has a rising spectra.

This analysis shows that all of the radio-quiet sources have rising spectra. They are unlikely to be starforming galaxies, as starforming galaxies only have rising spectra at 15~GHz at $z \gtrsim 3$ \citep{2014arXiv1412.5677M}, and although our data is very deep it is not deep enough to detect starforming galaxies at $z > 3$ unless they have a prodigious amount of star formation. These sources are therefore likely to be dominated by AGN emission.

\section{Radio source properties}\label{section:properties_z}

In this section, the radio properties of the sources are considered in light of the redshift values derived in Section~\ref{section:lephare}. Redshift values are available for 78 sources, nearly a third of which are spectroscopic reshifts.  The large errors on some of the redshift estimates, along with the significant discrepancies between catalogues in some cases, mean that photometric methods cannot be used to produce a reliable redshift for any individual source. They can, however, be used to provide valuable information about the properties of the population as a whole.

\subsection{Summary of the radio properties of the sample}

The radio properties of the sources in this sample (along with those of the full 10C Lockman Hole sample) were presented in Paper~I. Flux densities are available at a range of frequencies including 1.4 and 15.7~GHz, and radio spectral indices are calculated for all sources but one (an upper limit is available for the one remaining source). The sources were split into flat-spectrum sources, with $\alpha < 0.5$, and steep-spectrum sources with $\alpha > 0.5$.

\subsection{Redshift distribution}

\begin{figure}
\centerline{\includegraphics[width=\columnwidth]{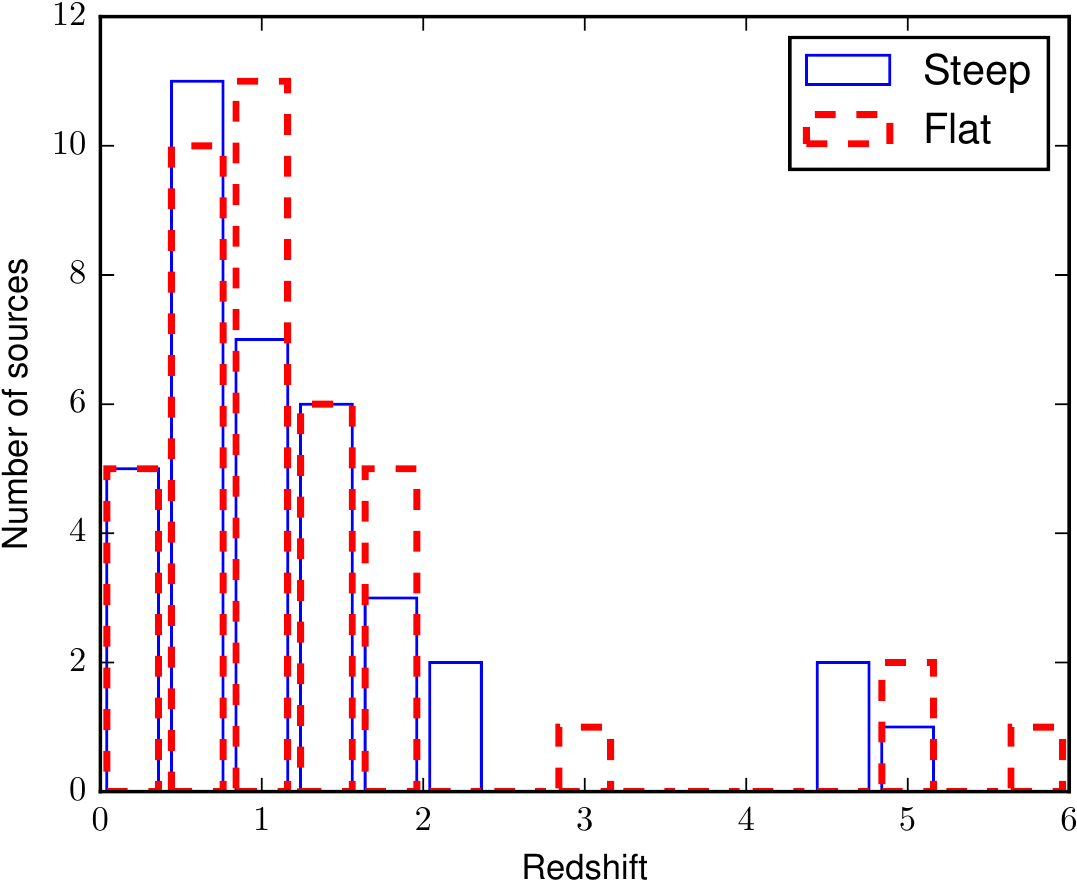}}
\caption{Redshift distribution for all sources with a redshift value or estimate, with steep and flat spectrum sources shown separately.}\label{fig:z-dist-properties}
\end{figure}

There are redshift values or estimates for 77 out of the 96 sources in this sample (Section~\ref{section:lephare}). The redshift distributions of steep and flat spectrum sources are shown separately in Fig.~\ref{fig:z-dist-properties}; the redshift distributions for the two samples appear fairly similar, although the distribution for flat spectrum sources peaks at a slightly higher redshift. A KS test was performed on the two samples and the probability of them being drawn from the same population was 0.16, indicating that the two distributions are not significantly different. 

\subsection{Luminosity distribution}\label{section:luminosity}

Luminosities were calculated for all sources with a redshift value or estimate. The luminosities were $k$-corrected based on their radio spectral index using the following expression:
\begin{equation}
L_\nu = 4 \pi d_L(z)^2 S_ \nu [(1 + z)^{\alpha - 1}]
\label{eqn:lum}
\end{equation}
where $d_L$ is the luminosity distance. 

\begin{figure}
\centerline{\includegraphics[width=\columnwidth]{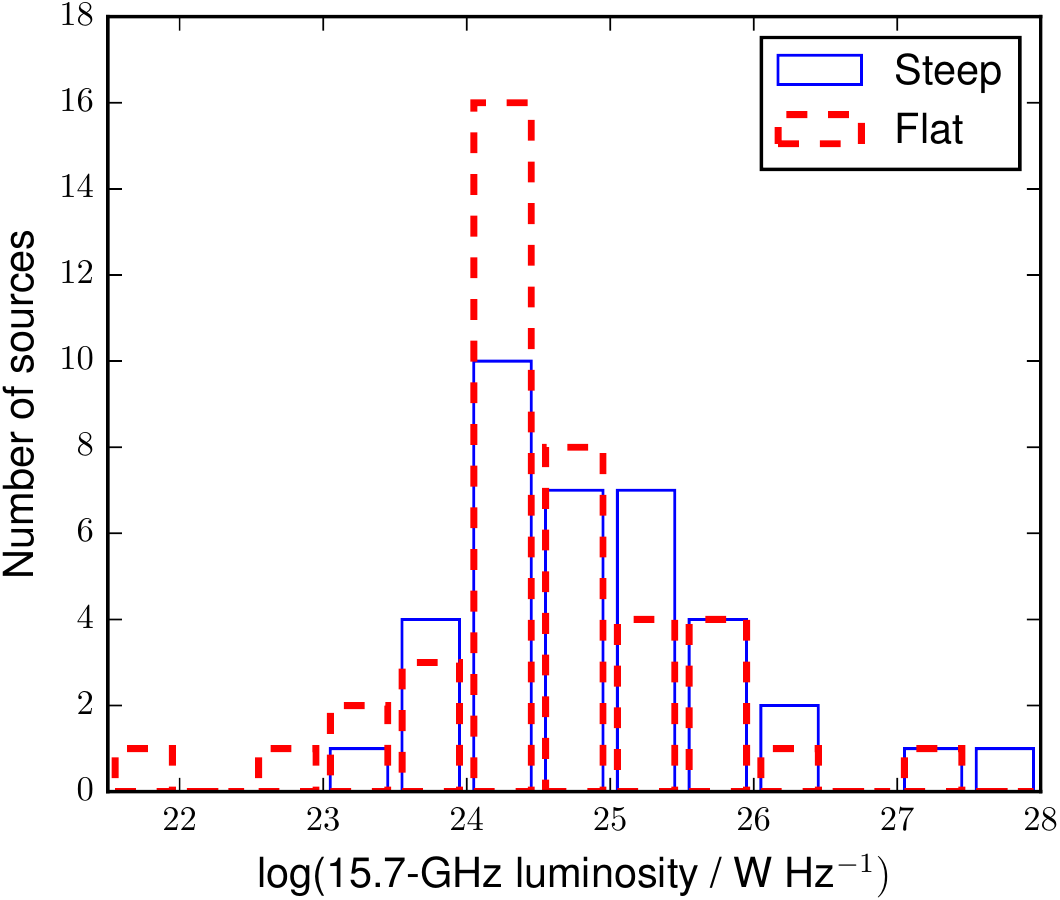}}
\caption{15.7-GHz luminosity distribution for all sources with a redshift estimate; the sample is divided into flat and steep spectrum sources.}\label{fig:lum15-dist}
\end{figure}

The 15.7-GHz luminosity distribution for all these sources is shown in Fig.~\ref{fig:lum15-dist}, with steep and flat spectrum sources shown separately. The distributions for the steep and flat-spectrum sources are very similar, suggesting there is no difference in luminosity between the two populations. The sources display a large range of 15.7-GHz luminosities, comparable with those of the powerful FRI and FRII radio galaxies \citep{1974MNRAS.167P..31F}.

\subsection{Linear size distribution}\label{section:sizes}

\begin{figure*}
\centerline{\includegraphics[width=\columnwidth]{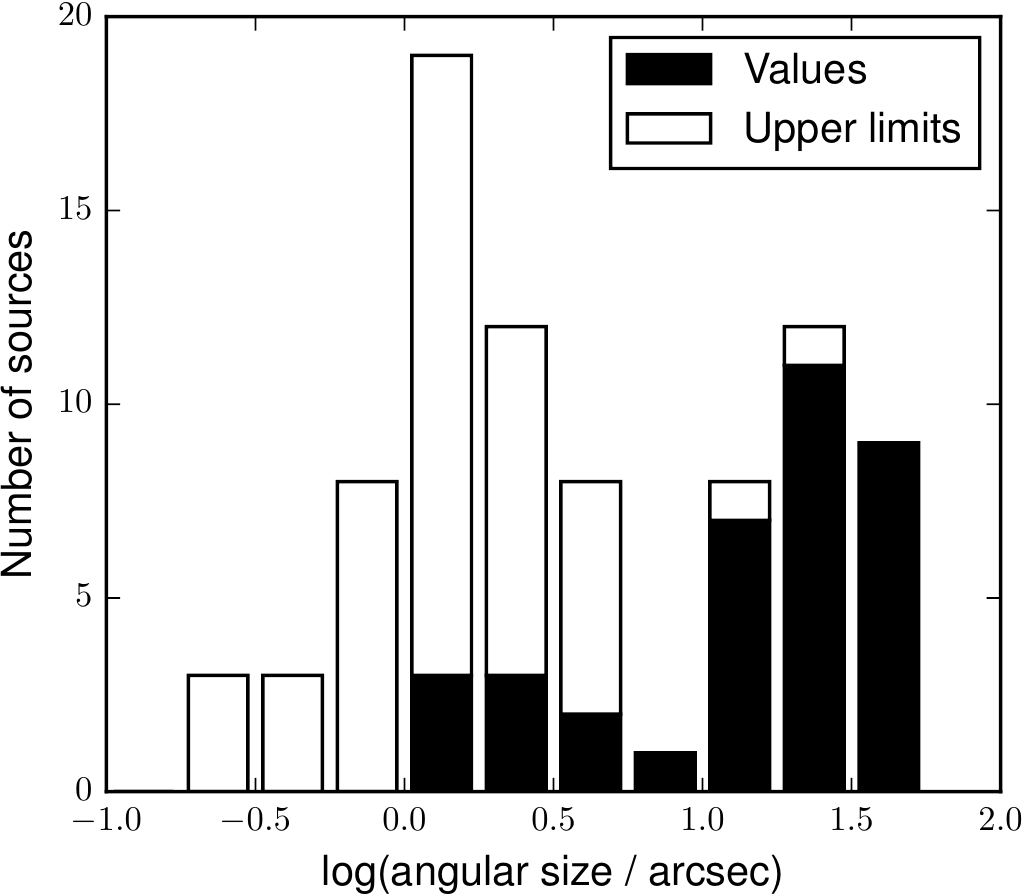}
            \quad
            \includegraphics[width=\columnwidth]{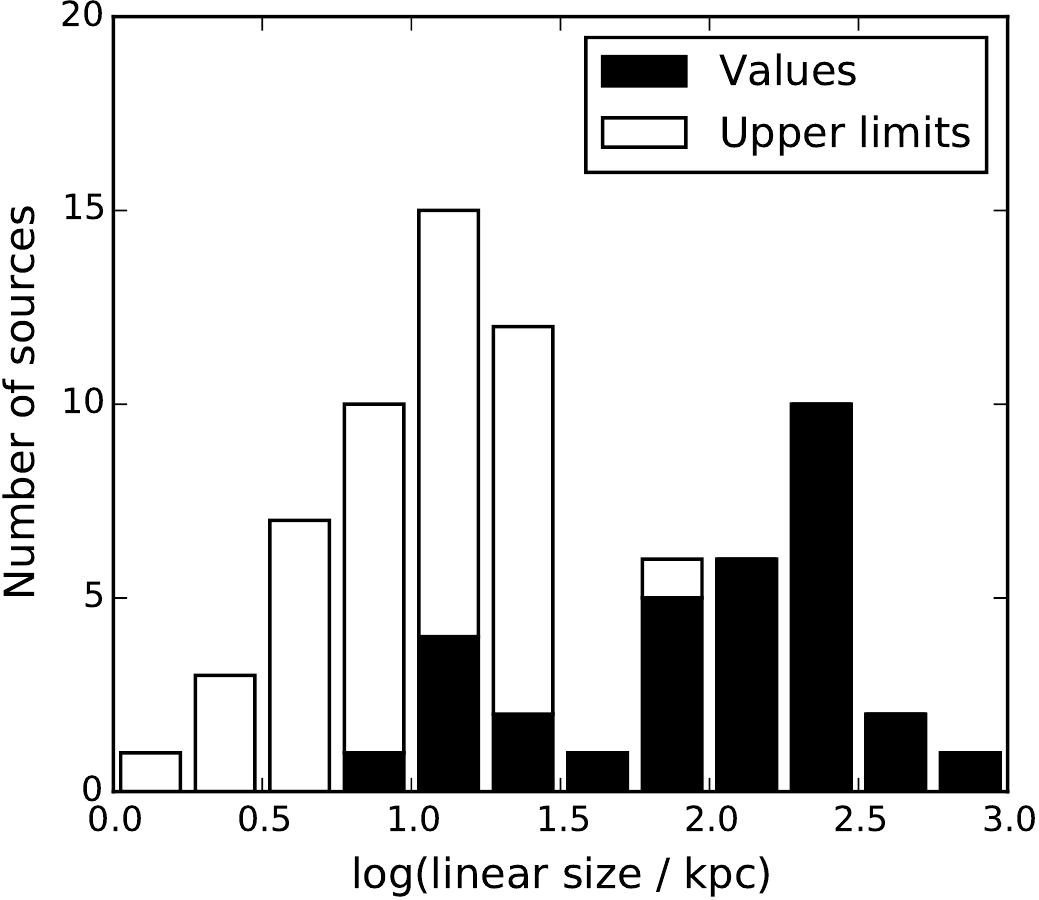}}
\caption{Angular and linear size distributions. The left panel shows angular sizes for all sources in the sample and the right panel shows linear sizes for the 77 sources with a redshift available. Upper limits are shown in white, and could move to the left.}\label{fig:ang_size}
\end{figure*}

To calculate linear sizes we need to first estimate angular sizes for the sources. Angular size information is available on a range of scales from the different radio catalogues available in the field. Best estimates of the angular sizes are compiled from these catalogues for the 96 sources in this sample. The two deep VLA surveys, BI2006 and OM2008, are the highest resolution surveys, with resolutions of 1.3 and 1.6 arcsec respectively, so provide information on the smallest angular scales. However, the angular sizes from these two catalogues may not be reliable for significantly extended sources, as some structure may be resolved out due to a lack of short baselines. Additionally, some sources may be resolved into multiple components which are then listed separately in the catalogue, so the size listed in the catalogue would be for only part of the source. For this reason, for any source which was classified as extended in Paper~I the angular size was estimated from the \citet{2008MNRAS.387.1037G} GMRT observations instead. The angular sizes for these sources were measured by hand, to avoid any problems caused by fitting Gaussians to extended sources with complicated structures. For the remaining sources (which were not classified as extended) the angular size was taken from the catalogue with the highest resolution available. The maximum angular size from the relevant catalogue is used, with the catalogues being used in the following order of preference: OM2008/BI2006, FIRST, WSRT (highest to lowest resolution). A value is flagged as an upper limit on the angular size if the size listed is less than the synthesised beam size for those observations.

Two sources required different treatment. The very extended source 10CJ105437+565922 is not detected in FIRST or GMRT as it has diffuse low-brightness structure; its size is measured from the NVSS map. The other source, 10CJ104927+583830, is only in the 10C catalogue, so for this source the 10C beam size (30 arcsec) is used as an upper limit on the angular size of the source. 

Linear sizes were then calculated from these angular sizes for the 78 sources with redshift information using:
\begin{equation}
D = \frac{\theta d_L}{(1+z)^2}
\label{eqn:size}
\end{equation}
where $D$ is the linear size of the source, $d_L$ is the luminosity distance and $\theta$ is the angular size of the source in radians. The angular and linear size distributions are shown in Fig.~\ref{fig:ang_size}; the left panel shows the angular sizes and the right panel shows the linear sizes. Note that the 18 sources without a redshift value are not included in the linear size plot. 

\begin{figure}
\centerline{\includegraphics[width=\columnwidth]{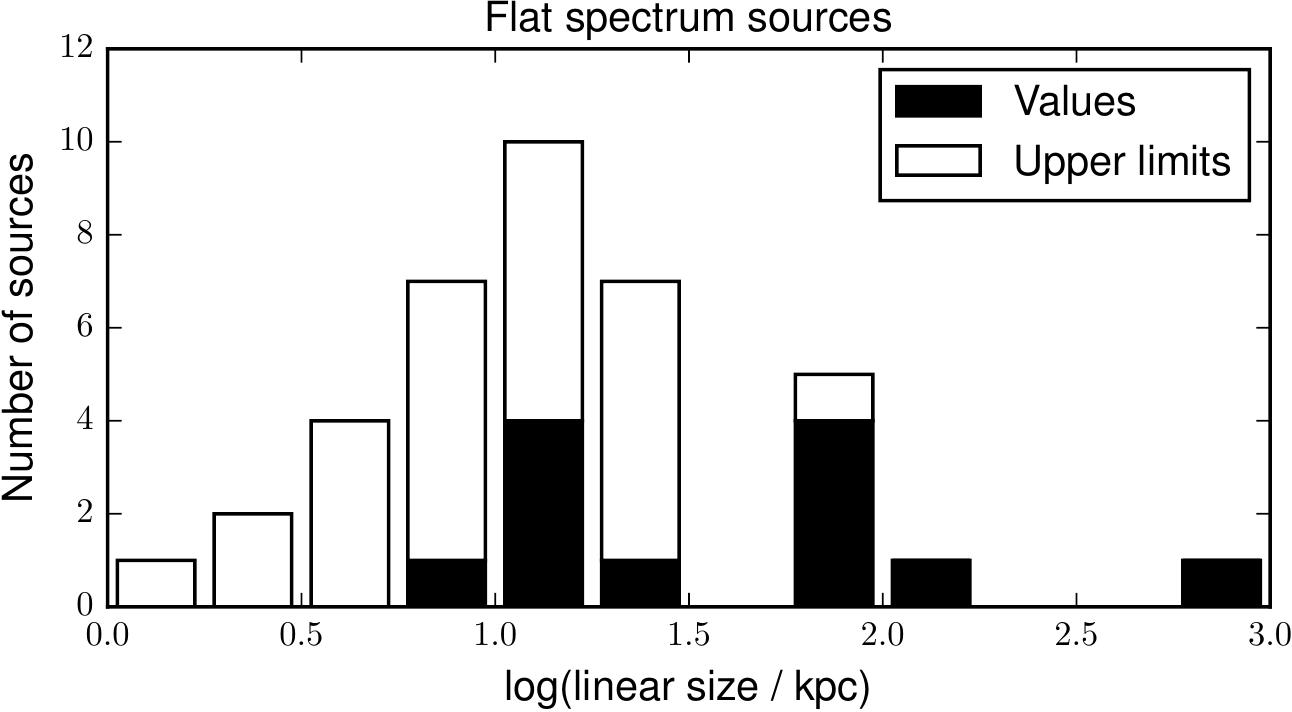}}
\smallskip
\centerline{\includegraphics[width=\columnwidth]{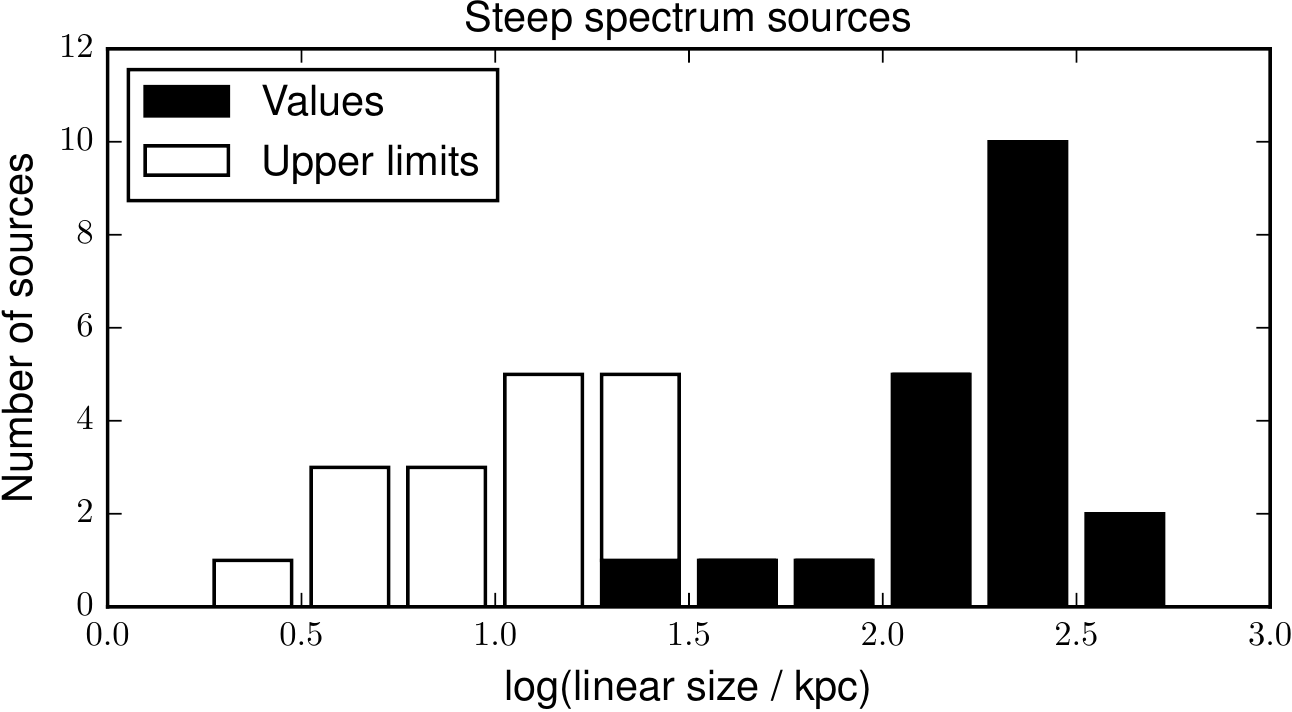}}
\caption{Linear size distribution for sources with redshift values. The top panel shows flat spectrum sources and the bottom panel shows sources with steep spectra. Upper limits are shown in white, and could move to the left on these plots.}\label{fig:linear_size_alpha}
\end{figure}

Fig.~\ref{fig:linear_size_alpha} shows the linear size distributions for flat and steep spectum sources. It is evident that the flat spectrum sources are on average smaller than the steep spectrum sources. This is expected as the extended lobes present in many radio galaxies have steep spectra due to optically-thin synchrotron emission, while the cores of radio galaxies generally have flat spectra due to self-absorption. The spectral index of a source therefore informs us about the relative contributions of the cores and lobes to the total flux density of the source; if a source has a steep spectrum its emission is dominated by the lobes, and it is therefore more likely to appear extended. If, however, a source has a flat spectrum it is dominated by emission from its core, so its lobes may be very weak or not visible at all, and the source appears more compact.

While this trend is true for the majority of the population, both steep and flat spectrum sources cover the full range of linear sizes ($1 \lesssim D \lesssim 1000~\rm kpc$), so there are a number of small steep spectrum sources and large flat spectrum sources. This is consistent with the
findings in Paper I and \citet{2014MNRAS.440...40W} which indicate that there are small populations of both extended, flat spectrum sources
and compact, steep spectrum sources in the 10C sample.

\begin{figure}
\centerline{\includegraphics[width=\columnwidth]{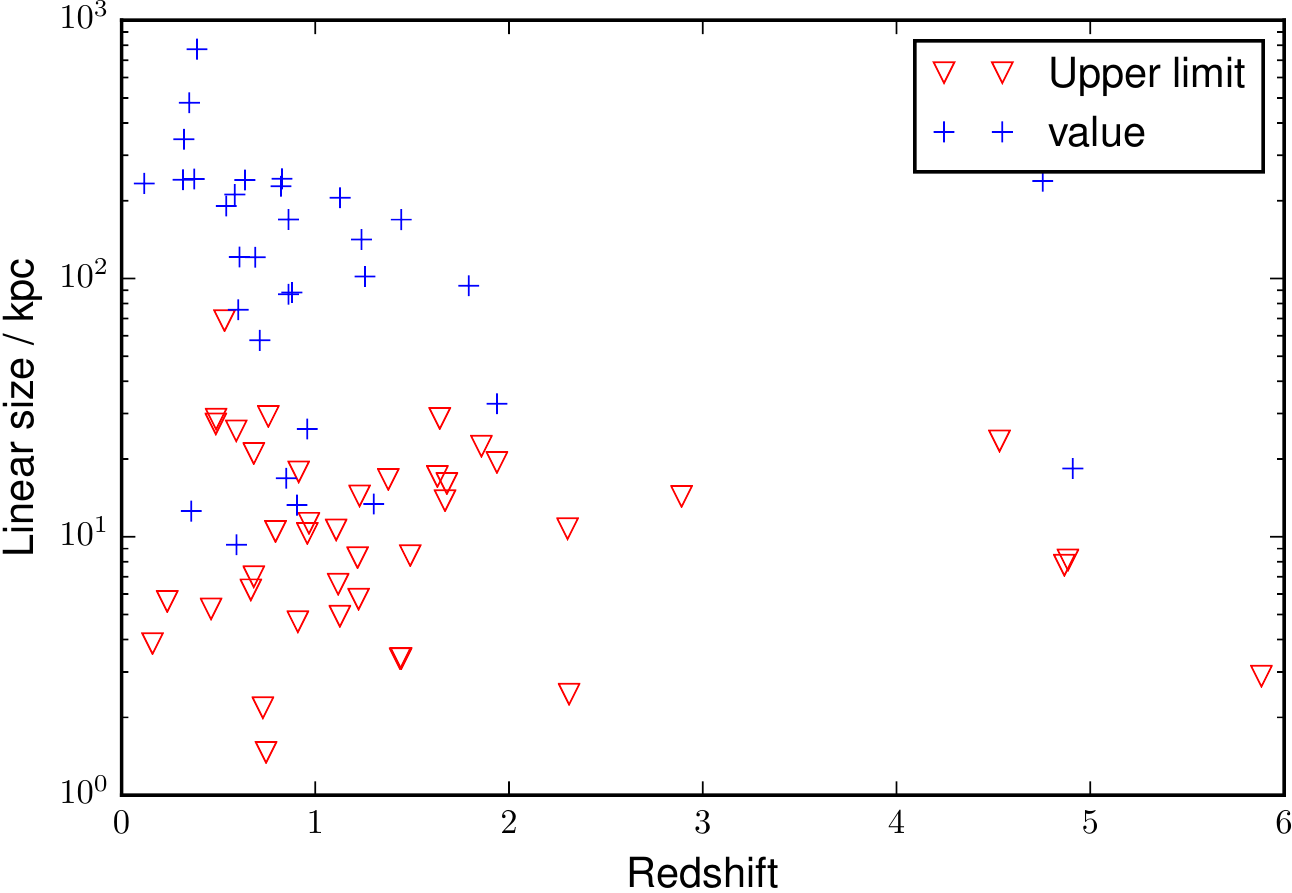}}
\caption{Linear size as a function of redshift. Triangles (red in the online version) are upper limits on size, crosses (blue in the online version) are values.}\label{fig:z_size}
\end{figure}

Fig.~\ref{fig:z_size} shows linear size as a function of redshift for all sources in the sample used in this paper. This shows that the sources with larger linear sizes tend to be a lower redshifts, and at $z > 2$ all but two sources are unresolved.

\section{Comparison with the SKADS Simulated Sky}\label{section:s3_z}

In Paper~I a sample of sources with $S_{18~\rm GHz} > 0.5$~mJy was selected from the S$^3$ catalogue; this sample should be directly comparable with the 10C sample. The radio properties of this S$^3$ sample were compared to the 10C sample, which showed that the simulation fails to accurately reproduce the spectral index distribution of the observed sample. The number of flat spectrum sources is massively underpredicted; there are essentially no sources in the simulated sample with $\alpha < 0.3$, while 40 percent of the 10C sample have $\alpha^{15.7}_{0.61} < 0.3$. 

The simulation predicts that the 10C sample should be dominated by FRI sources, making up 71 percent of the population, while FRII sources are the second-largest source type (13 per cent). Starforming sources make up seven per cent of the simulated sample. There is some indication that this may be an overestimate, as we find no starforming sources in the 10C sample. However, we are sampling a small area so cosmic variance is very high at low redshifts and could therefore be the cause of this discrepancy.

\begin{figure}
\centerline{\includegraphics[width=\columnwidth]{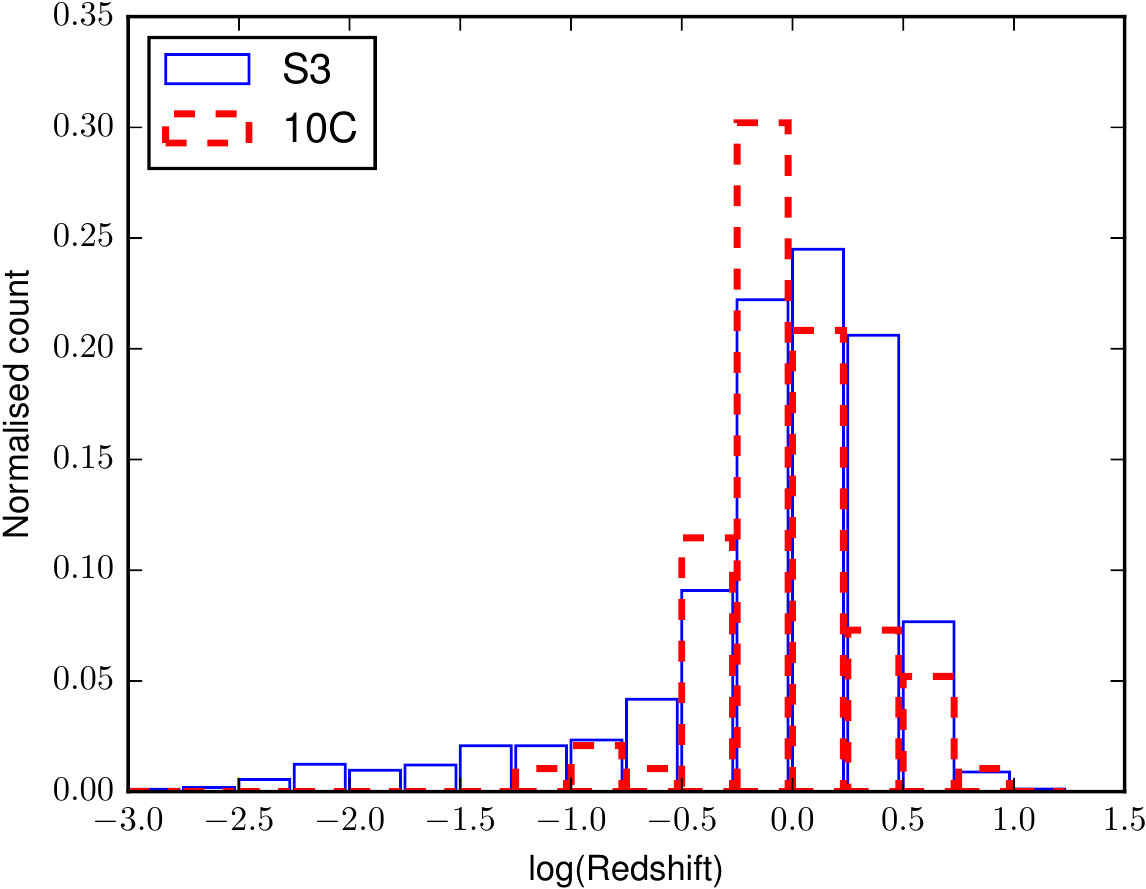}}
\caption{Normalised redshift distribution for sources in S$^3$ and in the 10C sample. Objects with no redshift value have been omitted, but included in the normalisation.}\label{fig:s_cubed_z}
\end{figure} 

Fig.~\ref{fig:s_cubed_z} shows the normalised redshift distribution for the S$^3$ and 10C samples. The 10C sample is normalised by the total sample size (96 sources), which includes the 19 sources with no redshift value available. The redshift distributions of the two samples are similar, although the observed sample displays a sharper peak and is missing the extensive low redshift tail in the simulated sample. Eleven percent of the sources in the simulated sample have $z < 0.2$, so we would expect to find approximately ten sources in the observed sample in this range if the two distributions are similar. However, there are only three sources in the observed sample with a redshift less than 0.2. The majority of the sources with $z<0.2$ in the simulation are starforming sources, so the lack of low-redshift sources in the 10C sample is consistent with the fact that there are very few starforming sources in this sample. 

It is also a possibility that some of the 18 sources which lack redshift information and are therefore missing from the observed sample have redshifts less than 0.2 and are therefore responsible for the discrepancy in the redshift distributions. It is unlikely, however, that any of the eight sources without a match are at low redshift, as they would have to be very faint in the optical to have $z < 0.2$ and not be detected in the optical or infrared observations. Only two of the 30 possible counterparts for the confused sources have $z<0.2$, so these sources cannot account for the missing sources in this redshift range.

The simulation therefore incorrectly predicts that there is a population of low-redshift starforming galaxies in the 10C sample. This could indicate that the spectra assumed for the starforming sources in the simulation is wrong, and they in fact have much steeper spectra, or that the extrapolation of the luminosity function for the starforming galaxies is not correct. Studies of the faint ($S_{1.4~\rm GHz} < 0.1$~mJy) source population at lower frequencies by \citet{2012MNRAS.421.3060S}, \citet{2014MNRAS.440.1527L} and \citet{2015arXiv150701144L} have also found fewer starforming galaxies than predicted by the simulation. These results support our suggestion that the faint end of the luminosity function for starforming galaxies in the simulation is not correct, with the number of starforming galaxies being overestimated.

The peak in the redshift distribution also appears to be shifted to slightly lower redshifts in the observed sample, with more sources in the bins $0.56 < z < 1.26$ ($-0.25 < {\rm log}(z) < 0.1$) but fewer sources in the bins $z > 1.8$ (${\rm log}(z) > 0.25$).  It is plausible that some of the sources without an optical counterpart have $z > 1.8$, and are therefore the cause of this difference.

\section{Comparison with other studies}\label{section:other-studies}

\citet{2008A&A...477..459M} investigated the properties of a complete sample of 131 radio sources with $S > 0.4$~mJy observed at 1.4 and 5~GHz as part of the Australia Telescope ESO Slice Project (ATESP) 5~GHz radio survey \citep{2006A&A...457..517P}. This sample provides a useful comparison as it has a comparable flux density limit to the 10C survey, albeit at a lower frequency. The ESO Deep Public Survey provides deep multi-colour (\textsl{UBVRIJK}) images which cover most of this field, and optical/near-infrared counterparts are found for 66 out of the 85 (78 percent) sources in the area covered. Estimates of redshift and optical object type are obtained for 56 of these 66 sources. These results showed that 78 percent of the ATESP 5~GHz sample with optical identifications had an active nucleus (i.e.\ they are either quasars or radio galaxies associated with early-type objects), significantly lower that the proportion of radio galaxies found in the 10C sample (94 percent). This is confirmed by looking at the radio-to-optical light ratios of the two samples; approximately $30$ percent of the full ATESP sample (131 sources) have $R < 1000$ (classifying them as radio quiet), compared to just six percent of the 10C sources. This suggests selection at a higher frequency at this flux density level preferentially selects radio galaxies, as the steep-spectrum starforming galaxies drop out of the sample.

\citeauthor{2008A&A...477..459M} find that those sources in the ATESP sample which have flat or inverted radio spectra and are associated with objects with early-type spectra are preferentially compact (with linear sizes $<$ 10 to 30 kpc). They suggest that these sources may be FRIs, due to their low radio powers ($P_{1.4~\rm GHz} \sim 10 ^{22-24}~\rm W\, Hz^{-1}$) and the absence of emission lines in their optical spectra. They do, however, note that they would expect FRI sources to have larger linear sizes and steeper spectra. As these sources have flat spectra, we would expect them to be present in significant numbers in the 15-GHz-selected 10C sample; these are likely to be the flat, core-dominated radio-loud sources which we observe in the 10C sample (but which are not present in, for example, the S$^3$ simulation).

\citet{2010A&A...510A..42P} followed up a sample of early-type galaxies selected from the ATESP 5~GHz survey at 4.8, 8.6 and 19~GHz to further investigate their properties. The main aim was to establish whether the AGN population of the sub-mJy sample is more closely related to efficiently accreting systems (such as radio-quiet quasars), or to systems with low accretion rates (such as FRI galaxies), or to low radiative efficiency accretion flows. They compare this AGN population to the much brighter ($>500$~mJy) 20~GHz AT20G Bright Source Sample \citep{2008MNRAS.384..775M} and find strong similarities in the radio spectra of the two samples. They therefore conclude that the ATESP AGN sources are lower luminosity counterparts of the AT20G FRII radio galaxies, and do not find any compelling evidence for a radio-quiet AGN population. This is consistent with the properties of the 10C sources, as the vast majority of the sources are radio loud. 

\section{Conclusions}

In this paper we have studied the multi-wavelength properties of a complete sample of 96 sources selected from the 10C survey. This sample was matched to the Data Fusion multi-wavelength catalogue, which contains up to ten photometric bands in the optical and mid-infrared, and counterparts were identified for 80 out of 96 10C sources. Spectroscopic redshifts are available for 24 sources.

Photometric redshifts were estimated for all sources with sufficient photometric information available using the \textsc{Le Phare} code. This produced redshift estimates for 70 of the 80 sources, albeit with large errors in some cases. The results are compared to two published photometric redshift catalogues (F12 and RR13), and are generally in good agreement, although there are some significant outliers. These catalogues were then combined to produce a final redshift catalogue, which contains redshift estimates for 77 (24 spectroscopic and 54 photometric) out of the 96 sources in the sample. The median redshift of the sample is 0.91 with an interquartile range of 0.84. The large errors on some of the redshift estimates, along with the significant discrepancies between catalogues in some cases, mean that photometric methods cannot be used to produce a reliable redshift for any individual source. They can, however, provide information about the properties of a population as a whole. 

The radio-to-optical and radio-to-infrared ratios (or lower limits) were calculated for all 96 sources using $i$-band magnitudes and 3.6-$\rm \muup m$ flux densities respectively. Using these ratios six sources are classified as radio quiet using at least one of the two values. Only one of these sources is significantly radio quiet; the other five lie close to the radio-loud/radio-quiet boundary. Therefore at least 90 out of the 96 sources (94 per cent) in the sample are radio loud, indicating that the 10C sample is dominated by radio galaxies. All six potentially radio-quiet sources have rising spectra, ruling out the possibility that they are starforming galaxies, so their emission is dominated by AGN activity.

These results confirm the conclusions of Paper I that the faint, flat-spectrum sources which are found to dominate the 10C sample below $\sim 1$~mJy are the cores of radio galaxies.

The overall radio properties of the sources in the sample are discussed in light of this redshift information; luminosities and linear sizes are derived for those sources with redshift estimates. There is a large range of 15.7-GHz luminosities, with values comparable to those of powerful FRI and FRII sources. There is no correlation between luminosity and spectral index.

The redshift distribution for sources in the sample is compared to the distribution of the S$^3$ catalogue; the samples have similar distributions, although the sources with $z<0.2$ which are predicted to be present by the simulation are missing from the 10C sample. These low-redshift sources in the simulated sample are starforming sources, so the fact that they are missing from the sample is consistent with the finding that there are essentially no starforming sources in the 10C sample.

The proportion of radio-loud sources in the 10C 15.7-GHz selected sample ($\gtrsim$ 94 percent) is significantly higher than the proportion in the ATESP 5-GHz selected sample ($\sim$ 60 percent), which has a comparable flux density range. High frequency surveys are therefore a very effective method of selecting sub-mJy radio-loud AGN, as the steep-spectrum starforming galaxies found in samples selected at lower frequencies are not present. The radio galaxies in this sample will be discussed further in a following paper, where they will be split into high-excitation and low-excitation radio galaxies and the properties of these two classes of sources are compared.

\section*{Acknowledgements}

IHW acknowledges a Science and Technology Facilities Council studentship. IHW, MJ, MV acknowledge support from the Square Kilometre Array South Africa project and the South African National Research Foundation. MV is supported by the European Commission Research Executive Agency FP7-SPACE-2013-1 Scheme (Grant Agreement 607254 - Herschel Extragalactic Legacy Project - HELP). This research has made use of the NASA/IPAC Extragalactic Database (NED) which is operated by the Jet Propulsion Laboratory, California Institute of Technology, under contract with the National Aeronautics and Space Administration. Opinions expressed and conclusions arrived at are those of the authors and not necessarily attributed to the SKA SA. We thank the anonymous referee for their helpful comments.

%
%

\setlength{\labelwidth}{0pt} 

\bsp

\appendix
\section{Multi-wavelength data}\label{app:A}

Table \ref{tab:all-sources} shows the multi-wavelength data and redshift estimates available for the 96 10C sources studied in this paper. Table \ref{tab:specz} lists the origins of the spectroscopic redshifts used.

%
%
\makeatletter
\let\@makecaption=\SFB@makefigurecaption
\makeatother
\setlength{\rotFPtop}{0pt plus 1fil}
\setlength{\rotFPbot}{0pt plus 1fil}
\def\vpad{{\Large$\mathstrut$}}

\begin{sidewaystable*}
\caption{Multi-wavelength properties of the 96 10C sources studied in this paper. Optical and infrared magnitudes or flux densities are listed, along with redshift values. Redshifts from the \textsc{Le Phare} photometric redshift fitting described in Section \ref{section:lephare} are given, as well as values from the \citet{2013MNRAS.428.1958R} and \citet{2012ApJS..198....1F} catalogues. The `best' redshift value, used in the analysis, is also listed. All magnitudes are AB magnitudes.}\label{tab:all-sources}
\small\rm
\renewcommand{\tabcolsep}{1.3mm}
\smallskip
\centering
\begin{tabular}{lccccccddddddcccccccc}\hline
10C ID & $g$ & $i$ & $r$ & $z$ & $J$ & $K$ & \dhead{SERVS1}    & \dhead{SERVS2}   & \dhead{SWIRE1}    & \dhead{SWIRE2}    & \dhead{SWIRE3}    & \dhead{SWIRE4}    & flag$^a$ & \textsc{Le Phare} $z$ & RR13 $z$ & F12 $z$ & Spec $z$ & Best $z^b$ & $z$ flag2$^c$\\
           &     &     &     &     &     &     & \dhead{$\muup$Jy} & \dhead{$\muup$Jy}& \dhead{$\muup$Jy} & \dhead{$\muup$Jy} & \dhead{$\muup$Jy} & \dhead{$\muup$Jy} &   &   &  &  &  &  & \\\hline
\vpad
  10C J104320+585621 & 21.17 & 19.62 & 20.01 &   &   & 17.94 & 207.74 & 168.12 & 175.33 & 152.47 & 118.58 & 96.27 & 2 & 0.51 & 0.30 &   & 0.35 & 0.35 & 1\\
  10C J104328+590312 &   &   & 19.40 &   &   & 17.42 & 303.22 & 232.28 & 264.97 & 206.65 & 115.63 & 94.74 & 5 & 0.53 & 0.24 &   &   & 0.24 & 2\\
  10C J104344+591503 &   &   &   &   &   &   & 164.69 & 188.75 & 140.07 & 170.76 & 154.28 & 79.02 & 5 & 1.60 & 0.91 &   &   & 0.91 & 2\\
  10C J104428+591540 &   &   &   &   & 21.69 & 20.22 & 69.74 & 60.02 & 64.62 & 61.74 & 55.49 & 34.17 & 6 & 1.13 &   &   & 0.36 & 0.36 & 1\\
  10C J104441+591949 &   &   &   &   & 20.89 & 19.79 & 111.46 & 103.01 & 100.31 & 97.79 & 65.54 & 56.97 & 6 & 1.22 & 1.30 &   &   & 1.30 & 2\\
  10C J104451+591929 &   &   &   &   & 20.41 & 19.50 & 98.60 & 67.75 & 91.08 & 66.04 & 34.65 &  & 1 & 0.87 & 0.96 &   &   & 0.96 & 2\\
  10C J104528+591328 &   &   &   &   & 21.77 & 21.06 & 34.37 & 45.52 & 31.73 & 44.83 & 86.0 & 200.26 & 5 & 3.28 & 1.79 &   & 2.31 & 2.31 & 1\\
  10C J104539+585730 &   &   & 20.25 &   & 18.54 & 17.90 & 192.83 & 152.48 & 168.3 & 140.06 & 96.94 & 65.49 & 1 & 0.49 & 0.38 &   & 0.39 & 0.39 & 1\\
  10C J104551+590838 &   &   & 24.03 &   &   & 22.77 & 5.53 & 4.77 & 3.83 &  &  &  & 5 & 4.52 & 0.75 &   &   & 0.75 & 2\\
  10C J104624+590447 &   &   &   &   &   &   & 2.93 & 2.89 & 3.37 &  &  &  & 1 &   & 1.86 &   &   & 1.86 & 2\\
  10C J104630+582748 &   & 17.41 & 17.57 &   &   & 16.10 & 1355.66 & 902.51 & 1072.43 & 742.61 & 477.94 & 374.37 & 1 & 0.79 &   &   & 0.12 & 0.12 & 1\\
  10C J104633+585816 &   &   & 21.58 &   & 20.00 & 19.44 & 160.06 & 119.52 & 134.25 & 99.1 & 110.45 & 88.94 & 5 & 4.72 & 0.79 &   & 0.85 & 0.85 & 1\\
  10C J104648+590956 &   &   &   &   &   & 21.59 & 36.78 & 41.80 & 28.45 & 35.53 & 48.72 & 36.16 & 6 & 5.88 &   &   &   & 5.88 & 4\\
  10C J104700+591903 &   &   &   &   &   &   &   &   &  &  &  &  & 4 &   &   &   &   &   &  \\
  10C J104710+582821 &   & 20.51 & 21.29 &   &   & 18.33 & 357.52 & 378.50 & 351.19 & 396.96 & 536.34 & 711.65 & 1 & 2.68 & 0.59 &   &   & 0.59 & 2\\
  10C J104718+585119 &   &   & 22.86 &   & 20.75 & 20.12 & 42.82 & 29.45 & 35.68 & 28.47 & 32.65 &  & 2 & 0.88 & 0.64 &   &   & 0.64 & 2\\
  10C J104719+582114 &   & 18.84 & 18.42 &   &   & 18.40 & 325.51 & 443.00 & 319.68 & 434.31 & 620.55 & 973.67 & 5 & 0.56 &   &   & 1.22 & 1.22 & 1\\
  10C J104733+591244 &   &   &   &   &   & 21.87 & 10.41 & 14.53 & 11.44 & 14.26 &  &  & 6 &  &   &   &   &  & 4\\
  10C J104737+592028 &   &   &   &   & 20.62 & 19.53 & 131.04 & 90.74 & 114.0 & 78.66 & 68.31 &  & 5 & 4.66 & 0.79 &   &   & 0.79 & 2\\
  10C J104741+584811 &   &   &   &   &   &   &   &   &  &  &  &  & 3 &   &   &   &   &   &  \\
  10C J104742+585318 &   & 20.71 & 21.69 &   & 19.34 & 18.58 & 189.76 & 120.75 & 166.58 & 107.98 & 105.92 & 42.34 & 2 & 0.75 & 0.58 &   &   & 0.58 & 2\\
  10C J104751+574259 & 23.47 &   & 23.28 &   &   &   & 30.24 & 36.07 & 28.25 & 35.97 &  &  & 5 & 1.91 & 1.63 &   &   & 1.63 & 2\\
  10C J104802+574117 & 22.59 & 22.01 & 22.70 &   &   &   & 77.07 & 69.56 & 71.2 & 66.97 &  &  & 5 & 1.25 & 1.38 &   &   & 1.38 & 2\\
  10C J104822+582436 &   & 22.75 & 23.20 &   &   & 19.90 & 77.20 & 70.49 & 69.69 & 68.42 & 43.27 &  & 5 & 1.22 & 1.23 &   &   & 1.23 & 2\\
  10C J104824+583029 &   & 21.07 & 22.12 &   &   & 18.57 & 237.95 & 152.17 & 206.9 & 134.95 & 108.09 & 71.81 & 1 & 0.76 &   &   &   & 0.76 & 4\\
  10C J104826+584838 &   & 23.31 &   &   &   & 21.40 & 20.56 & 15.49 & 17.0 & 12.5 & 26.87 &  & 2 & 4.94 & 0.96 &   &   & 0.96 & 2\\
  10C J104836+591846 & 20.81 & 21.35 & 21.87 & 21.38 & 20.80 & 20.65 & 39.23 & 48.12 & 35.25 & 43.57 & 56.75 & 57.49 & 5 & 0.40 & 1.94 &   &   & 1.94 & 2\\
  10C J104844+582309 &   & 21.12 & 22.26 &   &   & 18.62 & 274.28 & 170.77 & 219.63 & 149.27 & 115.99 & 83.84 & 1 & 0.77 & 0.86 &   &   & 0.86 & 2\\
  10C J104849+571417 & 23.45 & 20.94 & 22.04 &   &   &   & 160.51 & 101.35 & 140.16 & 91.36 & 66.5 &  & 1 & 0.66 & 0.61 &   &   & 0.61 & 2\\
  10C J104856+575528 &   &   &   &   &   &   & 44.12 & 44.85 & 38.34 & 40.39 &  &  & 2 &  & 4.75 &   &   & 4.75 & 2\\
  10C J104857+584103 &   & 22.87 &   &   &   & 20.20 & 58.20 & 37.14 & 46.2 & 27.28 &  &  & 5 & 0.82 & 0.68 &   &   & 0.68 & 2\\
  10C J104906+571156 &   &   &   &   &   &   &   &   &  &  &  &  & 3 &   &   &   &   &   &  \\
  10C J104918+582801 &   & 20.10 & 20.05 &   &   & 19.00 & 124.20 & 142.39 & 117.07 & 138.94 & 210.61 & 270.57 & 1 & 3.16 &   &   & 2.30 & 2.30 & 1\\
  10C J104927+583830 &   &   &   &   &   &   &   &   &  &  &  &  & 6 &   &   &   &   &   &  \\
  10C J104934+570613 &   &   & 23.87 &   &   &   & 90.90 & 68.82 & 83.47 & 65.03 &  &  & 2 &  & 1.13 &   &   & 1.13 & 2\\
\hline
\end{tabular}
\end{sidewaystable*}

\clearpage

\begin{sidewaystable*}
\contcaption{}
\small\rm
\renewcommand{\tabcolsep}{1.3mm}
\smallskip
\begingroup\centering
\begin{tabular}{lccccccddddddccccccc}\hline
10C ID & $g$ & $i$ & $r$ & $z$ & $J$ & $K$ & \dhead{SERVS1}    & \dhead{SERVS2}   & \dhead{SWIRE1}    & \dhead{SWIRE2}    & \dhead{SWIRE3}    & \dhead{SWIRE4}    & flag$^a$ & \textsc{Le Phare} $z$ & RR13 $z$ & F12 $z$ & Spec $z$ & Best $z^b$ & $z$ flag2$^c$\\
           &     &     &     & \phantom{00.00} &     &     & \dhead{$\muup$Jy} & \dhead{$\muup$Jy}& \dhead{$\muup$Jy} & \dhead{$\muup$Jy} & \dhead{$\muup$Jy} & \dhead{$\muup$Jy} &   &   &  &  &  &  & \\\hline
\vpad
  10C J104939+583530 &   & 21.66 & 22.23 &   &   & 19.60 & 90.74 & 93.88 & 62.57 & 71.52 & 72.21 & 87.13 & 5 & 1.46 & 1.70 &   & 0.97 & 0.97 & 1\\
  10C J104943+571739 & 22.35 & 20.70 & 21.48 &   &   &   & 198.68 & 129.64 & 165.34 & 114.95 & 106.26 & 91.62 & 1 & 0.73 & 0.69 &   & 0.59 & 0.59 & 1\\
  10C J104954+570456 & 20.96 & 20.33 & 21.06 &   &   &   & 287.70 & 348.87 & 285.32 & 349.98 & 477.46 & 611.71 & 5 & 2.98 & 0.75 &   & 0.53 & 0.53 & 1\\
  10C J105000+585227 &   &   &   &   & 21.76 & 20.54 & 47.27 & 45.15 & 43.79 & 40.31 &  &  & 5 & 1.26 &   &   &   & 1.26 & 4\\
  10C J105007+572020 & 23.53 & 22.64 & 23.35 &   &   &   & 82.95 & 75.08 & 66.15 & 64.82 &  &  & 2 & 1.26 & 1.22 & 1.70 &   & 1.22 & 2\\
  10C J105007+574251 & 23.83 & 21.65 & 22.57 &   &   &   &   & 60.81 & 93.42 & 62.16 & 67.62 &  & 5 & 0.83 & 0.88 & 0.72 &   & 0.88 & 2\\
  10C J105009+570724 &   &   &   &   &   &   &   &   &  &  &  &  & 3 &   &   &   &   &   &  \\
  10C J105020+574048 &   & 21.20 & 22.49 &   &   &   & 125.74 & 76.71 & 106.66 & 67.01 & 56.93 & 38.51 & 1 & 0.70 & 0.71 & 0.72 &   & 0.71 & 2\\
  10C J105028+574522 & 18.16 & 17.28 & 17.64 &   &   &   & 1088.85 & 712.67 & 905.85 & 605.93 & 472.13 & 387.6 & 2 & 0.68 &   &   & 0.07 & 0.07 & 1\\
  10C J105034+572922 &   &   &   &   &   &   & 48.97 & 49.10 & 44.9 & 44.97 &  &  & 2 &   &   & 1.12 &   & 1.12 & 3\\
  10C J105039+572339 & 19.19 & 19.35 & 19.48 &   &   &   & 125.72 & 189.30 & 124.32 & 183.84 & 251.84 & 364.94 & 2 & 0.80 & 1.69 & 0.27 & 1.44 & 1.44 & 1\\
  10C J105039+574200 &   & 21.96 & 22.77 &   &   &   & 55.42 & 34.92 & 46.17 & 30.86 & 33.1 &  & 5 & 0.71 & 0.86 & 0.74 &   & 0.86 & 2\\
  10C J105039+585118 &   &   & 20.19 &   & 18.59 & 17.97 & 273.96 & 215.72 & 205.5 & 174.39 & 103.99 &  & 2 & 1.05 & 0.44 &   & 0.37 & 0.37 & 1\\
  10C J105040+573308 &   &   &   &   &   &   & 83^\ast  &   &  &  &  &  & 4 &   &   &   &   &   &  \\
  10C J105042+575233 &   &   &   &   &   &   & 55.44 & 44.15 & 48.29 & 40.4 &  &  & 1 &   & 1.24 &   &   & 1.24 & 2\\
  10C J105050+580200 & 23.28 & 21.09 & 22.12 &   &   & 18.90 & 147.43 & 113.71 & 152.26 & 127.77 & 146.55 & 176.16 & 1 & 1.00 & 0.68 &   &   & 0.68 & 2\\
  10C J105053+583233 &   &   &   &   &   & 21.67 & 18.46 & 21.91 & 16.99 & 19.75 & 28.79 & 54.56 & 2 & 4.53 &   &   &   & 4.53 & 4\\
  10C J105054+580943 &   &   &   &   &   &   &   &   &  &  &  &  & 4 &   &   &   &   &   &  \\
  10C J105058+573356 &   &   &   &   &   &   &   &   &  &  &  &  & 6 &   &   &   &   &   &  \\
  10C J105104+574456 &   &   &   &   &   &   &   &   &  &  &  &  & 3 &   &   &   &   &   &  \\
  10C J105104+575415 & 21.42 & 21.01 & 21.28 &   & 20.18 & 19.99 & 92.29 & 139.91 & 93.06 & 142.93 & 201.17 & 295.42 & 2 & 0.80 & 0.90 &   & 1.67 & 1.67 & 1\\
  10C J105107+575752 & 23.40 & 20.80 & 22.07 &   & 19.43 & 18.62 & 205.92 & 125.70 & 178.24 & 108.21 & 82.13 & 42.54 & 2 & 0.75 & 0.69 &   &   & 0.69 & 2\\
  10C J105115+573552 &   &   &   &   &   &   &   &   &  &  &  &  & 6 &   &   &   &   &   &  \\
  10C J105121+582648 &   &   &   &   & 21.55 & 20.47 & 70.98 & 81.04 & 59.12 & 69.24 & 52.42 &  & 2 & 1.44 &   &   &   & 1.44 & 4\\
  10C J105122+570854 &   &   &   &   &   & 22.96 & 5.07 & 8.54 & 4.14 & 10.25 &  &  & 5 &  &   & 1.13 &   & 1.13 & 3\\
  10C J105122+584136 &   &   &   &   & 22.08 & 21.08 & 31.53 & 28.64 & 24.34 & 21.85 &  &  & 5 & 4.74 & 1.68 &   &   & 1.68 & 2\\
  10C J105122+584409 &   &   & 23.67 &   & 22.09 & 21.33 & 26.21 & 33.52 & 23.85 & 31.51 & 46.68 & 61.91 & 5 & 4.52 & 1.79 &   &   & 1.79 & 2\\
  10C J105128+570901 & 22.25 & 20.02 & 20.94 &   & 18.82 & 18.14 & 307.34 & 204.97 & 247.26 & 172.61 & 134.61 & 77.27 & 1 & 0.84 & 0.52 & 0.55 & 0.54 & 0.54 & 1\\
  10C J105132+571114 & 20.68 & 18.96 & 19.46 &   & 17.89 & 17.32 & 383.11 & 297.39 & 322.12 & 256.47 & 158.6 & 98.2 & 2 & 0.56 & 0.32 & 0.40 & 0.32 & 0.32 & 1\\
  10C J105136+572944 &   &   &   &   & 21.21 & 20.13 & 109.31 & 108.41 & 92.49 & 100.3 & 83.38 & 70.92 & 1 & 4.91 &   &   &   & 4.91 & 4\\
  10C J105138+574957 &   &   &   &   &   &   & 5.60 & 9.61 & 4.11 & 9.2 &  &  & 1 &   &   &   &   &   &  \\
  10C J105139+580757 &   &   &   &   &   &   &   &   &  &  &  &  & 3 &   &   &   &   &   &  \\
  10C J105142+573447 & 23.50 & 21.30 & 22.24 &   & 19.83 & 19.06 & 133.90 & 79.67 & 114.69 & 69.03 & 45.1 &  & 2 & 0.75 & 0.73 & 0.58 &   & 0.73 & 2\\
  10C J105142+573557 &   &   &   &   & 22.48 & 21.19 & 31.68 & 36.43 & 28.08 & 32.17 &  &  & 2 & 4.90 & 1.44 & 1.73 &   & 1.44 & 2\\
  10C J105144+573313 &   &   &   &   &   &   &   &   &  &  &  &  & 4 &   &   &   &   &   &  \\
  \hline
\end{tabular}\\
\endgroup
\smallskip
$^\ast$ Value estimated from the SERVS image.
\end{sidewaystable*}

\clearpage

\begin{sidewaystable*}
\contcaption{}
\small\rm
\renewcommand{\tabcolsep}{1.3mm}
\smallskip
\begingroup\centering
\begin{tabular}{lccccccddddddccccccc}\hline
10C ID & $g$ & $i$ & $r$ & $z$ & $J$ & $K$ & \dhead{SERVS1}    & \dhead{SERVS2}   & \dhead{SWIRE1}    & \dhead{SWIRE2}    & \dhead{SWIRE3}    & \dhead{SWIRE4}    & flag$^a$ & \textsc{Le Phare} $z$ & RR13 $z$ & F12 $z$ & Spec $z$ & Best $z^b$ & $z$ flag2$^c$\\
           &     &     &     &     &     &     & \dhead{$\muup$Jy} & \dhead{$\muup$Jy}& \dhead{$\muup$Jy} & \dhead{$\muup$Jy} & \dhead{$\muup$Jy} & \dhead{$\muup$Jy} &   &   &  &  &  &  & \\\hline
\vpad
  10C J105148+573245 & 23.84 & 22.55 & 23.21 &   & 20.84 & 19.96 & 109.27 & 145.91 & 99.44 & 146.91 & 212.26 & 344.21 & 2 & 4.30 & 0.93 & 1.00 & 0.99 & 0.99 & 1\\
  10C J105206+574111 &   &   &   &   & 21.78 & 20.90 & 39.35 & 42.88 & 33.4 & 35.74 &  & 31.8 & 5 & 4.91 &   & 1.53 & 0.46 & 0.46 & 1\\
  10C J105215+581627 &   & 22.57 & 23.36 &   & 20.74 & 19.91 & 66.99 & 44.09 & 56.39 & 35.09 &  &  & 5 & 0.86 & 0.91 &   &   & 0.91 & 2\\
  10C J105220+585051 &   &   &   &   & 21.36 & 20.44 & 67.33 & 58.80 & 50.97 & 48.14 &  &  & 1 & 4.94 & 1.94 &   &   & 1.94 & 2\\
  10C J105225+573323 & 22.87 & 20.58 & 21.59 & 20.10 & 19.22 & 18.51 & 214.07 & 133.80 & 179.69 & 115.75 & 88.42 & 51.99 & 1 & 0.75 &   & 0.63 & 0.61 & 0.61 & 1\\
  10C J105225+575507 &   &   &   &   & 21.74 & 20.93 & 32.02 & 37.17 & 30.11 & 35.91 &  & 53.81 & 5 & 4.87 &   &   &   & 4.87 & 4\\
  10C J105237+573058 &   &   &   &   &   &   &   &   &  &  &  &  & 3 &   &   &   &   &   &  \\
  10C J105240+572322 &   & 22.56 & 23.58 & 21.68 & 20.58 & 19.76 & 92.68 & 68.39 & 81.49 & 60.31 & 46.19 & 44.06 & 5 & 0.97 &   & 1.11 &   & 1.11 & 3\\
  10C J105243+574817 &   & 22.92 &   &   & 20.91 & 19.85 & 86.32 & 68.38 & 83.12 & 70.87 & 41.07 &  & 5 & 1.02 & 0.67 &   &   & 0.67 & 2\\
  10C J105327+574546 &   & 23.40 &   &   & 21.41 & 20.80 & 29.49 & 19.50 & 28.31 & 15.57 &  &  & 2 & 0.88 &   & 0.82 &   & 0.82 & 3\\
  10C J105341+571951 &   & 22.32 & 23.96 & 21.78 & 20.69 & 19.83 & 85.41 & 61.69 & 76.56 & 57.85 &  &  & 5 & 0.95 &   & 0.91 &   & 0.91 & 3\\
  10C J105342+574438 & 23.35 & 21.21 & 22.26 & 20.80 & 19.90 & 19.19 & 107.85 & 71.37 & 99.56 & 72.29 & 66.84 & 45.27 & 1 & 0.83 & 0.83 & 0.73 &   & 0.83 & 2\\
  10C J105400+573324 &   &   &   &   &   & 22.37 & 10.48 & 11.75 & 7.0 &  &  &  & 5 &  &   & 1.49 &   & 1.49 & 3\\
  10C J105425+573700 & 20.64 & 19.52 & 19.81 & 19.30 & 18.43 & 17.79 & 407.64 & 480.33 & 408.28 & 505.96 & 696.96 & 1104.53 & 2 & 0.64 &   & 0.33 & 0.32 & 0.32 & 1\\
  10C J105437+565922 &   &   &   &   &   &   &   &   &  &  &  &  & 4 &   &   &   &   &   &  \\
  10C J105441+571640 &   &   &   &   &   &   &   &   &  &  &  &  & 3 &   &   &   &   &   &  \\
  10C J105510+574503 &   & 22.64 & 23.17 & 22.10 & 20.53 & 19.69 & 93.77 & 68.16 & 85.7 & 64.98 & 43.26 &  & 2 & 0.95 & 1.14 &   &   & 1.14 & 2\\
  10C J105515+573256 &   &   &   &   &   &   &   &   &  &  &  &  & 3 &   &   &   &   &   &  \\
  10C J105520+572237 &   &   &   &   & 21.39 & 20.15 & 96.70 & 134.44 & 93.44 & 138.76 & 169.92 & 244.46 & 5 & 0.16 &   &   &   & 0.16 & 4\\
  10C J105527+571607 & 22.20 & 20.16 & 20.86 & 19.90 & 18.91 & 18.25 & 190.86 & 136.30 & 167.5 & 121.29 & 102.93 &  & 1 & 0.69 &   &   & 0.49 & 0.49 & 1\\
  10C J105535+574636 & 23.97 & 23.63 & 23.74 &   & 22.18 & 21.04 & 33.48 & 41.71 & 32.06 & 40.03 & 44.2 & 59.97 & 5 & 2.78 & 2.89 &   &   & 2.89 & 2\\
  10C J105550+570407 & 22.24 & 20.25 & 20.95 & 20.10 &   &   & 170.58 & 120.56 & 150.25 & 106.31 & 73.9 &  & 1 & 0.57 & 0.83 &   & 0.49 & 0.49 & 1\\
  10C J105604+570934 &   &   &   &   & 21.86 & 20.64 & 52.48 & 44.65 & 45.26 & 42.3 &  &  & 5 & 4.88 &   &   &   & 4.88 & 4\\
  10C J105627+574221 & 21.93 & 21.84 & 21.54 &   & 21.30 & 20.75 & 43.12 & 57.66 & 38.13 & 56.03 & 88.28 & 114.76 & 5 & 1.08 & 1.64 &   &   & 1.64 & 2\\
  10C J105653+580342 &   & 21.04 & 21.97 &   & 19.56 & 18.73 & 233.12 & 158.00 & 169.81 & 136.17 & 136.91 &  & 5 & 0.90 & 0.56 &   & 0.60 & 0.60 & 1\\
  10C J105716+572314 &   &   &   &   &   &   & 9.95 & 11.53 & 7.89 & 10.08 &  &  & 2 &   &   &   &   &   &  \\
  \hline
\end{tabular}\\
\endgroup
\smallskip
Notes:\\
a) Optical matching flag. 1 = extended, probable match; 2 = extended, possible match; 3 = extended, confused; 4 = extended, no match;\\
5 = compact, match; 6 = compact, no match; 7 = 10C position only.\\
b) Final redshift value, description is given in Section \ref{section:z-used}.\\
c) Origin of final redshift value. 1 = spectroscopic (see Table \ref{tab:specz}) 2 = RR13, 3 = F12, 4 = \textsc{Le Phare}.\\
\end{sidewaystable*}

\clearpage

\begin{table*}
\caption{References for the spectroscopic redshifts used in this work.}\label{tab:specz}
\begin{center}
\begin{tabular}{lcl}\hline
AMI ID & Spec. $z$ & Reference\\\hline
  10C J104320+585621 & 0.35 & NED$^a$\\
  10C J104428+591540 & 0.36 & NED\\
  10C J104528+591328 & 2.31 & NED\\
  10C J104539+585730 & 0.39 & NED\\
  10C J104630+582748 & 0.12 & CfA HectoSpec Spitzer Follow-Up by Huang, Rigopoulou et al. (in prep)\\
  10C J104633+585816 & 0.85 & NED\\
  10C J104719+582114 & 1.22 & NED\\
  10C J104918+582801 & 2.30 & CfA HectoSpec Spitzer Follow-Up by Huang, Rigopoulou et al. (in prep)\\
  10C J104939+583530 & 0.97 & SDSS-DR12, \citet{2015arXiv150100963A}\\
  10C J104943+571739 & 0.59 & CfA HectoSpec Spitzer Follow-Up by Huang, Rigopoulou et al. (in prep)\\
  10C J104954+570456 & 0.53 & CfA HectoSpec Spitzer Follow-Up by Huang, Rigopoulou et al. (in prep)\\
  10C J105028+574522 & 0.07 & NED\\
  10C J105039+572339 & 1.44 & NED\\
  10C J105039+585118 & 0.37 & SDSS-DR12$^b$, \citet{2015arXiv150100963A}\\
  10C J105104+575415 & 1.67 & SDSS-DR12, \citet{2015arXiv150100963A}\\
  10C J105128+570901 & 0.54 & SDSS-DR12, \citet{2015arXiv150100963A}\\
  10C J105132+571114 & 0.32 & NED\\
  10C J105148+573245 & 0.99 & NED\\
  10C J105206+574111 & 0.46 & NED\\
  10C J105225+573323 & 0.61 & SDSS-DR12, \citet{2015arXiv150100963A}\\
  10C J105425+573700 & 0.32 & ITP 2010 HerMES Follow-Up by Perez-Fournon, Page et al. (in prep.)\\
  10C J105527+571607 & 0.49 & SDSS-DR12, \citet{2015arXiv150100963A}\\
  10C J105550+570407 & 0.49 & SDSS-DR12, \citet{2015arXiv150100963A}\\
  10C J105653+580342 & 0.60 & SDSS-DR12, \citet{2015arXiv150100963A}\\
\hline
\end{tabular} 
\end{center}
\smallskip
\begin{flushleft}
Notes: \\
a) NASA/IPAC Extragalactic Database (extracted June 2013).\\
b) Sloan Digital Sky Survey Data Release 12.\\
\end{flushleft}
\end{table*}

\label{lastpage}

\begin{thebibliography}{}

 \bibitem[Alam et al.(2015)Alam et al.]{2015arXiv150100963A} 
  Alam S., et al., 2015, arXiv, arXiv:1501.00963 

 \bibitem[AMI Consortium: Davies et al.(2011)Davies et al.]{2011MNRAS.415.2708D} 
  AMI Consortium: Davies, et al., 2011, MNRAS, 415, 2708

 \bibitem[AMI Consortium: Franzen et al.(2011)Franzen et al.]{2011MNRAS.415.2699F} 
  AMI Consortium: Franzen, et al., 2011, MNRAS, 415, 2699 

 \bibitem[Appleton et al.(2004)Appleton et al.]{2004ApJS..154..147A} 
  Appleton P.~N., et al., 2004, ApJS, 154, 147 

 \bibitem[Arnouts et al.(1999)Arnouts et al.]{1999MNRAS.310..540A} 
  Arnouts S., Cristiani S., Moscardini L., Matarrese S., Lucchin F., Fontana A., Giallongo E., 1999, MNRAS, 310, 540 

 \bibitem[Biggs \& Ivison(2006)Biggs \& Ivison]{2006MNRAS.371..963B} 
  Biggs A.~D., Ivison R.~J., 2006, MNRAS, 371, 963 


 \bibitem[Brunner et al.(2008)Brunner et al.]{2008A&A...479..283B} 
  Brunner H., Cappelluti N., Hasinger G., Barcons X., Fabian A.~C., Mainieri V., Szokoly G., 2008, A\&A, 479, 283 

 \bibitem[Bruzual \& Charlot(2003)Bruzual \& Charlot]{2003MNRAS.344.1000B} 
  Bruzual G., Charlot S., 2003, MNRAS, 344, 1000 

 \bibitem[Calzetti et al.(2000)Calzetti et al.]{2000ApJ...533..682C} 
  Calzetti D., Armus L., Bohlin R.~C., Kinney A.~L., Koornneef J., Storchi-Bergmann T., 2000, ApJ, 533, 682 

 \bibitem[Condon(1980)Condon]{1980ApJ...242..894C} 
  Condon J.~J., 1980, ApJ, 242, 894 

 \bibitem[Condon et al.(1998)Condon et al.]{1998AJ....115.1693C} 
  Condon J.~J., Cotton W.~D., Greisen E.~W., Yin Q.~F., Perley R.~A., Taylor G.~B., Broderick J.~J., 1998, AJ, 115, 1693 

 \bibitem[de Zotti et al.(2010)de Zotti et al.]{2010A&ARv..18....1D} 
  de Zotti G., Massardi M., Negrello M., Wall J., 2010, A\&ARv, 18, 1 

 \bibitem[Fanaroff \& Riley(1974)Fanaroff \& Riley]{1974MNRAS.167P..31F} 
  Fanaroff B.~L., Riley J.~M., 1974, MNRAS, 167, 31P 

 \bibitem[Fotopoulou et al.(2012)Fotopoulou et al.]{2012ApJS..198....1F} 
  Fotopoulou S., et al., 2012, ApJS, 198, 1 

 \bibitem[Garn et al.(2008)Garn et al.]{2008MNRAS.387.1037G} 
  Garn T., Green D.~A., Riley J.~M., Alexander P., 2008, MNRAS, 387, 1037 

 \bibitem[Garn \& Alexander(2009)Garn \& Alexander]{2009MNRAS.394..105G} 
  Garn T., Alexander P., 2009, MNRAS, 394, 105 

 \bibitem[Garn et al.(2010)Garn et al.]{2010BASI...38..103G} 
  Garn T.~S., Green D.~A., Riley J.~M., Alexander P., 2010, BASI, 38, 103 

 \bibitem[Gonz{\'a}lez-Solares et al.(2011)Gonz{\'a}lez-Solares et al.]{2011MNRAS.416..927G} 
  Gonz{\'a}lez-Solares E.~A., et al., 2011, MNRAS, 416, 927 

 \bibitem[Guglielmino et al.(2012)Guglielmino et al.]{2012rsri.confE..22G} 
 Guglielmino G., Prandoni I., Morganti R., Heald G., 2012, in \emph{Resolving The Sky - Radio Interferometry: Past, Present and Future}, available online at http://pos.sissa.it/cgi-bin/reader/conf.cgi?confid=163, id.22

 \bibitem[Ibar et al.(2008)Ibar et al.]{2008MNRAS.386..953I} 
  Ibar E., et al., 2008, MNRAS, 386, 953 

 \bibitem[Ilbert et al.(2006)Ilbert et al.]{2006A&A...457..841I} 
  Ilbert O., et al., 2006, A\&A, 457, 841 

 \bibitem[Ilbert et al.(2009)Ilbert et al.]{2009ApJ...690.1236I} 
  Ilbert O., et al., 2009, ApJ, 690, 1236 

 \bibitem[Ishisaki et al.(2001)Ishisaki et al.]{2001PASJ...53..445I} 
  Ishisaki Y., Ueda Y., Yamashita A., Ohashi T., Lehmann I., Hasinger G., 2001, PASJ, 53, 445 

 \bibitem[Lawrence et al.(2007)Lawrence et al.]{2007MNRAS.379.1599L} 
  Lawrence A., et al., 2007, MNRAS, 379, 1599 


 \bibitem[Lindsay et al.(2014)Lindsay et al.]{2014MNRAS.440.1527L} 
  Lindsay S.~N., et al., 2014, MNRAS, 440, 1527 

 \bibitem[Lockman et al.(1986)Lockman, Jahoda, \& McCammon]{1986ApJ...302..432L} 
  Lockman F.~J., Jahoda K., McCammon D., 1986, ApJ, 302, 432 

 \bibitem[Lonsdale et al.(2003)Lonsdale et al.]{2003PASP..115..897L} 
  Lonsdale C.~J., et al., 2003, PASP, 115, 897 

 \bibitem[Luchsinger et al.(2015)Luchsinger et al.]{2015arXiv150701144L} 
  Luchsinger K.~M., et al., 2015, arXiv, arXiv:1507.01144 

 \bibitem[Machalski \& Condon(1999)Machalski \& Condon]{1999ApJS..123...41M} 
  Machalski J., Condon J.~J., 1999, ApJS, 123, 41 


 \bibitem[Marleau et al.(2007)Marleau et al.]{2007ApJ...663..218M} 
  Marleau F.~R., Fadda D., Appleton P.~N., Noriega-Crespo A., Im M., Clancy D., 2007, ApJ, 663, 218 

 \bibitem[Massardi et al.(2008)Massardi et al.]{2008MNRAS.384..775M} 
  Massardi M., et al., 2008, MNRAS, 384, 775 

 \bibitem[Mauduit et al.(2012)Mauduit et al.]{2012PASP..124..714M} 
  Mauduit J.-C., et al., 2012, PASP, 124, 714 

 \bibitem[McAlpine et al.(2013)McAlpine, Jarvis, \& Bonfield]{2013MNRAS.436.1084M} 
  McAlpine K., Jarvis M.~J., Bonfield D.~G., 2013, MNRAS, 436, 1084 


 \bibitem[Mignano et al.(2008)Mignano et al.]{2008A&A...477..459M} 
  Mignano A., Prandoni I., Gregorini L., Parma P., de Ruiter H.~R., Wieringa M.~H., Vettolani G., Ekers R.~D., 2008, A\&A, 477, 459 

 \bibitem[Murphy et al.(2014)Murphy et al.]{2014arXiv1412.5677M} 
  Murphy E.~J., et al., 2014, arXiv:1412.5677 


 \bibitem[Owen \& Morrison(2008)Owen \& Morrison]{2008AJ....136.1889O} 
  Owen F.~N., Morrison G.~E., 2008, AJ, 136, 1889 

 \bibitem[Owen et al.(2009)Owen et al.]{2009AJ....137.4846O} 
  Owen F.~N., Morrison G.~E., Klimek M.~D., Greisen E.~W., 2009, AJ, 137, 4846 

 \bibitem[Padovani et al.(2009)Padovani et al.]{2009ApJ...694..235P} 
  Padovani P., Mainieri V., Tozzi P., Kellermann K.~I., Fomalont E.~B., Miller N., Rosati P., Shaver P., 2009, ApJ, 694, 235 



 \bibitem[Polletta et al.(2007)Polletta et al.]{2007ApJ...663...81P} 
  Polletta M., et al., 2007, ApJ, 663, 81 

 \bibitem[Prandoni et al.(2006)Prandoni et al.]{2006A&A...457..517P} 
  Prandoni I., Parma P., Wieringa M.~H., de Ruiter H.~R., Gregorini L., Mignano A., Vettolani G., Ekers R.~D., 2006, A\&A, 457, 517 

 \bibitem[Prandoni et al.(2010a)Prandoni et al.]{2010A&A...510A..42P} 
  Prandoni I., de Ruiter H.~R., Ricci R., Parma P., Gregorini L., Ekers R.~D., 2010, A\&A, 510, A42 

\bibitem[Prandoni(2010b)Prandoni]{2010arXiv1008.4918P} 
 Prandoni I., 2010b, \textsl{Proceedings of the ISKAF2010 Science Meeting. June 10 -14 2010. Assen, the Netherlands.} Published online at http://pos.sissa.it/cgi-bin/reader/conf.cgi?confid=112, p.47

 \bibitem[Prevot et al.(1984)Prevot et al.]{1984A&A...132..389P} 
  Prevot M.~L., Lequeux J., Prevot L., Maurice E., Rocca-Volmerange B., 1984, A\&A, 132, 389 

 \bibitem[Rowan-Robinson et al.(2008)Rowan-Robinson et al.]{2008MNRAS.386..697R} 
  Rowan-Robinson M., et al., 2008, MNRAS, 386, 697 

 \bibitem[Rowan-Robinson et al.(2013)Rowan-Robinson et al.]{2013MNRAS.428.1958R} 
  Rowan-Robinson M., Gonzalez-Solares E., Vaccari M., Marchetti L., 2013, MNRAS, 428, 1958 

 \bibitem[Salvato et al.(2009)Salvato et al.]{2009ApJ...690.1250S} 
  Salvato M., et al., 2009, ApJ, 690, 1250 

 \bibitem[Seymour et al.(2008)Seymour et al.]{2008MNRAS.386.1695S} 
  Seymour N., et al., 2008, MNRAS, 386, 1695 

 \bibitem[Simpson et al.(2012)Simpson et al.]{2012MNRAS.421.3060S} 
  Simpson C., et al., 2012, MNRAS, 421, 3060 


 \bibitem[White et al.(1997)White et al.]{1997ApJ...475..479W} 
  White R.~L., Becker R.~H., Helfand D.~J., Gregg M.~D., 1997, ApJ, 475, 479 

 \bibitem[Whittam et al.(2013)Whittam et al.]{2013MNRAS.429.2080W} 
  Whittam I.~H., et al., 2013, MNRAS, 429, 2080 (Paper I)

 \bibitem[Whittam et al.(2014)Whittam, Riley, \& Green]{2014MNRAS.440...40W} 
  Whittam I.~H., Riley J.~M., Green D.~A., 2014, MNRAS, 440, 40 


 \bibitem[Wilman et al.(2008)Wilman et al.]{2008MNRAS.388.1335W} 
  Wilman R.~J., et al., 2008, MNRAS, 388, 1335 

 \bibitem[Wilman et al.(2010)Wilman et al.]{2010MNRAS.405..447W} 
  Wilman R.~J., Jarvis M.~J., Mauch T., Rawlings S., Hickey S., 2010, MNRAS, 405, 447 

 \bibitem[Wright et al.(2010)Wright et al.]{2010AJ....140.1868W} 
  Wright E.~L., et al., 2010, AJ, 140, 1868 

 \bibitem[Zwart et al.(2008)Zwart et al.]{2008MNRAS.391.1545Z} 
  Zwart J.~T.~L., et al., 2008, MNRAS, 391, 1545 


\end{thebibliography}
\end{document}